\documentclass{aa}
\usepackage{graphicx,epsfig,psfrag}
\usepackage{txfonts}
\usepackage{natbib}
\bibpunct{(}{)}{;}{a}{}{,}

\def \Ha {H$\alpha$ }
\def \Hi {H{\sc i}\,}
\def \Hei {He{\sc i}\,}
\def \Heii {He{\sc ii}\,}
\def \Heiii {He{\sc iii}\,}
\def \Ciii {C{\sc iii}\,}
\def \Civ {C{\sc iv}\,}
\def \Oii {O{\sc ii}\,}
\def \Oiv {O{\sc iv}\,}
\def \Siiv {Si{\sc iv}\,}
\def \Feii {Fe{\sc ii}\,}

\def \Feiv {Fe{\sc iv}\,}
\def \Fev {Fe{\sc v}\,}
\def \Fevi {Fe{\sc vi}\,}

\def \taur {\tau_{\rm R}}

\def \Rstar {$R_\star$\,}
\def \Rsun  {$R_\odot$\,}
\def \Msun  {$M_\odot$\,}
\def \Teff {$T_{\rm eff}$\,}
\def \logg {$\log g$\,}
\def \Yhe {$Y_{\rm He}$\,}

\def \vinf {$v_\infty$\,}
\def \Mdot {$\dot M$\,}

\def\kms {km~s$^{-1}$}
\def\Mdu{$\cdot 10^{-6} {\rm M_{\odot}/yr}$}

\def \beq{\begin{equation}}
\def \eeq{\end{equation}}
\def \beqa{\begin{eqnarray}}
\def \eeqa{\end{eqnarray}}
\def \disp{\displaystyle}

\def \rarr{\rightarrow}
\def \Trad{T_{\rm rad}}
\def \Tradone {\displaystyle {T_{\rm r,1}}}
\def \Tradm   {\displaystyle {T_{\rm r,m}}}

\def \Tradonej {\displaystyle {T_{\rm r,1j}}}
\def \Tradonem {\displaystyle {T_{\rm r,1m}}}
\def \Tradij   {\displaystyle {T_{\rm r,ij}}}
\def \Tradmj   {\displaystyle {T_{\rm r,mj}}}
\def \Tmj   {\displaystyle {T_{{\rm r,m}_j}}}
\def \Te {\displaystyle {T_{\rm e}}}
\def \mj {m_j}
\def \CSj {C_{S''_j}}

\def \alo{\Lambda_{\nu}^{\ast}}
\def \dJ{\Delta J_{\nu}}

\def \Rik {R_{i \kappa}}
\def \Rki {R_{\kappa i}}

\def \Rstare {R_\star}
\def \Rmaxe {R_{\rm max}}
\def \Teffe {T_{\rm eff}}
\def \vinfe {v_\infty}
\def \Mdote {\dot M}

\def \dd{{\rm d}}
\def \vthm{v_{\rm th}^{\rm m}}
\def \<{\langle}
\def \>{\rangle}

\def \chim{\<\chi^{\rm tot}\>}
\def \chic{\chi_\nu^{\rm cont}}
\def \chict{\chi_\nu^{\rm c,true}}
\def \chilm{\<\chi_{\rm L}\>}
\def \chil{\chi_{\rm L}}
\def \chilj{\chi_{{\rm L}j}}

\def \Sl {S_{\rm L}}
\def \etact{\eta_\nu^{\rm c,true}}
\def \sige{\sigma_{\rm e}}

\def \fnth{f_{\rm nth}}
\def \fth{f_{\rm th}}
\begin{document}
\title{Atmospheric NLTE-Models for the Spectroscopic Analysis of Blue Stars
with Winds} 
\subtitle{II. Line-Blanketed Models}

\author{J. Puls\inst{1} 
        \and M. A.  Urbaneja\inst{2}
	\and R. Venero\inst{3}
        \and T. Repolust\inst{1}
	\and U. Springmann\inst{4}
        \and A. Jokuthy\inst{1} 
        \and M. R. Mokiem\inst{5} }

\offprints{J. Puls}

\institute{Universit\"{a}ts-Sternwarte M\"{u}nchen,
	Scheinerstrasse 1, D-81679 M\"unchen, Germany\\
        \email{uh101aw@usm.uni-muenchen.de}
        \and 
	Institute for Astronomy, University of Hawaii at Manoa, 
	2680 Woodlawn Drive, Honolulu, Hawaii 96822\\  
        \email{urbaneja@ifa.hawaii.edu}
	\and 
        Facultad de Ciencias Astron\'omicas y Geof\'{\i}sicas, Universidad
        Nacional de La Plata,\\
	Paseo del Bosque s/n, B1900FWA La Plata, Argentina\\
        \email{roberto@fcaglp.unlp.edu.ar}
        \and 
        BT (Germany) GmbH \& Co. oHG, Barthstr. 22, D-80339 M\"unchen\\
        \email{uwe@springmann.net}
	\and
	Astronomical Institute ``Anton Pannekoek'', Kruislaan 403, NL-1098 SJ
        Amsterdam\\
	\email{mokiem@science.uva.nl}}

\date{Received; accepted }

\abstract{
We present new or improved methods for calculating NLTE, line-blanketed
model atmospheres for hot stars with winds (spectral types A to O), with 
particular emphasis on a {\it fast performance}. These
methods have been implemented into a previous, more simple version of the
model atmosphere code {\sc Fastwind} (\citealt{sph97}) and allow to
spectroscopically analyze rather large samples of massive stars in a
reasonable time-scale, using state-of-the-art physics. Although this updated
version of the code has already been used in a number of recent
investigations, the corresponding methods have not been explained in detail
so far, and no rigorous comparison with results from alternative codes has
been performed. This paper intends to address both topics.

In particular, we describe our (partly approximate) approach to solve the
equations of statistical equilibrium for those elements which are primarily
responsible for line-blocking and blanketing, as well as an approximate
treatment of the line-blocking itself, which is based on a simple statistical
approach using suitable means for line opacities and emissivities. Both
methods are validated by specific tests. Furthermore, we comment on our
implementation of a consistent temperature structure.

In the second part, we concentrate on a detailed comparison with
results from those two codes which have been used in alternative 
spectroscopical investigations, namely {\sc cmfgen} (\citealt{hil98}) and
{\sc wm}-Basic (\citealt{Paul01}). All three codes predict almost identical
temperature structures and fluxes for $\lambda >$ 400~\AA, whereas at lower
wavelengths a number of discrepancies are found. Particularly in the \Heii
continua, where fluxes and corresponding numbers of ionizing photons react
extremely sensitively to subtle differences in the models, we consider any
uncritical use of these quantities (e.g., in the context of nebula
diagnostics) as being dangerous. Optical H/He lines as synthesized by {\sc
fastwind} are compared with results from {\sc cmfgen}, obtaining a remarkable 
coincidence, except for the \Hei\, singlets in the temperature range
between 36,000 to 41,000~K for dwarfs and between 31,000 to 35,000~K for
supergiants, where {\sc cmfgen} predicts much weaker lines. Consequences
due to these discrepancies are discussed.

Finally, suggestions are presented how to adequately parameterize 
model-grids for hot stars with winds, with only one additional 
parameter compared to standard grids from plane-parallel, hydrostatic models. 

}             
\authorrunning{J. Puls et al.}
\titlerunning{Line-blanketed NLTE model atmospheres}

\maketitle
%

\section{Introduction}
\label{intro}

During the last years, the quantitative spectroscopy of massive stars with
winds has made enormous progress due to the development of NLTE (non-local
thermodynamic equilibrium) atmosphere codes which allow for the treatment of
metal-line blocking and blanketing. With respect to both spectral range
(from the extreme ultraviolet, EUV, to the infrared, IR) and metallicity of
the analyzed objects (from SMC-abundances to Galactic center stars), a wide
range in parameters can now be covered. Presently, five different codes are
in use which have been developed for specific objectives, but due to
constant improvements they can be applied in other contexts as well. In
particular, these codes are {\sc cmfgen} (\citealt{hil98}), the
``Potsdam-group'' code developed by W.R.~Hamann and collaborators (for a
status report, see \citealt{Graef02}), the ``multi-purpose model atmosphere
code'' {\sc phoenix} (\citealt{hau99}), {\sc wm}-Basic (\citealt{Paul01}) and
{\sc fastwind}, which will be described here (see also \citealt{sph97} and
\citealt{h02} for previous versions). 

The first three of these codes are the most ``exact'' ones, since {\it all}
lines (including those from iron-group elements) are treated in the comoving
frame (CMF), which of course is a very time-consuming task. Moreover, since
the first two of these codes have originally been designed for the analysis
of the very dense winds from Wolf-Rayet stars, the treatment of the
photospheric density stratification is approximative (constant photospheric
scale-height). For several analyses this problem has been cured by
``coupling'' {\sc cmfgen} with the plane-parallel, hydrostatic code {\sc
tlusty} developed by \citet{hub95} (e.g., \citealt{bouret03}).

The multi-purpose code {\sc phoenix} is mainly used for the analysis of
supernovae and (very) cool dwarfs, but also a small number of hotter objects
have been considered, e.g., the A-type supergiant Deneb
(\citealt{aufdenberg02}). Due to this small number a detailed comparison
with corresponding results is presently not possible, and, therefore, we
will defer this important task until more material becomes available.

In contrast to all other codes which use a pre-described mass-loss rate and
velocity field for the wind structure, the model atmospheres from {\sc
wm}-Basic are calculated by actually solving the hydrodynamical equations
(with the radiative line-pressure being approximated within the
force-multiplier concept, cf. \citealt {CAK, PPK}) deep into the
photosphere. Thus, this code provides a more realistic stratification of
density and velocity, particularly in the transonic region (with the
disadvantage that the slope of the velocity field cannot be manipulated if 
the wind does not behave as theoretically predicted). Since {\sc wm}-Basic
aims mainly at the prediction of EUV/UV fluxes and profiles, the bound-bound
radiative rates are calculated using the Sobolev approximation (including continuum
interactions), which yields ``almost'' exact results except for
those lines which are formed in the transonic region (e.g.,
\citealt{sph97}). Moreover, line-blocking is treated in an effective way (by
means of opacity sampling throughout a first iteration cycle, and
``exactly'' in the final iterations), so that the computational time is
significantly reduced compared to the former three codes. 

{\sc fastwind}, finally, has been designed to cope with optical and IR
spectroscopy of ``normal'' stars with $\Teffe \ga 8,500$~K\footnote{i.e,
molecules do not play any role and hydrogen remains fairly ionized.}, i.e.,
OBA-stars of all luminosity classes and wind strengths.  

Since the parameter space investigated for the analysis of one
object alone is large, comprising the simultaneous derivation of effective
temperature \Teff, gravity \logg, wind-strength parameter $Q =
\Mdote/(\Rstare \vinfe)^{1.5}$ (cf. Sect.~\ref{modelgrids}), velocity field
parameter $\beta$, individual abundances (most important:
helium-abundance \Yhe) and also global background metallicity $z$, 
much computational effort is needed to calculate the vast amount of
necessary models. This is one of the reasons why the samples which have been
analyzed so far by both {\sc cmfgen} and {\sc wm}-Basic 
are not particularly large\footnote{From here on, we will concentrate
on the latter two codes because of our objective of analyzing ``normal''
stars, whereas the ``Potsdam''-code has mainly been used to analyze
WR-stars.}, comprising typically five to seven objects per analysis (e.g.,
\citealt{Hillier03,bouret03,martins04} for recent {\sc cmfgen}-analyses and
\citealt{fulli00,bg02,gb04} for recent {\sc wm}-Basic analyses).

The reader may note that although the number of fit-parameters gets 
smaller when the wind-strength becomes negligible, a certain difference
between the results from ``wind-codes'' and plane-parallel, hydrostatic
model atmospheres still remains: More or less independent of the actual
mass-loss rate, there will always be an enhanced probability of photon
escape from lines in regions close to the sonic point and above, if a
super-sonic velocity field is present. A prime example illuminating the
consequences of this enhanced escape is the \Heii ground-state depopulation
in O-stars (\citealt{Gabler89}), even though it is diminished by
line-blocking effects compared to the original case studied by means of pure
H/He atmospheres (see also Sect.~\ref{test_nlteapprox}).

With the advent of new telescopes and multi-object spectrographs, the number
of objects which can be observed during one run has significantly increased 
(e.g., {\sc flames} attached to the VLT allows for an observation of roughly
120 objects in parallel). An analysis of those samples will definitely 
result in more reliable parameters due to more extensive statistics but
remains prohibitive unless the available codes are considerably fast.

This was and still is the motivation which has driven the development of
{\sc fastwind}. We have always considered a fast performance to be of
highest priority. The required computational efficiency is obtained by
applying appropriate physical approximations to processes where high
accuracy is not needed (regarding the objective of the analysis - optical/IR
lines), in particular concerning the treatment of the metal-line background
opacities. 

Meanwhile, a number of analyses have been performed with our present version
of {\sc fastwind}, with significant sample sizes, of the order of 10 to 40
stars per sample (e.g., \citealt{urban03, trundle04, urban04, Repo04, massey04,
massey05}). Although the code has been carefully tested and first
comparisons with results from {\sc cmfgen} and {\sc tlusty} have been
published (\citealt{h02}), a detailed description of the code and an 
extensive comparison have not been presented so far. Particularly the latter
task is extremely important, because otherwise it is almost impossible to
compare the results from analyses performed using different codes and 
to draw appropriate conclusions. An example of this difficulty is the
discrepancy in stellar parameters if results from optical and UV analyses
are compared. Typically, UV-spectroscopy seems to result in lower values for
\Teff than a corresponding optical analysis, e.g., \citet{massey05}. Unless
the different codes have been carefully compared, no one can be sure
whether this is a problem related to either inadequate physics or 
certain inconsistencies within the codes.

This paper intends to answer part of these questions and is organized as
follows: In Sect.~\ref{basics} we give a quick overview of the basic
philosophy of the code, and in Sect.~\ref{atomdat} we describe the atomic
data used as well as our treatment of metallicity regarding the
flux-blocking background elements. Sects.~\ref{nlteapprox} and
\ref{lineblock} give a detailed description of our approach to obtain the
fast performance desired: Sect.~\ref{nlteapprox} details on the approximate
NLTE solution for the background elements (which is applied if {\it no}
consistent temperature structure is aimed at), and Sect.~\ref{lineblock}
describes our present method to tackle the problem of line-blocking.  Both
sections include important tests which have convinced ourselves of the 
validity of our approach, particularly after a comparison with results from {\sc
wm}-Basic. Sect.~\ref{treatinv} covers the problem of level inversions and
how to deal with them, and Sect.~\ref{tempstrat} comprises the calculation
of a consistent temperature structure. In Sect.~\ref{cmfgencomp}, a detailed
comparison with results from a grid of {\sc cmfgen} models\footnote{as
recently calculated by \citet{len04}} is performed, and
Sect.~\ref{modelgrids} suggests how to
parameterize model-grids adequately and reports on first progress. In
Sect.~\ref{summary}, finally, we present our summary and an outlook
regarding future work.

\section{Basic philosophy of the code}
\label{basics}

In the following, we will summarize the basic features of {\sc fastwind},
before we describe in detail the methods used. The first version of the code
(unblocked atmosphere/line formation) has been introduced by 
\citet[ hereafter Paper~I]{sph97}, and has been significantly improved
meanwhile. Let us first mention that we distinguish between two groups of
elements, namely the so-called {\it explicit} ones and the {\it background}
elements.

The explicit elements (mainly H, He, but also C, N, O, Si, Mg in the
B-star range, see below) are those which are used as diagnostic tools and
are treated with high precision, i.e., by detailed atomic models and by 
means of CMF transport for the bound-bound transitions. In order to allow
for a high degree of flexibility and to make use of any improvements in
atomic physics calculations, the code is atomic data driven with respect to
these ions, as explained in Paper~I: the atomic models, all necessary
data and the information on how to use these data are contained in a user
supplied file (in the so-called {\sc detail} input form, cf.
\citealt{ButGid85}) whereas the code itself is independent of any specific
data.

The background ions, on the other hand, are those allowing for the effects
of line-blocking/blanketing. The corresponding data originate from
\citet{Paul98, Paul01} and are used as provided, i.e., in a certain, fixed
form.

{\sc fastwind} follows the concept of ``unified model atmospheres'' (i.e., a
smooth transition from a pseudo-hydrostatic photosphere to the wind) along
with an appropriate treatment of line-broadening (Stark, pressure-) which is
a prerequisite for the analysis of O-stars of different luminosity classes
covering a variety of wind densities. Particularly and as already described
in Paper~I, the photospheric density consistently accounts for the
temperature stratification and the actual radiation pressure, now by including
both the explicit {\it and} the background elements. 

The corresponding occupation numbers and opacities (of the
background-elements) can be derived in two alternative ways: 
\begin{enumerate}
\item[a)]
in those cases, where the temperature stratification is calculated by means
of NLTE Hopf parameters (see below), we apply an approximate NLTE solution
for {\it all} background elements following the principal philosophy
developed by \citet{ab85}, \citet{schmutz91}, \citet{schaerer94} and
\citet{puls00}, where important features have now been improved (cf.
Sect.~\ref{nlteapprox}). Particularly, the equations of approximate
ionization equilibrium have been re-formulated to account for the actual
radiation field as a function of depth and frequency, and a consistent
iteration scheme regarding the coupling of the rate equations and the
radiation field has been established to avoid the well-known convergence
problems of a pure Lambda Iteration (Sect.~\ref{ali}).
\item[b)]In the other case, when the T-stratification shall 
be calculated from first principles, the complete set of rate equations is
solved almost ``exactly'' for the most abundant background elements (C, N,
O, Ne, Mg, Si, S, Ar, Fe, Ni, if not included as explicit ions), employing
the Sobolev approximation for the net radiative rates (with actual
illuminating radiation field). The remaining background elements, on the
other hand, remain to be treated by the approximation as outlined in a).
\end{enumerate}
In order to account for the effects of line-blocking, we use suitable means
for the line opacities, averaged over a frequency interval of the order of
1,000{\ldots} 1,500 \kms, and appropriate emissivities (Sect.~\ref{lineblock}).

Finally, the temperature stratification can be calculated in two different
ways. As long as one is exclusively interested in an optical analysis, the
concept of NLTE-Hopf parameters (cf. Paper~I) is still sufficient, if the
background elements are accounted for in a consistent way, i.e., have been
included in the particular models from which these parameters are derived. 
Since this method is flux-conservative, the correct amount of
line-blanketing is ``automatically'' obtained. Note that for optical depths
$\taur \la 0.01$ a lower cut-off temperature is defined, typically at
$T_{\rm min} = 0.6 \Teffe$. 

Alternatively, the new version of {\sc fastwind} allows for the calculation
of a consistent\footnote{Note, however, that non-radiative heating processes
might be of importance, e.g., due to shocks.} temperature, utilizing a
flux-correction method in the lower atmosphere and the thermal balance of
electrons in the outer one (Sect.~\ref{tempstrat}). As has been discussed,
e.g., by \citet{kubat99}, the latter method is advantageous compared to
exploiting the condition of radiative equilibrium in those regions where the
radiation field becomes almost independent on $\Te$. Particularly for the
IR-spectroscopy, such a consistent T-stratification is important, since the
IR is formed above the stellar photosphere in most cases and depends
critically on the run of $\Te$ in those regions, where our first method is
no longer applicable. 

\section{Atomic Data and Metallicity}
\label{atomdat}

\paragraph{Explicit elements.} In order to obtain reliable results also in
the IR, we have significantly updated our H- and He-models compared to those 
described in Paper~I. Our present H and \Heii\, models consist of 20 levels each (vs. 10
and 14 in the previous version, respectively), and \Hei\, includes levels
until $n=10$, where levels with $n=8{\ldots} 10$ have been packed (previous
version: 8 levels, packed from 5{\ldots} 8). Further information concerning
cross-sections etc. can be found in \citet{jok02}. Present atomic models for
metals have been accumulated from different sources, mainly with respect to
an analysis of B-stars, i.e., for ionization stages {\sc ii} and {\sc
iii}, except for Mg ({\sc i,ii}) and Si ({\sc ii, iii, iv}). Information
on our Si atomic model can be found in \citet{trundle04}, and on the
other metals incorporated so far (C, N, O, Mg) in \citet{urban04}.

\paragraph{Background elements.} The atomic data for background elements
originate from \citet{Paul98, Paul01}, who have given a detailed description
on the various approaches and sources. These data comprise the elements 
from hydrogen to zinc (except Li, Be, B and Sc which are too rare to
affect the background opacity) with ionization stages up to {\sc viii}.  The
number of connecting lines (lower and upper level present in the rate
equations) is of the order of 30,000, and the number of lines where only the
lower level is present is $4.2\cdot 10^6$. The former group of 
lines is used to solve the rate equations, whereas the latter is used to 
derive the metal-line background opacities (cf.  Sect~\ref{lineblock}). In
addition to bound-free cross-sections and $gf$-values, there is also
detailed information about the collision-strengths for the most important
collisional bound-bound transitions in each ion. 

\paragraph{Metallicity.} The abundances of the background elements are taken
from the solar values provided by \citet[ and references
therein]{Grevesse98}\footnote{Of course, the user is free to change these
numbers.}. For different ``global'' metallicities, $z=Z/Z_\odot$, these
abundances are scaled proportionally with respect to {\it mass} ratios,
e.g., by 0.2 for the SMC and by 0.5 for the LMC (although these values are
certainly disputable, e.g. \citealt{massey04} and references therein).

A particular problem (independent on the actual value of $z$) appears in
those cases when the He/H ratio becomes non-solar. In this case, we retain
the specific relative mass fractions of the other elements, which of course
has a significant effect on the {\it number} ratios.  Although this
procedure is not quite right, it preserves at least the overall mass
fraction of the metals, particularly the {\it unprocessed} iron group
elements, which are most important for the line-blocking. Further comments
on the validity of this procedure have been given by \citet{massey04}. Let
us briefly mention a comparison to evolutionary calculations
from \citet{schaerer93} performed by P. Massey (priv. comm.).

For the 120 \Msun track at $Z=0.008$ (roughly the LMC metallicity), $Z$
stays essentially unchanged in the core until the end of core H burning,
even though the mass fraction of C and N are going up while O is going down:
at a number ratio \Yhe = 2 (i.e., the mass ratio $Y$ has changed from 0.265
to 0.892), the value for $Z$ has changed insignificantly from 0.0080 to
0.0077, and even more interestingly, the mass fraction of the sum of C, N,
O, and Ne has essentially changed in the same way (0.0075 to 0.0070), even
though the actual mass fraction of N has more than doubled.

\section{Background elements: approximate NLTE occupation numbers}
\label{nlteapprox}

In order to save significant computational effort, the occupation numbers of
the background elements are calculated by means of an approximate solution
of the NLTE rate equations. Such an approach has been successfully applied
in a variety of stellar atmosphere calculations, e.g., to derive the radiative
acceleration of hot star winds (\citealt{ab85, lucyab93}) and for the
spectroscopy of hot stars (\citealt{schmutz91, schaerer94}) and Supernova
remnants (\citealt{mazlucy93, lucy99, maz00}). \citet{puls00} have used this
method for an examination of the line-statistics in hot star winds,
by closely following a procedure discussed by \citet{springmann97} which in
turn goes back to unpublished notes by L.~Lucy. 

One might argue that such an approximate approach can barely handle all the
complications arising from sophisticated NLTE effects. However, in the
following we will show that the approximate treatment is able to match
``exact'' NLTE calculations to an astonishingly high degree, at least if 
some modifications are applied to the original approach. Moreover,
the calculated occupation numbers will {\it not} be used to synthesize
line-spectra, but serve ``only'' as lower levels for the line-opacities
involved in the blocking calculations.

Actually, the major weakness of the original approach is the assumption of a
radiation field with {\it frequency independent} radiation temperatures
$\Trad$. Since solely the difference in radiation temperatures at strong
ionization edges is responsible for a number of important effects, we
have improved upon this simplifiction by using consistent radiation
temperatures (taken from the solution of the equations of radiative
transfer). As we will see in the following, this principally minor
modification requires a number of additional considerations.

\subsection{Selection of levels}

One of the major ingredients entering the approximate solution of the rate
equations is a careful selection of participating atomic levels. In
agreement with the argumentation by \citet{ab85} only the following levels
are used: 

\begin{itemize}
\item the ground-state level 
\item all meta-stable levels (from equal and different spin systems), 
denoted by ``M''
\item all excited levels which are coupled to the ground-state via 
one single {\it permitted} transition where this transition
is the strongest among all possible downward transitions; in the following
denoted as subordinate levels ``N''
\item all excited levels coupled to one of the meta-stable levels
$m \in M$ in a similar way (subordinate levels ``S'').
\end{itemize}

\noindent In the above definition, the term ``strongest'' refers to the
Einstein-coefficients $A_{ji}$. All other levels are neglected, since their
population is usually too low to be of importance and cannot be approximated
by simple methods.

\subsection{Ionization equilibrium}

In order to allow for a fast and clearly structured algorithm, we
allow only for ionizations to and recombinations from the ground-state of the
next higher ion, even if this is not the case in reality. Due to this
restriction and by summing over all line-processes an ``exact'' rate
equation connecting two neighboring ions is derived which exclusively 
consists of ionization/recombination processes. In the following, we will
further neglect any collisional ionization/recombination processes, which is
legitimate in the context considered here, namely in the NLTE-controlled
atmospheric regime of hot stars. (In the lowermost, LTE dominated part of the
atmosphere, $\tau_{\rm R} > 2/3$, we approximate the occupation numbers a
priori by LTE conditions).

\begin{figure*}
\begin{minipage}{8.8cm}
\resizebox{\hsize}{!}
   {\includegraphics{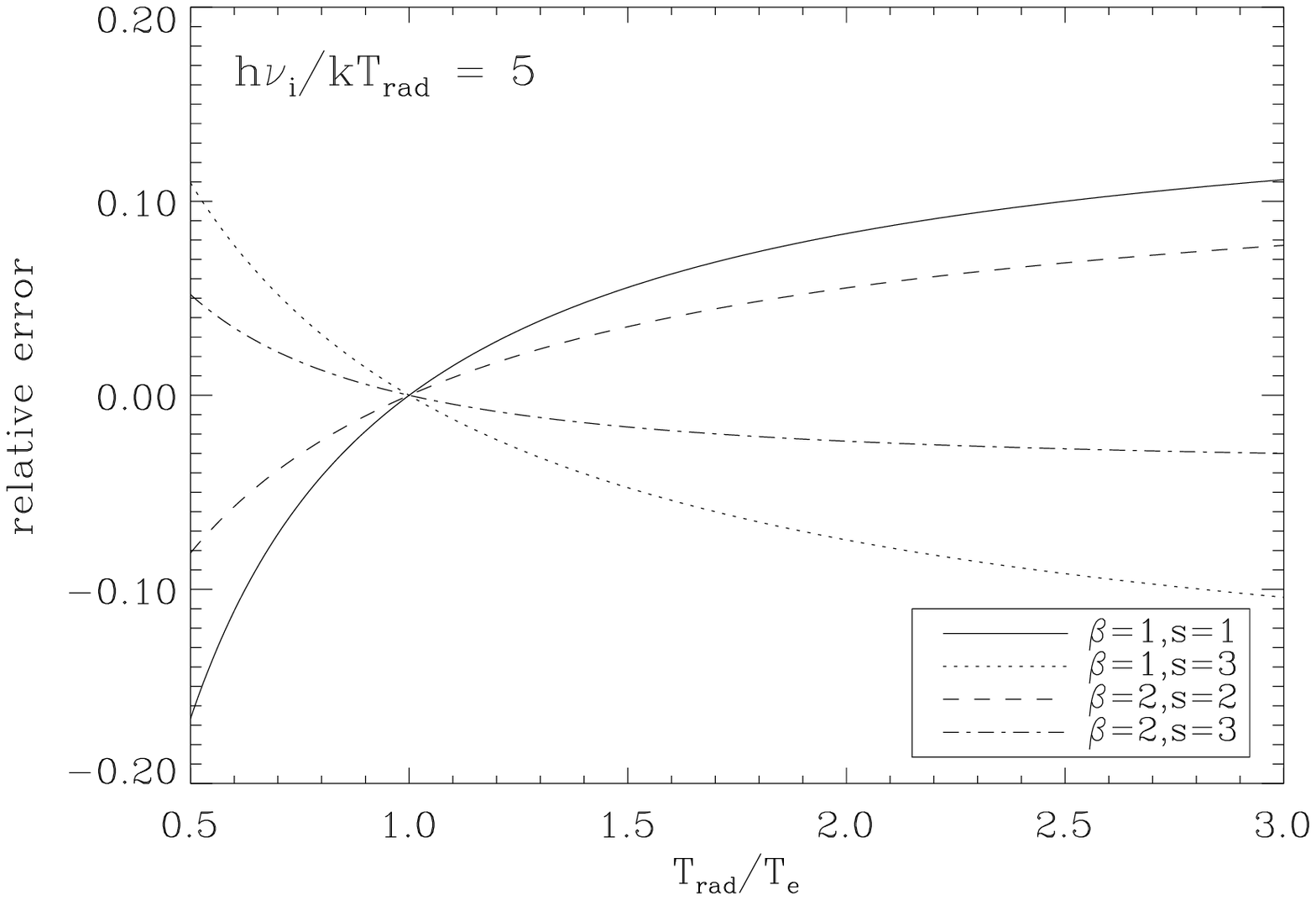}}
\end{minipage}
\hfill
\begin{minipage}{8.8cm}
   \resizebox{\hsize}{!}
      {\includegraphics{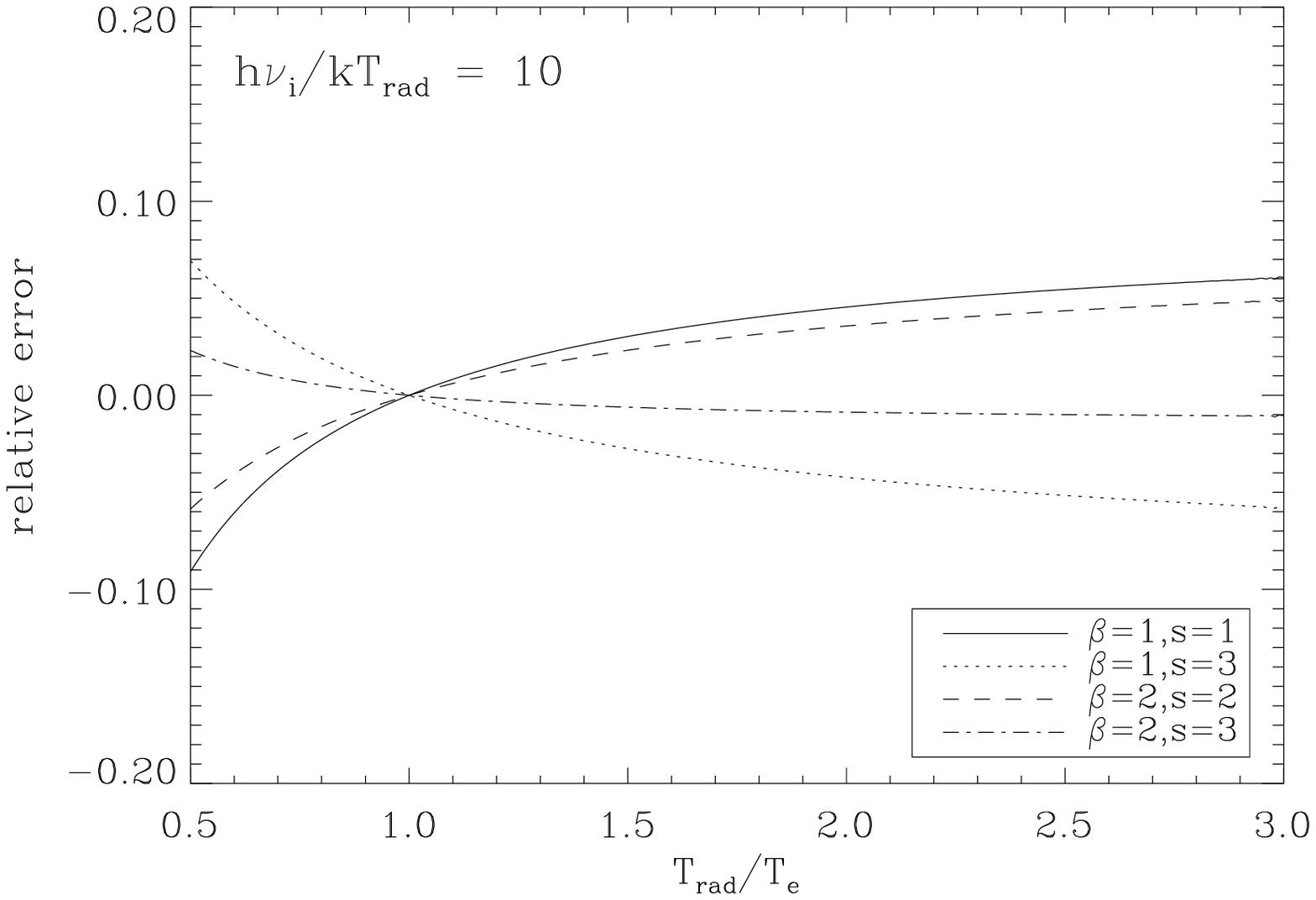}}
\end{minipage}
\caption{Ratio of ionization to recombination rate coefficients: relative error 
between ``exact'' ratios (Eq.~\ref{ratex}) and approximate ones 
(Eq. ~\ref{ratapp}, with $\beta = 1$ and $s
= 2$) as a function of $T_{\rm rad}/T_{\rm e}$, for different combinations of
($\beta, s$ ). The error decreases for even higher ionization energies.} 
\label{photo}
\end{figure*}

At first, let us consider an ion with only one spin system, e.g.,
a hydrogenic one. In this case, the ionization equilibrium becomes
\beq
\label{ionis0}
\sum_i n_{i} R_{i \kappa} = n_{\kappa} 
            \sum_i \left(\frac{n_i}{n_\kappa} \right)^* R_{\kappa i} =
            \sum_i n_i^* R_{\kappa i},
\eeq
with $n_{i}$ the occupation numbers of the lower ionization stage,
$n_{\kappa}$ the (ground-state!) occupation number of the higher ion, the
asterisks denoting LTE-conditions (at actual electron density, 
$(n_i/n_{\kappa})^* = n_e \Phi(T)$, cf. \citealt[ Sect.~5]{Mihalas}) and
ionization/recombination rate coefficients 
\beq 
R_{i \kappa} = \int_{\nu_i}^\infty \frac{4 \pi J_\nu}{h \nu}
                       a(\nu)\, \dd \nu
\eeq
\beq
R_{\kappa i}  = \int_{\nu_i}^\infty \frac{4 \pi}{h \nu} 
                       \left( \frac{2h \nu^3}{c^2} + J_\nu \right)
                       {\rm e}^{-h \nu/kT_{\rm e}}  a(\nu)\,\dd \nu.
\eeq
$J_\nu$ is the mean intensity, $a(\nu)$ the ionization cross-section and all
other symbols have their usual meaning. Once more, within our above
approximation (ionization to ground state only), Eq.~\ref{ionis0} is
``exact'' {\it and does not depend on any assumption concerning the
bound-bound processes} (radiative or collisional; optically thick or thin) 
since the corresponding rates drop out after summation. 

By introducing the
recombination coefficient $\alpha_i$ defined in the conventional way,
\beq
\label{rec_coeff}
n_\kappa n_e \alpha_i  = n_i^* R_{\kappa i},
\eeq
the ionization equilibrium can be reformulated
\beq
\label{rec_coeff1}
\sum_i n_{i} R_{i \kappa} = n_{\kappa} n_e \sum_i \alpha_i, 
\eeq
and we extract all quantities referring to the ground-state of the lower ion,
\beq
n_{\kappa} n_e = \frac{1}{\sum_i \alpha_i} n_1 R_{1 \kappa} 
     \left(1 \,+\,\frac{\sum_{i>1} n_i R_{i \kappa}}{n_1 R_{1 \kappa}}\right).
\eeq
Finally, inserting the ground-state recombination coefficient $\alpha_1$ (cf.
Eq.~\ref{rec_coeff}) on the {\it rhs}, 
we obtain the ionization equilibrium expressed as the 
ratio of two neighboring ground-states,
\beq
\label{ionis}
\frac{n_\kappa}{n_1} = \left(\frac{n_\kappa}{n_1} \right)^*
                        \frac{R_{1 \kappa}}{R_{\kappa 1}} 
			\frac{\alpha_1}{\sum_i \alpha_i} 
                 \left\{ 1 + \sum_{i \in M,N,S} \frac{n_{i}}{n_{1}}
                     \frac{R_{i \kappa}}{R_{1 \kappa}} \right\}. 
\eeq
Note that this ratio depends on the radiation field, the
actual electron density and temperature and on the {\it excitation} within
the lower ion, which will be discussed in the next subsection. Note also 
that all what follows is ``only'' a simplification of this equation. 

So far, our derivation and the above result are identical to previous
versions of the approximate approach. From now on, however, we will include
the frequency dependence of the radiation field. To this end, we describe
the ionization cross-sections by the {\it Seaton-approximation} 
(\citealt{Seaton58}), which is not too bad for most ions,
\beq
a(\nu) = a_i \left( \beta \left( \frac{\nu_i}{\nu}\right)^s
        + (1 - \beta)\left( \frac{\nu_i}{\nu}\right)^{s+1}\right).
\eeq
Writing the mean intensity as $J_{\nu}(r) = W(r) B_{\nu}(\Trad(\nu, r))$ with
dilution factor $W$  and neglecting the stimulated emission in the
recombination integral (valid for all important ionization edges), we obtain
(radial dependence of all quantities suppressed in the following)
\beqa 
\label{rik}
R_{i \kappa} &=& \frac{8 \pi W}{c^2} \left(\frac{k T_{\rm r,i}}{h}\right)^3
                a_i\, \mathcal{F}(x_{\rm r,i};\beta,s) \\
R_{\kappa i} &=& \frac{8 \pi}{c^2} \left(\frac{k T_{\rm e}}{h}\right)^3
                a_i\, \mathcal{F}(x_{\rm e,i};\beta,s)
\eeqa
with $x_{\rm r,i}=h\nu_i/kT_{\rm r,i}$, $x_{\rm e,i}=h\nu_i/kT_{\rm e}$ and
\beq
\mathcal{F}(x;\beta,s) = \beta\, x^s\, \Gamma(3-s,x) + 
                     (1-\beta)\,x^{1+s}\,\Gamma(2-s,x).
\eeq
We have assumed $\Trad(\nu) =: T_{\rm r,i}$ to be constant 
over the decisive range of the ionizing continuum $\nu \ga \nu_i$, since
only those frequencies close to the edge are relevant. In other
words, each transition is described by a {\it unique} radiation temperature. 
In the above equation, the incomplete Gamma-function $\Gamma(a,x)$ has been
generalized to include also negative parameters, $a \le 0$.
The ratio of ground-state ionization/recombination rate coefficients is thus 
given by
\beq 
\label{ratex}
	\frac{R_{1 \kappa}}{R_{\kappa 1}} = 
        W\,\left(\frac{\Tradone}{T_{\rm e}}\right)^3
         \frac{\mathcal{F}(x_{\rm r,1};\beta,s)}
	{\mathcal{F}(x_{\rm e,1};\beta,s)},
\eeq
i.e., is independent of the actual value of the cross-section at the threshold,
$a_i$. Although this expression is rather simple, it requires
the somewhat time-consuming evaluation of the incomplete Gamma-functions. To
keep things as fast as possible, we generally 
use the parameter set ($\beta=1, s=2$) instead of the actual parameters 
which results in a particularly simple function $\mathcal{F}$,
\beq
\mathcal{F}(x;1,2)=x^2 \exp(-x).
\eeq
Note that these parameters do {\it not} correspond to the hydrogenic
cross-section, which would be described by $s=3$. Using this parameter set,
the ionization/recombination rates simplify to
\beqa
\label{def_rki}
R_{i \kappa} &=& \frac{8 \pi W}{c^2}
	\left(\frac{k T_{\rm r,i}}{h}\right)
                    a_i \nu_i^2 {\rm e}^{-h \nu_i/k T_{\rm r,i}} \nonumber\\
        R_{\kappa i} &=& \frac{8 \pi}{c^2}\left(\frac{kT_{\rm e}}{h}\right)
                    a_i \nu_i^2 {\rm e}^{-h \nu_i/kT_{\rm e}},
\eeqa
and the ratio $R_{1 \kappa}/R_{\kappa 1}$ becomes
\beq 
\label{ratapp} 
\frac{R_{1 \kappa}}{R_{\kappa 1}} = W \,
        \frac{\Tradone}{T_{\rm e}}\,
        \exp \left[ -\frac{h \nu_1}{k}
        \left(\frac{1}{\Tradone} - \frac{1}{T_{\rm e}} \right)\right].
\eeq
We have convinced ourselves that this approximation leads 
to acceptable errors of the order of 10\%, cf. Fig.~\ref{photo}.
Furthermore, we define the following quantities, where $\zeta$ is just the
ratio of ground-state to total recombination coefficient,
\beq
\label{def_zeta}
\frac{\alpha_1}{\sum_i \alpha_i} = \zeta,
\sum_{i \in {\rm M(N,S)}} \frac{n_{i}}{n_{1}} \frac{R_{i \kappa}}{R_{1 \kappa}}
\equiv C_{M(N,S)}.
\eeq
Let us point out that any ratio $\alpha_i/\alpha_j$ 
(particularly, the case $j=1$ and thus $\zeta$) is independent of the
temperature and depends exclusively on atomic quantities, namely
cross-section, transition frequency and statistical weight, a fact which 
follows from Eqs.~\ref{rec_coeff} and \ref{def_rki}:
\beqa
\label{aiaj}
\frac{\alpha_i}{\alpha_j} & = & 
  \left(\frac{n_i}{n_j} \right)^* \frac{R_{\kappa i}}{R_{\kappa j}} 
  = \left(\frac{n_i}{n_j} \right)^* \frac{a_i}{a_j} 
  \left(\frac{\nu_i}{\nu_j} \right)^2 
  \exp \left[ -\frac{h(\nu_i - \nu_j)}{kT_{\rm e}} \right] = \nonumber \\
  &=& \frac{a_i}{a_j} \left(\frac{\nu_i}{\nu_j} \right)^2 \frac{g_i}{g_j}.  
\eeqa
Collecting terms, our approximate ionization equilibrium finally reads 
\beqa
\label{nkn1}
\frac{n_\kappa}{n_1} & = & \left(\frac{n_\kappa}{n_1} \right)^* 
        W\,\frac{\Tradone}{T_{\rm e}}\,
        \exp \left[ -\frac{h \nu_1}{k}
        \left(\frac{1}{\Tradone} - \frac{1}{T_{\rm e}} \right)\right]
        \times \nonumber \\
&\times& \zeta\,(1+C_N+C_M+C_S) \nonumber \\
&=& \left(\frac{n_\kappa}{n_1} \right)^*_{\Tradone} W 
\sqrt{\frac{T_{\rm e}}{\Tradone}}\, \zeta \,(1+C_M+C_N+C_S),
\eeqa
where the second variant uses the LTE ratio evaluated at actual
electron-density and radiation temperature of the ionizing continuum.

\subsection{Excitation}
\label{excitation}
The remaining step concerns the term in the bracket above, i.e., the 
approximate calculation of the excitation inside the lower ion (which, of
course, is also required in order to calculate the partition functions). For
consistency, frequencies (energies) are still defined with respect to the
ionization threshold, i.e., line frequencies have to be calculated from
$\nu_{ij} = \nu_i - \nu_j >0$ instead of the usual definition (upper -
lower) which would refer to excitation energies.

\subsubsection{Meta-stable levels}
\label{ex_meta}
We begin with the occupation numbers of meta-stable levels which can be
populated via excited levels or via the continuum (see also \citealt{ab85}).

\begin{figure}
      \begin{center}
        \psfrag{j}[l][l]{j}
        \psfrag{m}[l][l]{m}
        \psfrag{1}[l][l]{1}
        \psfrag{m1}[c][c]{$\nu_{1j} \approx \nu_{mj}$, $\Tradonej \approx \Tradmj$}
        \psfrag{m2}[l][l]{$\nu_{mj} < \nu_{1j} \approx \nu_{1m}$, $\Tradonej \approx \Tradonem$}
        \epsfig{file=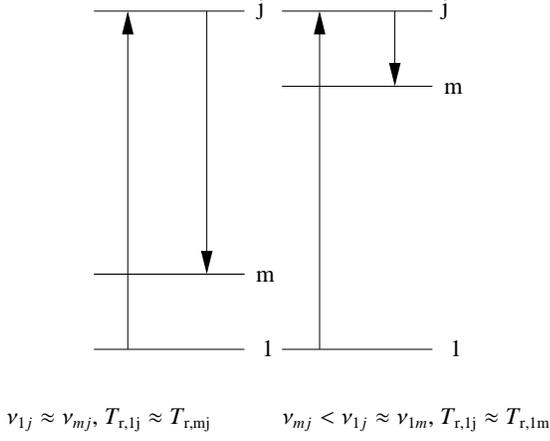, scale=0.5}
        \caption{Population of meta-stable levels via excited ones (see text).}
        \label{metaplot}
      \end{center}
\end{figure}

\paragraph{Population via excited levels.} Denoting the excited level
by $j$, considering the fact that this excited level is fed by
the ground state (otherwise it would not exist in our level hierarchy) and
neglecting collisional processes, the population can be approximated by
\beq
\label{meta}
\frac{n_m}{n_1} = \frac{W \left(\frac{n_j}{n_1} \right)^*_{\Tradonej}}
			{W \left(\frac{n_j}{n_m} \right)^*_{\Tradmj}},
			\qquad m \in {\rm M} \quad (j \in {\rm N} > m)
\eeq
(see also Eq.~\ref{normal} with $\delta \approx 0$), where the dilution factor
cancels out. In the following, we have to distinguish between two cases: 
the meta-stable level lies either close to the ground-state or 
close to the excited level, cf. Fig.~\ref{metaplot}.

\noindent {\it Case a: low lying meta-stable level.} The transition
frequencies of both transitions are fairly equal, $\nu_{1j} \approx
\nu_{mj}$, i.e, $\Tradonej \approx \Tradmj$, and we find
\beq
\label{meta_low}
\frac{n_m}{n_1} \approx 
 \frac{g_m}{g_1} \exp\left(-\frac{h\nu_{1m}}{k\Tradonej}\right)=
\left(\frac{n_m}{n_1} \right)^*_{\Tradonej}.
\eeq
Note that the population is controlled by the radiation field
$\Tradonej$, i.e., from frequencies much larger than the ``excitation
energy'', $\nu_{1m}$. 

\noindent{\it Case b: high lying meta-stable level.} Now we have
$\nu_{mj} < \nu_{1j} \approx \nu_{1m}$,

\beq
\label{meta_high}
\frac{n_m}{n_1} \approx \frac{g_m}{g_1} 
\exp \left( -\frac{h\nu_{1j}}{k\Tradonej} \right) \approx
\frac{g_m}{g_1} \exp \left( -\frac{h\nu_{1m}}{k\Tradonem} \right)=
\left(\frac{n_m}{n_1} \right)^*_{\Tradonem}.
\eeq
and the population depends on the radiation field at (or close to) the excitation
energy.

\paragraph{Population via continuum.} The third case comprises a population
via the continuum which will only be treated in a crude approximation,
where a correct evaluation will be given later (Sect.~\ref{diffspin}). If we
neglect for the moment the influence of any meta-stable and excited levels,
we find from Eq.~(\ref{nkn1}) with $\zeta \rarr 1, C_{N(M,S)} \rarr 0$
\beqa
\label{meta_cont}
\frac{n_m}{n_1} = \frac{\left(\frac{n_{\kappa}}{n_1} \right)}
     {\left(\frac{n_{\kappa}}{n_m} \right)} &\approx&
   \frac{\left(\frac{n_{\kappa}}{n_1} \right)^*_{\Tradone}}
	{\left(\frac{n_{\kappa}}{n_m} \right)^*_{\Tradm}} 
	\sqrt{\frac{\Tradm}{\Tradone}} \nonumber \\
&=&   \frac{g_m}{g_1}\, \frac{\Tradone}{\Tradm}\,
      \exp \left[ -\frac{h}{k}
   \left(\frac{\nu_1}{\Tradone} - \frac{\nu_m}{\Tradm} \right)\right].
\eeqa
Note that all three cases converge to the {\it identical} result
\beq
\label{meta_cont_tradconst}
\frac{n_m}{n_1} \rarr \left(\frac{n_m}{n_1}\right)^*_{\Trad} \quad {\rm for}
\quad \Trad = {\rm const} 
\eeq
which is quoted by \citet{ab85}. 

In order to continue our calculation of 
$C_M$, we find from Eqs.~\ref{def_rki} and \ref{aiaj} 
\beq 
\frac{R_{m \kappa}}{R_{1 \kappa}} =
        \frac{\Tradm}{\Tradone}\,\frac{\alpha_m}{\alpha_1}\,\frac{g_1}{g_m}\,
        \exp \left[ -\frac{h}{k}
        \left(\frac{\nu_m}{\Tradm} - \frac{\nu_1}{\Tradone} \right)\right].
\eeq
Multiplying by $n_m/n_1$ we find that for the three cases Eqs.~\ref{meta_low}, 
\ref{meta_high} and \ref{meta_cont} 
\beq
\left(\frac{n_m R_{m \kappa}}{n_1 R_{1 \kappa}}
       \right)_{m \in {\rm M}} = 
       \left\{
       \begin{array}{l}
       \disp \frac{\alpha_m}{\alpha_1}\frac{\Tradm}{\Tradone}
        \exp \left[ -\frac{h}{k}
        \left(\frac{\nu_m}{\Tradm} - \frac{\nu_1}{\Tradone} + 
	\frac{\nu_{1m}}{\Tradonej} \right)\right] \\
       \disp \frac{\alpha_m}{\alpha_1}\frac{\Tradm}{\Tradone}
        \exp \left[ -\frac{h}{k}
        \left(\frac{\nu_m}{\Tradm} - \frac{\nu_1}{\Tradone} + 
	\frac{\nu_{1m}}{\Tradonem} \right)\right]\\
       \disp \frac{\alpha_m}{\alpha_1},
       \end{array}
       \right.
\eeq
respectively. As mentioned before, the result for the third case (population
over continuum) is only a crude approximation, which is also evident from
the fact that it depends only on atomic quantities but not on any
radiation temperature.

\subsubsection{Subordinate levels}

Due to our definition of subordinate levels their population can
be approximated by a two-level-atom Ansatz (between ground-state $j=1$ and
subordinate level $i \in N$ or between meta-stable level $j \in M$  
and subordinate level $i \in S$), such that the population can be expressed by 
\beq 
\label{normal}
\frac{n_i}{n_j} = W(1-\delta)\left( \frac{n_i}{n_j}
		\right)^*_{\Tradij}
		+ \delta\left( \frac{n_i}{n_j} \right)^*_{\Te}, 
		\quad i \in {\rm N (S)},\,\, j \in {\rm 1 (M)}
\eeq
where $\delta$ is the parameter expressing the competition between
thermalization ($\delta \rarr 1$) and local escape (in Sobolev approximation),
\beq 
\label{deldef}
\delta = \frac{\epsilon}{\epsilon(1-\beta) + \beta}.
\eeq
$\epsilon$ is the usual LTE parameter in a two-level atom,
\beq
\epsilon = \frac{C_{ji}}{A_{ji} + C_{ji}},
\eeq
with collisional de-excitation rate $C_{ji}$ and Einstein-coefficient $A_{ji}$.
$\beta$ is the local escape probability in Sobolev approximation,
\beq
\beta  =  \frac{1}{2} \int_{-1}^{1}
	  \frac{1 - {\rm e}^{-\tau_S(\mu)}}{\tau_S(\mu)}\, \dd \mu, 
\label{betasob}
\eeq
and the illuminating radiation field is approximated by
\beq
\beta_c I_c = \frac{1}{2} \int_{\mu_{\ast}}^{1} I_c(\mu)
	  \frac{1 - {\rm e}^{-\tau_S(\mu)}}{\tau_S(\mu)}\, \dd \mu
	  \approx W B_{\nu}(\Tradij)\,\beta.
\eeq
Note, that our approximation (\ref{normal}) neglects 
any coupling to the continuum inside the resonance zone. 
By means of Eq.~(\ref{aiaj}), the individual terms
comprising $C_N$ can be calculated from
\beqa
\left(\frac{n_{i}}{n_{1}} \frac{R_{i \kappa}}{R_{1 \kappa}}
       \right)_{i \in {\rm N}} &=&
        \frac{\alpha_i}{\alpha_1} 
        \frac{T_{\rm r,i}}{T_{\rm r,1}}\,
        \exp \left[ -\frac{h}{k}
        \left(\frac{\nu_i}{T_{\rm r,i}} - \frac{\nu_1}{T_{\rm r,1}} 
	\right)\right] \times \\
       &\times& \left[W (1-\delta_{1i}) 
       \exp \left(-\frac{h \nu_{1i}}{kT_{\rm r,1i}}\right) \,+ \delta_{1i} 
       \exp \left(-\frac{h \nu_{1i}}{kT_{\rm e}}\right) \right]  \nonumber
\eeqa
whereas the components of $C_S$ are described by
\beqa
\left(\frac{n_{i}}{n_{1}} \frac{R_{i \kappa}}{R_{1 \kappa}}
       \right)_{i \in {\rm S}} &=&
  \left(\frac{n_{m}}{n_{1}}\right)_{m \in {\rm M}} \cdot
  \left(\frac{n_{i}}{n_{m}} 
       \frac{R_{i \kappa}}{R_{1 \kappa}} \right)_{i \in {\rm S}} = \\
  & &  \left(\frac{n_m}{n_1}/\frac{g_m}{g_1}\right) \times    
        \frac{\alpha_i}{\alpha_1} 
        \frac{T_{\rm r,i}}{T_{\rm r,1}}\,
        \exp \left[ -\frac{h}{k}
        \left(\frac{\nu_i}{T_{\rm r,i}} - \frac{\nu_1}{T_{\rm r,1}} 
	\right)\right] \times \nonumber \\
       &\times& \left[W (1-\delta_{mi}) 
       \exp \left(-\frac{h \nu_{mi}}{kT_{\rm r,mi}}\right) \,+ \delta_{mi} 
       \exp \left(-\frac{h \nu_{mi}}{kT_{\rm e}}\right) \right]  \nonumber
\eeqa
(with $(n_m/n_1)$ taken from Eqs.~\ref{meta_low},
\ref{meta_high} and \ref{meta_cont}, respectively). Obviously, the population
of subordinate levels is controlled by at least three different radiation
temperatures (ionization from the considered level, ionization 
from the connected lower level and excitation due to line processes). 

\subsection{Limiting cases}

In the following, we will consider some limiting cases which have to be
reproduced by our approach.

\paragraph{Constant radiation temperature, no collisional excitation} are
the assumptions underlying the description by \citet{springmann97} and 
\citet{puls00} on the basis of Lucy's unpublished notes. 
With $\Trad = {\rm const}$, meta-stable levels are populated via
 \beq
\left(\frac{n_{m}}{n_{1}} \frac{R_{m \kappa}}{R_{1 \kappa}}
       \right)_{m \in {\rm M}}  = \frac{\alpha_m}{\alpha_1} \qquad
      (\nu_m - \nu_1 + \nu_{1m} = 0!), 
\eeq
independent of the actual feeding mechanism. With $\delta = 0$ (only
radiative line processes), we thus obtain
for the population of subordinate levels (both $i \in N$ and $i \in S$!)
\beqa
\left (\frac{n_{i}}{n_{1}} \frac{R_{i \kappa}}{R_{1 \kappa}}
       \right)_{i \in {\rm N, S}} & = & W \frac{\alpha_i}{\alpha_1}.
\eeqa
Thus, for constant radiation temperatures, it does not play any role how
the meta-stable levels are populated, and whether subordinate levels are
connected to the ground-state or to a meta-stable level. Only the corresponding
recombination coefficient is of importance and the fact that subordinate
levels suffer from dilution (since they are fed by a diluted radiation
field), whereas for meta-stable levels this quantity cancels out (cf.
\citealt{ab85}). In total, our simplified ionization equilibrium then becomes
\beq
\frac{n_\kappa}{n_1}  = 
      \left(\frac{n_\kappa}{n_1} \right)^*_{T_{\rm rad}} W 
\sqrt{\frac{T_{\rm e}}{T_{\rm rad}}}\, \zeta \, \left(1\,+\,
    \sum_{i \in {\rm M}} \frac{\alpha_i}{\alpha_1} \,+\,
  W \sum_{i \in {\rm N,S}} \frac{\alpha_i}{\alpha_1} \right).
\eeq
If we define $\eta$ as the fraction of recombination coefficients 
for all meta-stable levels, 
\beq
\eta=\frac{\sum_{i \in M} \alpha_i}{\sum_{i} \alpha_i}
\eeq
we find
\beq 
\label{sumM}
\sum_{i \in M} \frac{\alpha_i}{\alpha_1}
	= \frac{1}{\zeta} \frac{\sum_{i \in M}
	\frac{\alpha_i}{\alpha_1}}{\sum_i \frac{\alpha_i}{\alpha_1}}
	= \frac{\eta}{\zeta}	
\eeq
and
\beq 
\label{sumNS}
\sum_{i \in N,S} \frac{\alpha_i}{\alpha_1} = \frac{1-\eta-\zeta}{\zeta},
\eeq
and the ionization equilibrium can be described in a very compact way,
\beq 
\label{app_lucy}
 \frac{n_\kappa}{n_1} = \left(\frac{n_\kappa}{n_1}\right)^*_{T_{\rm rad}}
 W \sqrt{\frac{T_{\rm e}}{T_{\rm rad}}}\;
 \bigl(\,\zeta + \eta + W (1-\eta - \zeta) \,\bigr),
\eeq
which indeed is the result of the previous investigations mentioned above.
If we further prohibit all ionizations from meta-stable and subordinate
levels, i.e. allow for

\paragraph{Ionization/recombination only from and to the ground-state,} 
we find with $\zeta =1$ and  $\eta =0$ 
\beqa
 \frac{n_\kappa}{n_1} & = & \left(\frac{n_\kappa}{n_1}\right)^*_{T_{\rm rad}}
 W \sqrt{\frac{T_{\rm e}}{T_{\rm rad}}} = \nonumber \\
& = & \frac{2 g_\kappa}{g_1} \frac{1}{n_{\rm e}} 
   \left(\frac{2 \pi m_{\rm e} kT_{\rm rad}}{h^2}\right)^{3/2}\,
   \exp \left(-\frac{h \nu_1}{kT_{\rm rad}}\right)
   W \sqrt{\frac{T_{\rm e}}{T_{\rm rad}}},
\eeqa
which is a well-known result and also valid for the case where all lines
are optically thick and in detailed balance, e.g., \citet{ab82}. The 

\paragraph{LTE-case} is recovered independently from the specific
values of $\zeta$ and $\eta$ in the lowermost atmosphere, when the dilution
factor approaches unity, $W=1$, and the radiation field becomes Planck,
$\Trad \rarr \Te$. In this case, the ionization balance becomes
\beq 
 \frac{n_\kappa}{n_1} = \left(\frac{n_\kappa}{n_1}\right)^*_{T_{\rm e}}
 (\,\zeta + \eta + (1-\eta - \zeta) \,) = \left(\frac{n_\kappa}{n_1}\right)^*_{T_{\rm e}}
\eeq
and for the excitation we have
\beq
\frac{n_i}{n_1} = \left( \frac{n_i}{n_1} \right)^*_{\Te}, 
		\quad i \in {\rm M, N, S}.
\eeq

\subsection{Different spin systems}

\label{diffspin}
The last problem to overcome is the presence of different spin systems, a
problem already encountered for \Hei. Our approximation is to consider
the different systems as completely decoupled (except if strong
inter-combination lines are present, see below), since a coupling via
collisional inter-combination is effective only at high densities (i.e.,
in or close to LTE, which is treated explicitely in our procedure anyway).

Then for each of the separate multiplets, the ionization equation can be
calculated independently. The different subsystems are defined in 
the following way

\begin{itemize}
\item[$\bullet$] 
the first subsystem includes all levels coupled to the ground-state
plus those meta-stable levels fed from higher lying (subordinate) levels
(case a/b in Sect.~\ref{ex_meta}). In this way, we include also systems of
different spin which are connected to the ground-state system via {\it
strong} inter-combination lines, a condition which is rarely met. 
In total, the ground-state subsystem includes the levels $i \in$ 1, N, M', S', where
M' comprises all case a/b  meta-stable levels and S' those excited levels
which are coupled to M'. For reasons of convenience, 
we will denote this set of levels by (1, N').
\item[$\bullet$] a second group of $j$ subsystems comprises
\begin{itemize}
\item systems of different spin decoupled from the ground-state; 
\item ``normal'' meta-stable levels populated via continuum processes
(poorly approximated so far) and excited levels coupled to those.
\end{itemize}
Both groups can be treated in a similar way and are also identified in a
similar manner, namely from the condition that the lowest state of these
systems is meta-stable and {\it not fed} from higher lying levels.  Each
subsystem comprises the ``effective'' ground state $m_j \in$ M'' (either
different spin or fed by continuum) and coupled levels, $i \in$ S''$_j$.
\end{itemize}
Once more, $j$ is the number of meta-stable levels per ion which are {\it
not} fed by higher lying levels. Note that for a {\it single} spin-system
with meta-stable levels, there are now $1 +j$ different subsystems if $j$
continuum fed meta-stable levels were present. Note also that by using this
approach we neglect a possible coupling of two or more non-ground-state
multiplets via strong inter-combination lines, if there were any.

Because of the assumed decoupling, for each subsystem we can 
write down the appropriate ionization equation. For the ground-state 
system, we have
\beqa
\label{gs_system}
\frac{n_\kappa}{n_1} & = &\left(\frac{n_\kappa}{n_1} \right)^*_{\Tradone} W 
\sqrt{\frac{T_{\rm e}}{\Tradone}}\, \zeta_1 \,(1+C_{N'}) \\
\zeta_1 & = & \frac{\alpha_1}{\sum_{(1,N')} \alpha_i},\qquad
\label{gs_system1}
C_{N'} = \sum_{i \in {\rm N'}} \frac{n_{i}}{n_{1}} 
\frac{R_{i \kappa}}{R_{1 \kappa}}
\eeqa
where, again, N' comprises the ``old'' levels $\in$ N, M' and S'. Note the
difference between $\zeta_1$ and $\zeta$ from Eq.~\ref{def_zeta}.

\noindent
For each of the $j$ additional subsystems, we obtain in analogy 
\beqa
\label{mj_system}
\frac{n_\kappa}{n_{\mj}} 
& = &\left(\frac{n_\kappa}{n_{\mj}} \right)^*_{\Tmj} W 
 \sqrt{\frac{T_{\rm e}}{\Tmj}}\,\frac{\alpha_{\mj}}{\alpha_1}\, 
 \zeta_{\mj}\,(1+\CSj) \\
\zeta_{\mj} & = & \frac{\alpha_1}{\sum_{({\mj},S''_j)} \alpha_i},\qquad
\CSj = \sum_{i \in {\rm S''_j}} \frac{n_{i}}{n_{\mj}} 
\frac{R_{i \kappa}}{R_{{\mj} \kappa}}
\eeqa
and S''$_j$ comprises all levels coupled to $m_j$. The individual components
of $C_{N'}$ and $\CSj$ are calculated as described in Sect.~\ref{excitation}.
Dividing Eq.~\ref{gs_system} by Eq.~\ref{mj_system},  we find for the ratios 
$(n_{\mj}/n_1)$ (required, e.g., for calculating the partition functions),

\beq
\frac{n_{\mj}}{n_1}  = 
\frac{(n_\kappa/n_1)^*_{\Tradone}}{(n_\kappa/n_{\mj})^*_{\Tmj}} 
\sqrt{\frac{\Tmj}{\Tradone}}\, \frac{\alpha_1 \zeta_1}
{\alpha_{\mj} \zeta_{\mj}} \, \left(\frac{1+C_{N'}}{1+\CSj}\right),
\eeq
or, explicitly written,
\beq
\label{nmn1}
\frac{n_{\mj}}{n_1}  = \frac{g_{\mj} \Tradone}{g_1 \Tmj}
      \exp \left[ -\frac{h}{k}
   \left(\frac{\nu_1}{\Tradone} - \frac{\nu_{\mj}}{\Tmj} \right)\right]
\frac{\alpha_1 \zeta_1}
{\alpha_{\mj} \zeta_{\mj}} \left(\frac{1+C_{N'}}{1+\CSj}\right).
\eeq
The last equation is the ``correct approximation'' for continuum fed 
meta-stable levels. On the one hand, if the ion consists of the
ground-state plus a number of meta-stable levels alone, we would have $C_{N'} =
\CSj = 0$, $\zeta_1 = 1$ and $\zeta_{\mj} = \alpha_1 / \alpha_{\mj}$. In
this case, Eq.~\ref{nmn1} and Eq.~\ref{meta_cont} would give identical results,
which shows that both approaches are consistent under the discussed
conditions. But as already pointed out,
Eq.~\ref{meta_cont} is highly approximative if a variety of levels are
involved, and the occupation numbers should be calculated according to
Eq.~\ref{nmn1} always.

The major difference to our former approach (one spin system only) 
is the following. In approach ``one'', the ground-state population,
$n_{\kappa}/n_1$, is affected
by {\it all} meta-stable levels, whereas in approach ``two'' only those
meta-stable levels have an influence which are coupled to the
ground-state system via higher levels.

\paragraph{Constant radiation temperature, no collisional excitation.} 
Concerning the limiting case where $\Trad = {\rm const}$ and $\delta = 0$,
Eq.~\ref{app_lucy} remains valid if we account for the different
``normalization'', i.e., if we replace $\zeta$ by $\zeta_1$ and include
only those meta-stable levels into $\eta$ which are populated via excited levels:
\beq 
 \frac{n_\kappa}{n_1} = \left(\frac{n_\kappa}{n_1}\right)^*_{T_{\rm rad}}
 W \sqrt{\frac{T_{\rm e}}{T_{\rm rad}}}\;
 \bigl(\,\zeta_1 + \eta_1 + W (1-\eta_1 - \zeta_1) \,\bigr)
\label{nkn1_sp}
\eeq
with 
\beq
\zeta_1= \frac{\alpha_1}{\sum_{(1,N')} \alpha_i}, \qquad 
\eta_1=\frac{\sum_{i \in M'} \alpha_i}{\sum_{(1,N')} \alpha_i}.
\eeq
Inside the individual sub-systems we then obtain
\beq 
 \frac{n_\kappa}{n_{\mj}} = \left(\frac{n_\kappa}{n_{\mj}}\right)^*_{T_{\rm rad}}
 W \sqrt{\frac{T_{\rm e}}{T_{\rm rad}}}\;
 \bigl(\,\zeta' + W (1-\zeta') \,\bigr), \quad
\zeta'=\frac{\alpha_{\mj}}{\alpha_1}\zeta_{mj}
\label{nknm_sp}
\eeq
which immediately indicates the correct thermalization for $W=1$ and 
$\Trad \rarr \Te$. After dividing Eq.~\ref{nkn1_sp} by \ref{nknm_sp}, 
we find for the population of $(n_{\mj}/n_1)$ in the same limit 
\beq
\frac{n_{\mj}}{n_1}  = 
\frac{g_{\mj}}{g_1}
      \exp \left( -\frac{h \nu_{1\mj}}{k\Trad}\right)
\left(\frac{\zeta_1 + \eta_1 + W (1-\eta_1 - \zeta_1)}
{\zeta' + W (1-\zeta')}\right).
\eeq
This expression reveals two things. First, we obtain the correct population
in LTE when $W \rarr 1$. Second, the difference to our crude approximation in
Sect.~\ref{ex_meta} becomes obvious: the quasi-LTE ratio
(\ref{meta_cont_tradconst}) has to be multiplied by the last factor in the
above equation to obtain consistent populations.  This factor (which can be
lower or larger than unity) becomes unity only when $W \rarr 1$ (i.e., in the
lower atmosphere) or for $\zeta_1 = \zeta' = 1$, i.e., in those cases where
only ground-state and meta-stable levels are present, as already discussed
above.

\subsection{Accelerated Lambda Iteration}
\label{ali}

To overcome the well-known problems of the Lambda-iteration when coupling
the rate-equations with the equation of radiative transfer, we apply the
concept of the Accelerated Lambda Iteration (ALI, for a review see
\citealt{hub92}) to obtain a fast and reliable convergence of the solution.
Since our rate-equations have been formulated in a non-conventional way and
since the radiation field is expressed in terms of local,
frequency-dependent radiation temperatures, the procedure has to be
modified somewhat, and we will describe the required re-formulations as
follows (for a comparable implementation see also \citealt{deKoteretal93}).

At first, assume that only {\it one} bound-free opacity is present, i.e.,
the radiation-field is controlled by the opacity of the considered
transition $i$ (no overlapping continua present).  In this case, the usual
ALI formulation for the mean intensity $J_{\nu}^n$ at
iteration cycle $n$ is given by
\beqa
J_{\nu}^n & \rarr & J_{\nu}^{n-1} + \alo (S_i^n - S_i^{n-1}) \nonumber \\
   & = & \dJ + \alo S_i^n \quad {\rm with} \quad \dJ = 
   J_{\nu}^{n-1} - \alo S_i^{n-1},
\eeqa
where $S_i$ is the continuum source-function for transition $i$ and $\alo$
the corresponding Approximate Lambda Operator (ALO), calculated in
parallel with the solution of the continuum transfer\footnote{including 
the pseudo-continua from the multitude of overlapping lines, cf.
Sect.~\ref{lineblock}}\,following the method suggested by \citet[ Appendix
A]{RybHum91}.

Substituting this expression into the rate equations, we find for the 
corresponding {\it effective} ionization/recombination rate coefficients
\beqa
\Rik & \rarr & \int_{\nu_i}^\infty \frac{4 \pi a_\nu}{h \nu} \dJ\, \dd \nu \\
\Rki & \rarr & \int_{\nu_i}^\infty \frac{4 \pi a_\nu}{h \nu} 
                       \left( \frac{2h \nu^3}{c^2} (1 - \alo) + \dJ \right)
                       \rm{e}^{-h \nu/kT_{\rm e}} \,\dd \nu,
\eeqa
i.e., the problematic, optically thick part of the radiation field has been
canceled analytically. Neglecting again stimulated emission 
(the $\dJ$-term in the recombination rate coefficient above), approximating
$S_i^{n-1} = B_{\nu}/b_i^{n-1}$ with Planck-function $B_{\nu}$ and
NLTE-departure coefficient $b_i^{n-1}$, and using the radiation temperature
at the threshold, $T_{\rm rad,i}$ along with Seaton parameters $\beta=1, s=2$, 
we have in analogy to Eq.~\ref{def_rki}
\beqa
\Rik & \rarr & \frac{8 \pi}{c^2} a_i \nu_i^2 \left\{
	W \frac{k T_{\rm r,i}^{n-1}}{h}
         {\rm e}^{-h \nu_i/k T_{\rm r,i}^{n-1}}
        - \frac{\alo}{b_i^{n-1}} \frac{k T_{\rm e}}{h}
        {\rm e}^{-h \nu_i/k T_{\rm e}} \right\} \nonumber\\
     & := & \Rik^{n-1} - \Rik' \\ 		    
\Rki & \rarr & \frac{8 \pi}{c^2} \frac{kT_{\rm e}}{h}
               a_i \nu_i^2 {\rm e}^{-h \nu_i/kT_{\rm e}} 
	      \left(1 - \alo\right) \,:=\, \Rki - \Rki'.
\eeqa
In those cases where an overlapping continuum is present, i.e., 
if different transitions contribute to the opacity, the ALO has to 
be modified according to
\beq
\alo \rarr \beta_i(\nu) \alo \quad {\rm with} \quad \beta_i =
\frac{\chi_i(\nu)}{\chi_{\rm tot}(\nu)}.
\eeq
$\chi_i$ is the opacity of the considered transition, $\chi_{\rm tot}$ the
total opacity and $\beta_i$ is assumed to be constant between two subsequent
iterations (cf. Paper~I). Let us mention explicitely that the opacities
used for the radiative transfer are calculated from their {\it actual}
Seaton parameters $(\beta,\, s)$, whereas the uniform values $(\beta=1,\,
s=2)$ are applied ``only'' to evaluate the approximate
ionization/recombination rates. 
\begin{figure*}
\begin{minipage}{8.8cm}
\resizebox{\hsize}{!}
   {\includegraphics{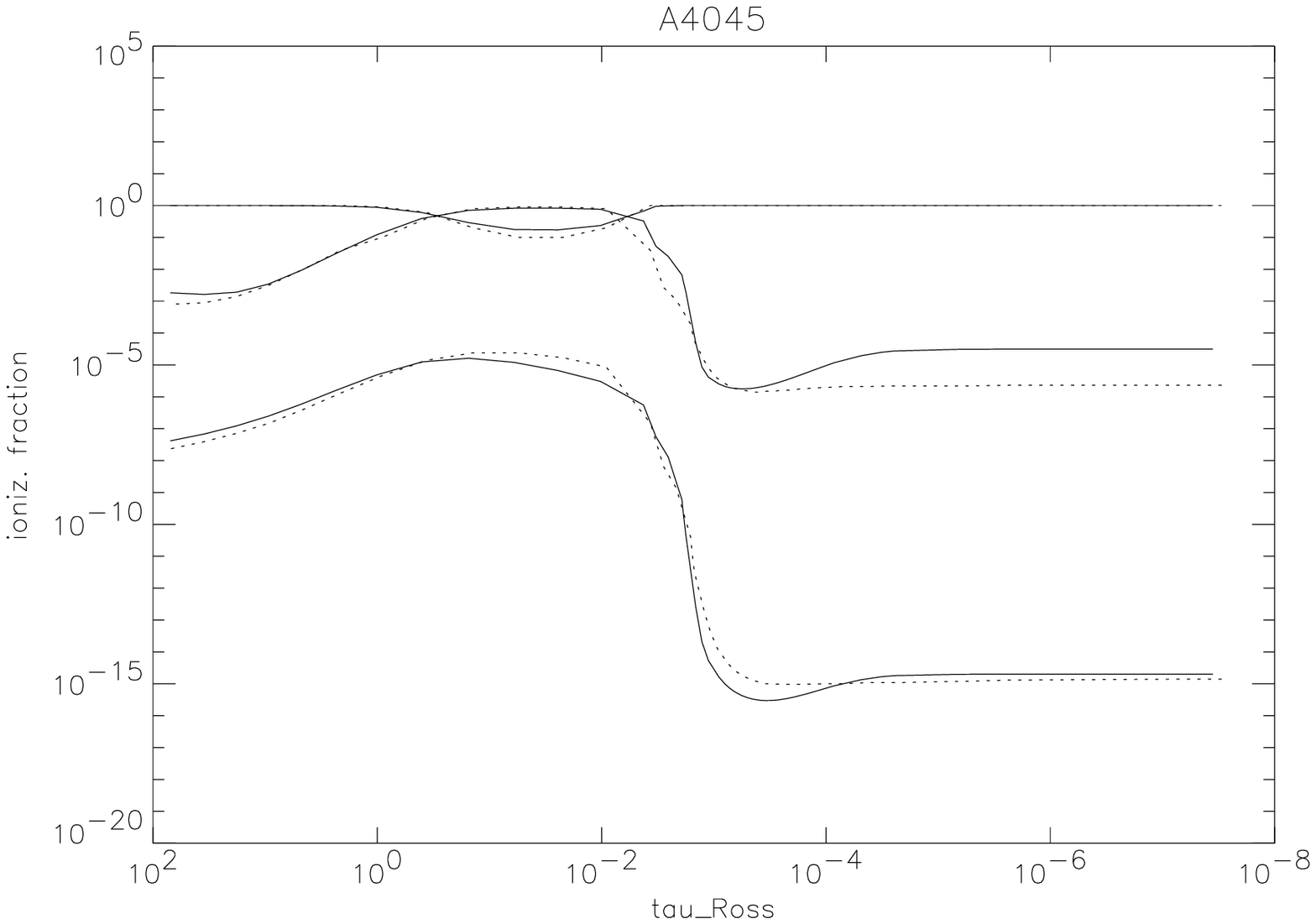}}
\end{minipage}
\hfill
\begin{minipage}{8.8cm}
   \resizebox{\hsize}{!}
      {\includegraphics{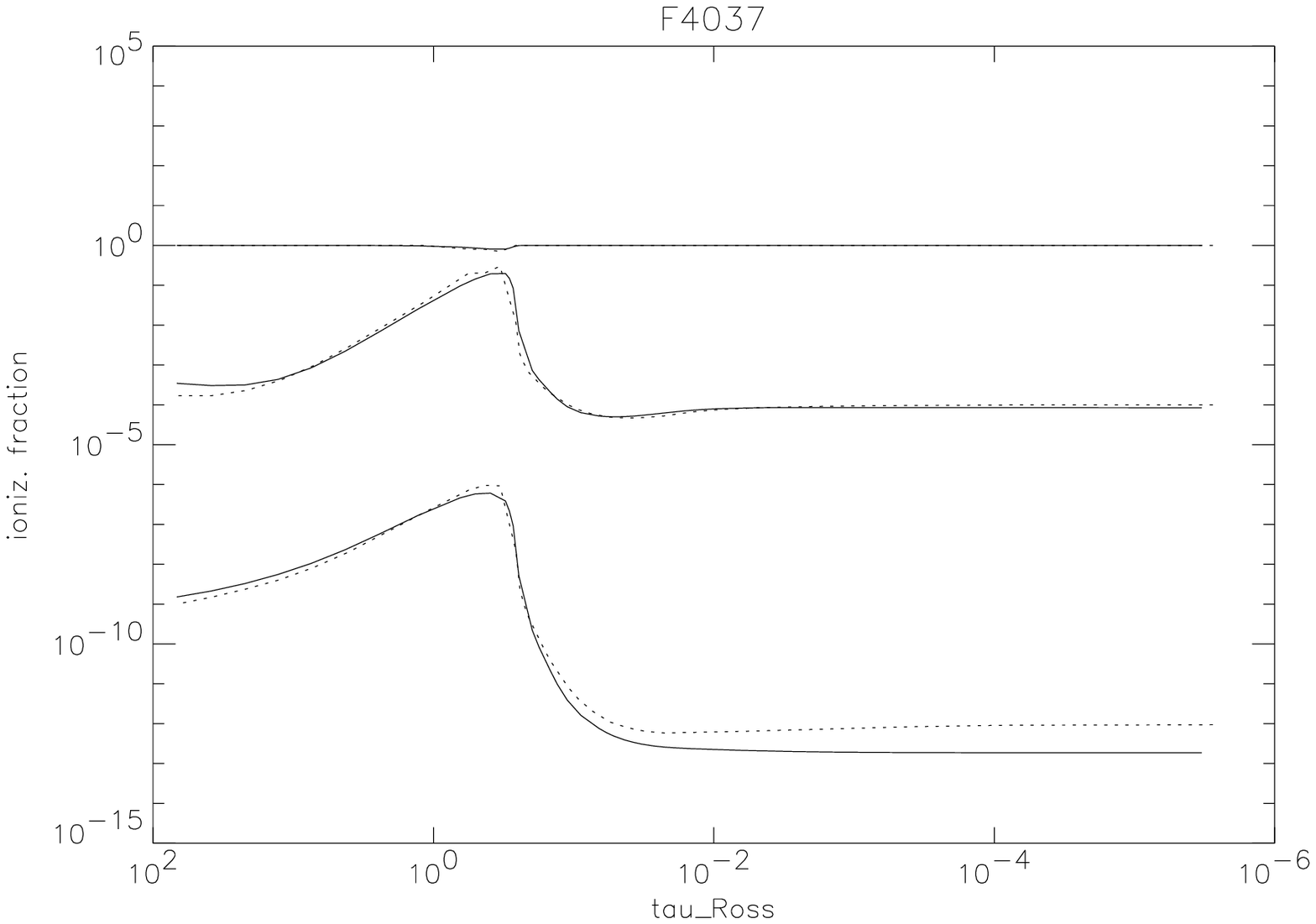}}
\end{minipage}
\caption{Approximate NLTE vs. the exact case: He ionization fractions (from
top to bottom: \Heiii, \Heii, \Hei) for pure H/He atmospheric models at
\Teff=40,000~K (left panel: dwarf with \logg=4.5 and thin wind; right panel:
supergiant with \logg=3.7 and thick wind). Bold: exact solution for
helium; dotted: He in approximate NLTE (see text).}
\label{nlteapp_hefrac}
\end{figure*}

Since the Lambda Iteration fails only in the optically thick case, we apply
the ALI-scheme exclusively for ground state transitions. Thus, by
substituting the effective rate coefficients $R_{1 \kappa}$ and $R_{\kappa 1}$
into Eqs.~\ref{rec_coeff}, \ref{rec_coeff1}, we have 
\beq
\frac{n_\kappa}{n_1} = \left(\frac{n_\kappa}{n_1} \right)^*
                       \frac{R_{1 \kappa}}{R_{\kappa 1}}		       
                       \frac{1 - \frac{R_{1 \kappa}'}{R_{1 \kappa}} +   
		        \sum_{i \in N'} 
                        \frac{n_i}{n_1} \frac{R_{i \kappa}}{R_{1 \kappa}}}
                        {1 - \frac{R_{\kappa 1}'}{R_{\kappa 1}} +   
		        \sum_{i \in N'}
			\left(\frac{n_i}{n_1} \right)^*
                        \frac{R_{\kappa i}}{R_{\kappa_1}}}.
\eeq
Using, again, Eq.~\ref{aiaj} and the definitions given in (\ref{gs_system1}), we
finally obtain
\beq
\frac{n_\kappa}{n_1} = \left(\frac{n_\kappa}{n_1} \right)^*
                       \frac{R_{1 \kappa}}{R_{\kappa 1}} 
		       \,\zeta_1 \, (1 + C_{N'}) \,	       
                       \frac{1 - \frac{1}{1 + C_{N'}} \frac{R_{1 \kappa}'}{R_{1 \kappa}}}
                        {1 - \zeta_1 \,\frac{R_{\kappa 1}'}{R_{\kappa 1}}}. 
\eeq
In the case of $\alo \equiv 0$ (implying $R_{1 \kappa}' = R_{\kappa 1}' =
0$), we immediately recover the original result, Eq.~\ref{gs_system}, since 
\beq
\left(\frac{n_\kappa}{n_1} \right)^* \frac{R_{1 \kappa}}{R_{\kappa 1}} =
\left(\frac{n_\kappa}{n_1} \right)^*_{\Tradone} 
W \sqrt{\frac{T_{\rm e}}{\Tradone}}
\eeq
by means of Eq.~\ref{ratapp}. If, on the other hand, the ALO is significant
(i.e., close to unity), we find
\beqa 
\frac{R_{1 \kappa}'}{R_{1 \kappa}}  & = & 
\frac{\alo}{W b_1^{n-1}} \, \frac{\Te}{\Tradone^{n-1}} \,
        \exp \left[ -\frac{h \nu_1}{k}
        \left(\frac{1}{\Te} - \frac{1}{\Tradone^{n-1}}\right)\right]\\
\frac{R_{\kappa 1}'}{R_{\kappa 1}} & = & \alo. 
\eeqa
Thus, the reformulated ALI-scheme collapses to a simple correction of the
original equation (\ref{gs_system}) for the ground-state population,
\beqa
\frac{n_\kappa}{n_1} & = &\frac{n_\kappa}{n_1}(\alo \equiv 0) \cdot
C_{\rm A} (\Tradone^{n-1},\,b_1^{n-1}), \quad \mbox{with factor} \\
C_{\rm A} & = & \frac{1 - \disp{\frac{\alo}{(1 + C_{N'}) W b_1^{n-1}}}
	\frac{\Te}{\Tradone^{n-1}} 
        \exp \left[ -\frac{h \nu_1}{k}
        \left(\frac{1}{\Te} - \frac{1}{\Tradone^{n-1}} \right)\right]}
	{1 - \zeta_1 \alo}. \nonumber
\eeqa	
The consistency of this scheme is easily proven, because after convergence we
would get (cf. Eq.~\ref{gs_system})
\beqa
\frac{1}{b_1} & = &\left(\frac{n_1^*}{n_1}\right) = 
\frac{n_\kappa}{n_1} \left(\frac{n_1}{n_\kappa} \right)^* \nonumber \\
& = & W \, \frac{\Tradone}{T_{\rm e}}\, \exp \left[ -\frac{h \nu_1}{k}
        \left(\frac{1}{\Tradone} - \frac{1}{T_{\rm e}} \right)\right]
\zeta_1 \,(1 + C_{N'}),
\eeqa
so that the ``ALO-correction factor'' $C_{\rm A}$ becomes unity.
Through\-out the iteration the correction factor can take 
values smaller or larger than unity, leading to a fast and reliable 
convergence.

\subsection{Test calculations}
\label{test_nlteapprox}

In order to check the accuracy of our approximate approach, we will present
two different test calculations. The first test aims at a clean
investigation of the methods outlined above, unaffected by additional
complications such as line-blocking/blanketing. To this end, we have
computed a pure H/He atmosphere at \Teff=40,000~K, for two different sets of
parameters: the first model (A4045 with \logg=4.5) corresponds to a dwarf with
thin wind, the second (F4037 with \logg=3.7) to a supergiant with thick
wind.\footnote{Concerning the nomenclature of our models, cf.
Sect.~\ref{modelgrids}}

For both models we have calculated an ``exact'' solution as described in
Paper~I, namely by solving for the H/He occupation numbers from the complete
rate equations, with all lines in the CMF and a temperature stratification
calculated from NLTE Hopf-parameters. In order to test our approach, we
calculated two additional models, with an exact solution for hydrogen only,
whereas helium has been treated by means of our approximate approach. (In the
standard version of our code, helium is always treated exactly.)

Fig.~\ref{nlteapp_hefrac} shows the very good agreement of the resulting
ionization fraction for helium in both cases. The small differences at large
optical depths (i.e., for LTE conditions) are due to the different atomic
models for helium used in both the exact and the approximate solution. (The
data-base applied to the approximate solution comprises a lower number of
levels for both \Hei\, and \Heii, so that the partition functions are somewhat
smaller than in the exact case, and consequently also the ionization
fractions. The occupation numbers of the levels {\it in
common} are identical though). 

\begin{figure}
\resizebox{\hsize}{!}
   {\includegraphics{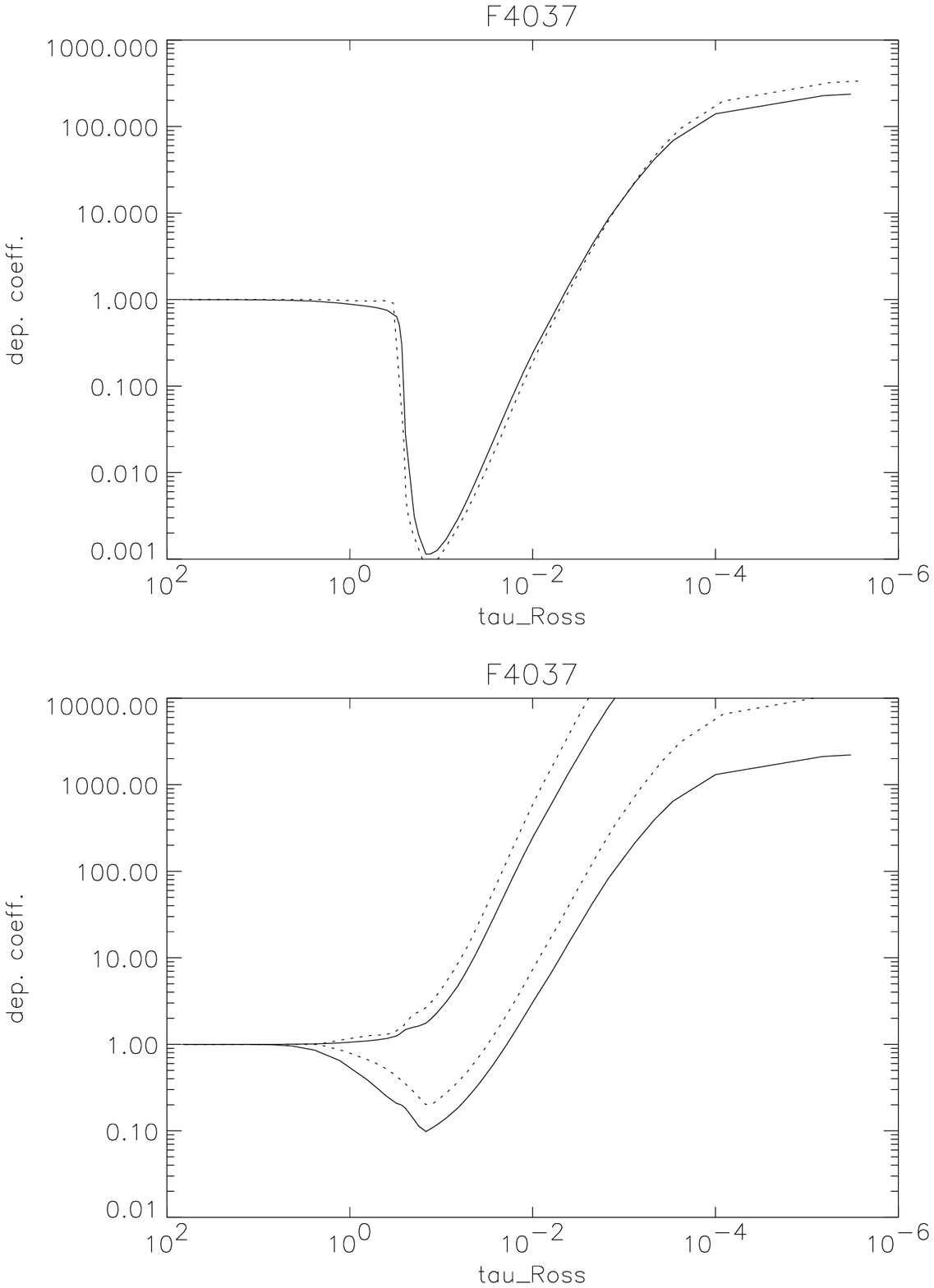}}
\caption{Approximate NLTE (dotted) vs. the exact case (bold): 
He departure coefficients for model F4037. Upper panel: \Heii ground-state
departure coefficient. Lower panel: \Hei triplet and singlet
``ground''-states (upper and lower curves, respectively).}
\label{nlteapp_hedep}
\end{figure}

Most intriguing is the excellent agreement of the He{\sc ii} ground state
departure coefficient as a function of depth (Fig.~\ref{nlteapp_hedep},
upper panel). The crucial feature is the depopulation of the \Heii
ground-state close to the sonic point, which is a sophisticate NLTE-effect
arising in unified model atmospheres and depends on a delicate balance
between the conditions at the \Heii ground-state, the $n=2$ ionization edge
and the \Heii Ly$_\alpha$ line (which in itself depends on the radiation
field at 303~\AA\, and the escape probabilities), cf. \citet{Gabler89}. The
comparison between exact and approximate solution shows clearly that our
approach, accounting for frequency dependent radiation temperatures and
important line transitions, is actually able to cope with such complicated
problems.\footnote{Actually, it was this feature that motivated us to
refrain from frequency independent radiation temperatures, since a first
comparison using the latter simplification gave extremely unsatisfactory
results.}

In the lower panel of the figure, we have displayed the ``ground''-state 
departure coefficients of \Hei, for the triplet and singlet
system (upper and lower curves, respectively). Although the precision is
not as excellent as for the \Heii ground-state, one has to consider that
\Hei\, at 40,000~K is an extremely rare ion, and that the major features
(depopulation of the singlet ground-state, no depopulation for the triplet
ground-state) are reproduced fairly well.

\begin{figure*}
\begin{center}
\begin{minipage}{16cm}
\resizebox{\hsize}{!}
   {\includegraphics{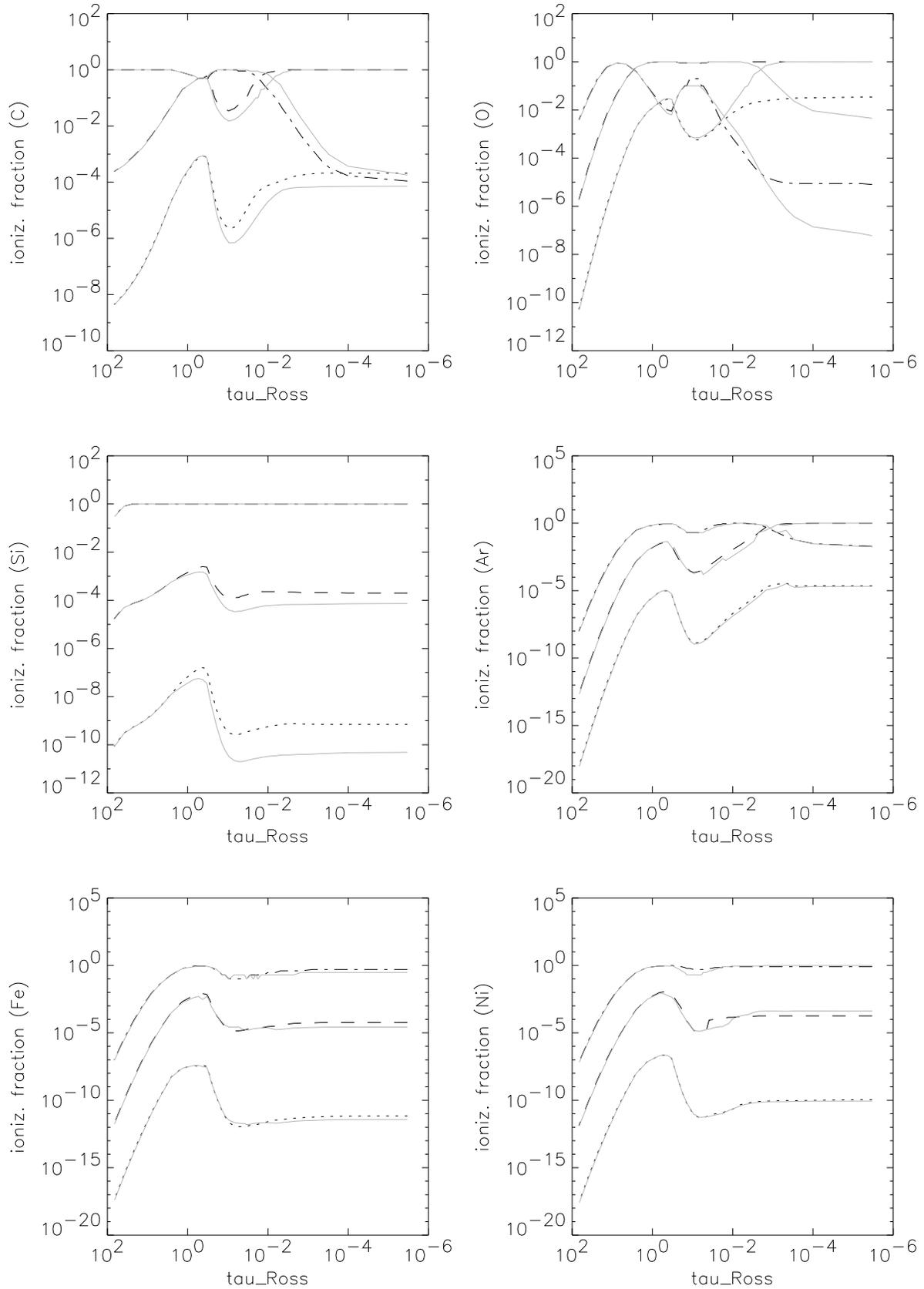}}
\end{minipage}
\end{center}
\caption{Approximate NLTE (grey) vs. the results of a solution of the
complete rate equations, using Sobolev line transfer (black): ionization 
fractions of important metals for model F4037. Displayed are the 
ionization stages {\sc iii, iv, v} (dotted, dashed and dashed-dotted,
respectively).}
\label{nlteapp_met_ifrac}
\end{figure*}

\begin{figure*}
\begin{center}
\begin{minipage}{16cm}
\resizebox{\hsize}{!}
   {\includegraphics{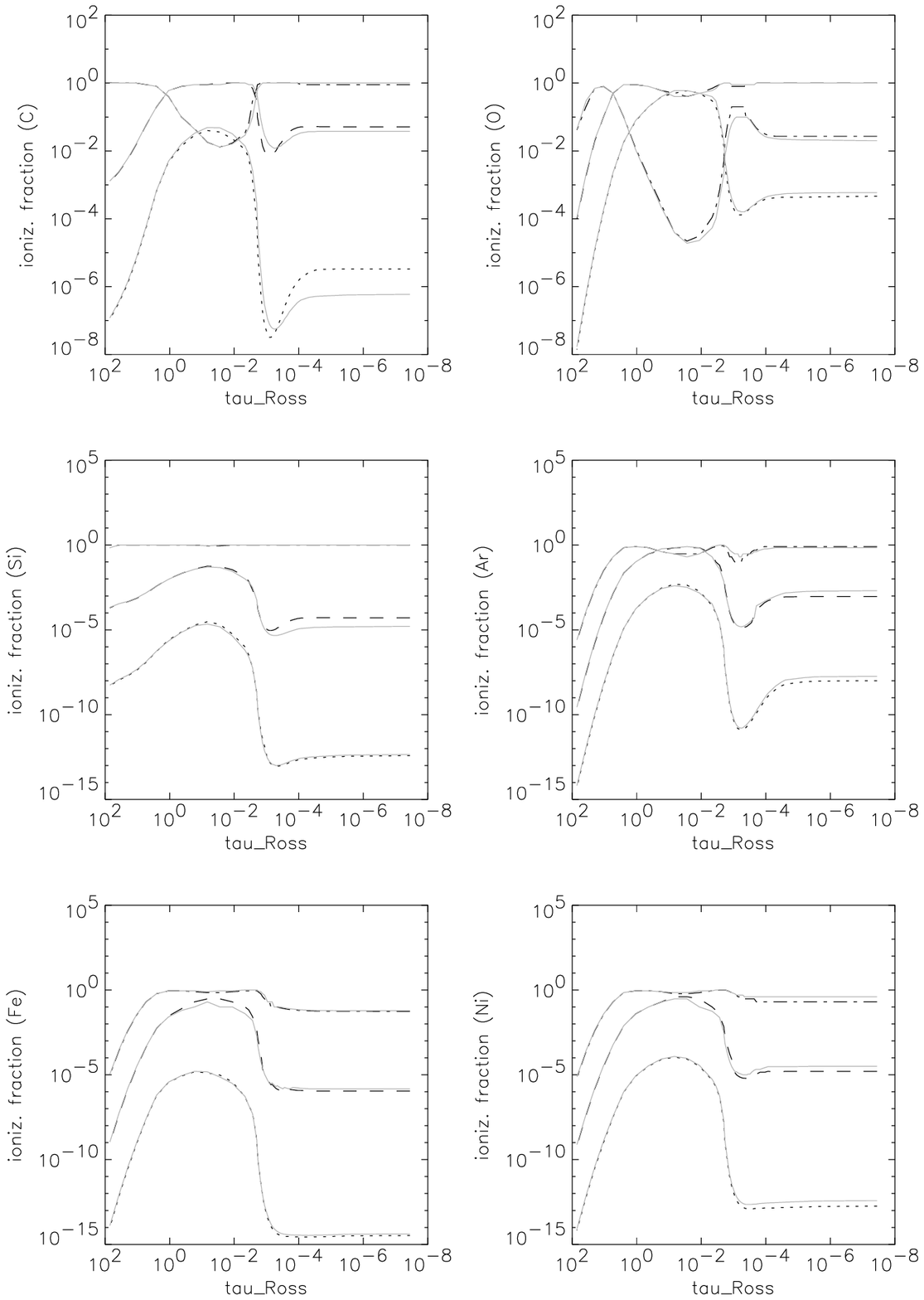}}
\end{minipage}
\end{center}
\caption{As Fig.~\ref{nlteapp_met_ifrac}, but for model A4045.}
\label{nlteapp_met_ifrac_A4045}
\end{figure*}

The second test investigates the behaviour of the metals. We compare the
results from the approximate method with results from an ``almost''
exact solution, for model F4037. As we will see in
Sect.~\ref{tempstrat}, the introduction of a consistent temperature
structure calculated in parallel with the solution of the rate equations
forced us to consider the most important elements (in terms of their
abundance) in a more precise way than described so far, at least if we
calculate the temperature from the electron thermal balance. In this case it
is extremely important that the occupation numbers from {\it all}
excited levels are known to a high precision in order to account for the
cooling/heating by bound-bound collisions in a concise way. Unfortunately,
this latter constraint cannot be fulfilled by our approximate method, simply
because not all excited levels are considered, and small deviations from the
exact solution (which are negligible for the effects of line-blocking,
see below) can have disastrous effects on the total cooling/heating rates.

Thus, for the most abundant elements the complete set of rate-equations has
to be solved for in any case, and this solution (which uses a Sobolev line
transfer, cf. Sect.~\ref{tempstrat}) is compared with our approximate
one in Fig.~\ref{nlteapp_met_ifrac}, for the ionization stages {\sc iii} to
{\sc v} of some important metals, namely C, O, Si, Ar, Fe and Ni. Note that
the comparison includes the effects of line-blocking on the radiation
field, where this radiation field has been calculated either
from the exact occupation numbers or from the corresponding approximate
values, respectively. Our comparison demonstrates three important points.
\begin{itemize}
\item The transition between LTE and NLTE (taking place at $\taur > 2/3$ in
our approximate approach) is described correctly.
\item The approximate treatment works particularly well for elements 
with complex electronic structure (Ar, Fe, Ni), i.e., our treatment of 
meta-stable levels is reasonable.
\item If there are differences, they occur predominantly in the outer wind.
\end{itemize}
In almost all considered cases, the principal run of the approximate
ionization fractions agrees reasonably or even perfectly well with the exact
result. The only exception is oxygen where the major/minor stages ({\sc
iv}/{\sc iii}) appear reversed in the outer wind (no problems have been
found for nitrogen and neon which are not displayed here). These differences
in the outer wind (see also C and Si) are partly due to two effects.
On the one hand, our approach becomes questionable in those cases when {\it
all} line transitions are optically thin so that the two-level-atom approach
fails to describe the excitation-balance of subordinate levels. If
only this effect were responsible this would imply (as suggested by our referee)
that the discrepancy should become worse for thinner winds. Thus, we
performed a similar comparison for model A4045, which has a considerably
lower wind density than model F4037, by a factor of almost 100. The
corresponding ionization fractions are shown in
Fig.~\ref{nlteapp_met_ifrac_A4045}. Note that the transition point between
photosphere and wind is located at lower values of $\taur$, compared to
model F4037, due to the weaker wind. Interestingly, the discrepancies
between approximated and ``exact'' ionization fractions in the outer wind
have remained at the same level as for model F4037, and in the case of
oxygen the situation is almost perfect now. Consequently, the effect
discussed above cannot be responsible alone for the observed discrepancy,
and we attribute it to a combination of various ingredients inherent to our
approximative approach.

For our models, however, this is of minor importance, since we are not
aiming at a perfect description of the occupation numbers in the outer wind 
unless we actually need it, i.e., when a consistent temperature structure
shall be derived. In this latter case, the occupation numbers are calculated
exactly anyway. 

\begin{figure}
\resizebox{\hsize}{!}
   {\includegraphics{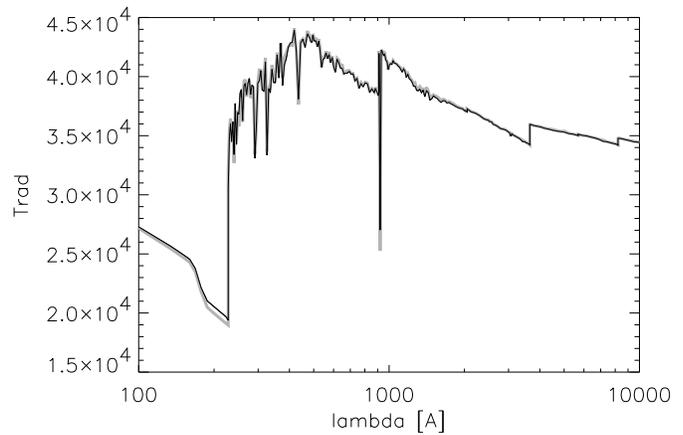}}
\caption{As Fig.~\ref{nlteapp_met_ifrac}. 
Comparison of radiation temperatures of {\it converged} models.}
\label{nlteapp_f4037_trad}
\end{figure}

Different occupation numbers influence the radiation field, which in turn
influences the occupation numbers, and so on. This is the second important 
process which might affect our final approximate solution.
Fig.~\ref{nlteapp_f4037_trad} compares the emergent fluxes (expressed as
radiation temperatures) for the converged models of F4037, calculated by
both alternative approaches.  

Due to the excellent agreement between the ionization fractions in the
line/continuum forming part of the atmosphere, also the fluxes agree very
well. The maximum differences, located between 200 to 400 \AA, are of the
order of $\pm$1,000~K, which translates to a typical
difference in population of $\pm 0.15\, dex$ in the outer wind.

Globally, however, the differences in flux are so small that we can consider
the two results as equivalent. Thus, the radiation field calculated in
parallel with the line-blocking background elements is insensitive to the
chosen approach (exact vs. approximate occupation numbers) which primarily 
differs in the precision (and presence) of subordinate levels.

\section{Approximate line-blocking}
\label{lineblock}

The most time-consuming part in the computation of {\it realistic} stellar
atmospheres is the calculation of the radiation field, realizing the
multitude of overlapping\footnote{both in the observer's and in the comoving
frame} lines with considerable opacity (see also the discussion by
\citealt{pp90} and \citealt{Paul01}). 

For {\sc cmfgen} as well as for the wind-code developed by the Potsdam group
(for a recent status report, see \citealt{Graef02}), this problem has been
tackled by performing a comoving-frame solution for the {\it complete} EUV/UV range.
Obviously, this approach is very time-consuming. A quick calculation shows
that the number of frequency points which must be treated is of the order of
900,000, if a range between 200 and 2,000~\AA\, and a typical resolution of
0.8~\kms is considered (i.e., ten points covering a thermal width of 8~\kms). 

In the approach followed by {\sc wm}-Basic, on the other hand, an observer's
frame solution is performed which requires ``only'' a couple of thousand
frequency points to be considered. The conservation of work, however,
immediately implies that in this case a lot of time has to be spent
on the resolution of the resonance zones of the
overlapping lines, a problem which is avoided a priori in a CMF calculation.

In order to solve the problem on a minimum time-scale, both a Monte-Carlo 
solution\footnote{which becomes costly as well if a detailed
description of all possible interactions between radiation field and plasma
is accounted for.} (e.g., \citet{schaerer94, schaerer97} , and a {\it
statistical} approach are feasible.

Since the number of metal lines to be treated is very large, the information
about the exact position of individual lines inside a (continuum transfer)
frequency grid interval becomes less important for obtaining a
representative mean background. As shown by \citet{wehrse98}, the {\it
Poisson Point Process} is well suited to describe such a line ensemble,
particularly because it is very flexible and can be described by relatively
few parameters.

The additional introduction of a {\it Generalized Opacity Distribution
Function} by \citet{baschek01} serves two purposes. First, additional
analytical insight is given into the effects of the vast amount of blocking
lines on the mean opacity in differentially moving media with line overlap.
Second, it is a fast tool to derive such mean backgrounds numerically. In
particular, it is able to ``solve problems that have been inaccessible up to
now as e.g. the influence of very many, very weak
lines''(\citealt{baschek01}), and to describe the transition from a static
to a moving configuration, since it is equally efficient in both cases.

In our opinion, this approach is very promising, and work adapting and 
applying the corresponding method is presently under way in our group.
Since it will take some time to finalize this approach (the most
cumbersome problem is the formulation of consistent emissivities), 
we have followed a somewhat simplified approach in the mean time, which relies
on similar arguments and has been developed by carefully comparing with
results from ``exact'' methods, mostly with the model grid calculated 
with {\sc wm}-Basic as described by \citet{Paul01}.

Again, the principal idea is to define suitably averaged quantities
which represent a mean background and which can be calculated easily and
fast. The multitude of lines will be approximated in terms of a pseudo
continuum (split into a ``true'' absorption and a scattering component), so
that the radiative transfer can be performed by means of a standard
continuum solution, for relatively few frequency points (see
below). Strongest emphasis has been given to the requirement that any
integral quantity calculated from the radiation field (such as the
photo-integrals) has to give good approximations compared to the exact case,
because these quantities (and not the frequential ones) are most decisive for
a correct description of the level populations and, in turn, for the blocked
radiation field. 

\subsection{Mean opacities}

To this end, we define a ``coarse grid'' with spacing $2 N \vthm$, where
$\vthm$ is a typical thermal velocity (say, of oxygen) including
micro-turbulence, and $2 N$ is an integer of the order of 100. (The reason to
define here $2N$ instead of $N$ will soon become clear.) Under typical
conditions, this grid has a resolution of 1,000-1,500 \kms and is used to
calculate appropriate averaged opacities. With respect to a simplified
approach, a mean constructed in analogy to the Rosseland mean is perfectly
suited, i.e., an average of the {\it inverse} of the opacity,
\beq
\label{chim}
\frac{1}{\chim} =: \frac {\disp{\int_{2N \vthm}\frac{\dd \nu}{\chi_\nu^{\rm
tot}}}} {\disp{\int_{2N \vthm}\dd \nu}}, 
\eeq
since it has the following advantageous properties:
\begin{enumerate}
\renewcommand{\labelenumi}{\alph{enumi})}
\item if no lines are present, the pure continuum opacity is recovered 
\item if one frequency interval is completely filled with non-overlapping, 
strong lines {\it of equal strength}, also the average opacity 
approaches this value, whereas 
\item in those cases when the interval has ``gaps'' in the opacity, these
gaps lead to a significant reduction of the mean, i.e., allow for an
appropriate escape of photons. Note that any {\it linear} average has 
the effect that {\it one} strong line alone (of typical width $2 \vthm$) 
would give rise to a rather large mean opacity (just a factor of 
$N$ weaker than in case b) and, thus, would forbid the actual escape.  
\item Finally, the average according 
to Eq.~\ref{chim} is consistent with the standard Rosseland mean in the
lowermost atmosphere (as long as $\partial B_{\nu}/\partial T$ is roughly 
constant over one interval), i.e., it is consistent with the diffusion 
approximation applied as a lower boundary condition in the equation 
of (continuum) radiative transfer.
\end{enumerate}
Because of the large number of contributing lines (typically $5 \cdot 10^5$
(O-type) to $10^6$ (A-type) lines if only ions of significant population are
considered\footnote{Remember, that our present data base comprises 
$4.2 \cdot 10^6$ lines in total.}), the calculation of this mean has to be
fast. 

First, assume that any velocity field effects (leading to Doppler-shift
induced line overlaps) are insignificant, i.e., assume a thin wind, so that
line blocking is essential only in the subsonic regions of the wind. The
generalization in order to approximate line-overlap in the wind will be
described later on.

Instead of evaluating the ``exact'' profile function, for each line we use a
box car profile of width $2 \vthm$. The frequential line opacity is, thus,
given by
\beqa
\label{boxprofile}
\chil(\nu) &=& \left\{
       \begin{array}{l} 
       \chil \quad {\rm for} \quad 
       \nu_0 + \Delta \nu_{\rm D} \,\le\, \nu \,\le\, \nu_0 - \Delta \nu_{\rm D} \\
      0 \qquad {\rm else}
       \end{array}
       \right. \\
\chil &=& \frac{1}{2 \Delta \nu_{\rm D}} \frac{\pi e^2}{m_{\rm e} c} gf 
\frac{n_l}{g_l}, \quad \Delta \nu_{\rm D} = \frac{\nu_0 \vthm}{c},
\eeqa
where stimulated emission has been neglected again. 
Due to this definition, at least the
frequency-integrated line opacity is correctly recovered.  The {\it coarse}
frequency grid is now divided into $N$ sub-intervals of width $\Delta \nu =
2 \Delta \nu_{\rm D}$. Inside each of these sub-intervals (``channels'') we
sum up any line opacity which has appropriate rest-wavelength. Insofar, we
account (approximatively) for any {\it intrinsic} (i.e., not wind-induced) 
line overlap. Inside each channel $i$, we thus have a (total) frequential
opacity 
\beq
\chi_{\nu,i}^{\rm tot} = \sum_{j} \chilj + \chic, \quad \chic = \chict +
\sige
\eeq
if lines $j$ are located inside channel $i$ and the continuum opacity is
assumed to be constant inside each coarse grid interval. $\chict$ is the
contribution by true absorption processes, and $\sige$ the contribution by
electron scattering. After replacing the integrals by appropriate sums and 
since all channels have the same width, the mean opacity (on the coarse
grid) is simply given by
\beq
\frac{1}{\chim} \approx \frac{\disp{\sum_{i=1}^N \frac{\Delta \nu}
                        {\chi_{\nu,i}^{\rm tot}}}}
                        {\disp{\sum_{i=1}^N \Delta \nu}} = 
			\frac{1}{N} \sum_{i=1}^N \frac{1}
			{\chi_{\nu,i}^{\rm tot}}.
\eeq
For later purposes we split this mean opacity into
the contribution from lines and continuum, respectively, where 
the line-contribution is given by
\beq
\label{chilm}
\chilm = \frac{\disp{N}}{\disp{\sum_{i} \frac{1}{\chi_{\nu,i}^{\rm tot}}}} 
- \chi_\nu^{\rm cont}
\eeq
and we have
\beq
\chim = \chilm + \chi_\nu^{\rm cont}.
\eeq
Note that both mean opacities, $\chilm$ and $\chim$, are frequency dependent
as a function of coarse grid index. In accordance with our reasoning from
above, Eq.~
(\ref{chilm}) implies that 
\begin{enumerate} 
\renewcommand{\labelenumi}{\alph{enumi})}
\item if $\chilj = 0$ for {\it all} lines
inside one interval, the correct result $\chilm = 0$ is obtained
\item if the same {\it total} line opacity $\chil(\nu)$ is present 
inside {\it each channel}, this value will also be obtained for the mean, 
$\chilm = \chil(\nu)$. 
\item if only one (strong) line is present, the mean line opacity 
is given by $\chilm \approx \chi^{\rm cont}/(N-1)$,
i.e., it will be  much smaller than the continuum opacity, since 
most of the flux can escape via the $(N-1)$ unblocked channels (according
to our present assumption that Doppler-induced line overlap is negligible).
\end{enumerate}
\noindent
Finally, let us point out that the opacities constructed in this way are
used also to calculate the photospheric line pressure, in analogy
to the description given in Paper~I (Eq.~3), however including the line 
contribution (cf. Fig.~\ref{s30grad}).\footnote{In our present version of {\sc
fastwind} we allow for deviations from the generalized Kramer-law
(Paper~I, Eq.~2) by simply including theses deviations as correction-factors
into the atmospheric structure equations. This method becomes important
for models at rather cool temperatures when hydrogen and 
background-metals are recombining (and become ionized again) 
in photospheric regions, which usually leads 
to some deviations from the above (power-) law.}

\subsection{Emissivities}
In order to calculate the corresponding emissivities, we assume that each 
transition can be described by means of a two-level atom, where the {\it
lower} occupation number is known from the solution (``exact'' or
approximate) of the rate equations.\footnote{Note that this approach is
equivalent to the typical assumption made if deriving the radiation field
via Monte-Carlo simulations.}

Although this assumption is hardly justified for (weak) recombination
lines, it is a fair representation for most of the stronger transitions 
arising from either the ground-state or a meta-stable level, particularly if
the level population itself is calculated from a multi-level atom. 

It might be argued that the two-level atom approach is superfluous for those
{\it connecting} transitions which are calculated from an exact NLTE
solution, since the occupation numbers for both levels and, thus, the
source-functions are already known. The maximum number of these lines is of
the order of 30,000, and therefore much lower than the total number of lines
we are using for our line-blocking calculations (cf. Sect.~\ref{atomdat}).
For the latter transitions, however, only the lower level is present in the
atomic models, so that the corresponding source-functions have to be
approximated in any case. 

Moreover, treating all lines (including the connecting transitions) in a
two-level way has the additional advantage that the contribution of
scattering and thermal processes can be easily split, which allows to
simulate their impact by means of a pseudo-continuum, so that the standard
continuum transfer can be applied without any modification. 

To keep things simple and as fast as possible and to be in accordance
with our assumption of box car profiles,  we replace 
the scattering integral inside the two-level source-function 
by mean intensities, i.e., we write
\beq
\Sl = \rho J_\nu + \delta B_\nu, \quad \rho = 1- \delta,
\eeq
where $\delta$ has been defined in Eq.~\ref{deldef} and is 
evaluated for the line-specific thermalization parameter and 
escape probability. The total source-function
(in channel $i$, before averaging) is then given by
\beq
S_{\nu,i} = \frac{\etact + \sige J_\nu + \sum_j \chilj \,(\rho_j J_\nu +
\delta_j B_\nu)}{\chic + \sum_j \chilj},
\eeq
with $\etact$ being the thermal component of the continuum emissivity.
Note that the frequential line-opacity $\chilj$ includes the ``profile
function'' $(2 \Delta\nu_{\rm D})^{-1}$, cf. Eq.~\ref{boxprofile}. 

In the
following, we will investigate how to average the above quantities in order
to be consistent with our definition of $\chim$ and $\chilm$. With respect
to the equation of transfer, which will be finally solved on the coarse
grid, we find that after integration over the subgrid-channels
\beq
\frac{1}{\chim} \frac{\dd}{\dd z} \<I_\nu\> = \<S_\nu\> - \<I_\nu\>,
\eeq
with $z$ being the depth variable along the impact parameter $p$ in the usual
$(p,z)$-geometry.
Strictly speaking, the first term in the above equation (i.e., the mean
inverse opacity) is given by 
\beq
\frac{1}{\chim} =  \frac{\int \frac{1}{\chi_\nu} \frac{\dd}{\dd z} 
I_\nu \dd \nu}{\int \frac{\dd}{\dd z} I_\nu \dd \nu}
\eeq
(where the denominator is equivalent to $\dd/\dd z \<I_\nu\>$, and all
integrals extend over the range $2N \vthm$), i.e., a different definition
applies when compared to the corresponding quantity in Eq.~\ref{chim}. 
Our crucial approximation is to equate both definitions, i.e., inside each 
coarse grid cell (of width $\approx$ 1,000{\ldots}1,500~\kms) we assume that
\[
\int \frac{1}{\chi_\nu} \frac{\dd}{\dd z}
I_\nu \dd \nu \, / \, \int \frac{\dd}{\dd z} I_\nu \dd \nu 
\quad \approx \quad
\int \frac{\dd \nu}{\chi_\nu} \, / \, \int \dd \nu.
\]
Let us frankly admit that this approximation can be justified only
if a) the spatial gradient of the specific intensity is a slowly varying
function of frequency (i.e., deep in the atmosphere) or b) the opacities
are similar for {\it most} of the sub-channels, i.e., either no lines
are present at all or the (summed) line-opacities do not vary too much. 
Additionally and most important, this approximation still works in
those cases when only a couple of channels are populated by large opacities
and the rest is filled by a weak background due to the inverse
relation between opacity and intensity: on the lhs, the high opacity 
channels do not contribute to the fraction because of the correspondingly 
low intensities in both the nominator and the denominator, whereas
on the rhs these channels drop out at least in the nominator because of the low
value of $1/\chi$.

There are, of course, a number of cases where the above
approximation is only poor. With respect to the results presented below 
and since we are {\it not} aiming at a perfect, highly resolved description
of the radiation field in the line-blocking EUV/UV regime, the
errors introduced by the above approximation (and the following one, which is
of similar quality) are acceptable though. 

In order to proceed with appropriate expressions for the emissivity, the
mean source-function, $\<S_\nu\>$, is given by 
\beq
\label{meansource}
\<S_\nu\> = \frac{\etact}{\chim} + \bigl(\frac{\sige}{\chim} +
\big\<\frac{\sum_j \chilj \rho_j}{\chi_\nu^{\rm tot}}\big\>\bigr)\<J_\nu\> + 
\big\<\frac{\sum_j \chilj \delta_j}{\chi_\nu^{\rm tot}}\big\>\,B_\nu,
\eeq
where the Planck-function $B_\nu$ is assumed to be constant within one
macro-grid interval. For those averages which are multiplied by $\<J_\nu\>$,
we have employed an approximation similar to the one discussed above. If we,
finally, denote the opacity dependent means of the third and fourth term by
$\fnth$ and $\fth$, respectively (i.e., {\it n}on-{\it th}ermal/{\it th}ermal),
the equation of radiative transfer for the averaged quantities becomes 
\beqa
\lefteqn{\frac{1}{\chim} \frac{\dd}{\dd z} \<I_\nu\> =} \nonumber \\
& = &\frac{\etact + 
\,(\sige + \chim \fnth)\,\<J_\nu\> + \chim \fth \,B_\nu}{\chim}
- \<I_\nu\>,
\eeqa
and can be solved in the conventional way (pure continuum transport). The 
resulting quantities for the radiation field are to be understood
as average quantities, in the sense that integral quantities such as
$\int J_\nu \dd \nu$ or $\int H_\nu \dd \nu$ are described correctly, 
at least in most cases. The coefficients $\fth$ and $\fnth$ 
can be calculated by summing over the sub-channels,
\beqa
\fth &=& \frac{1}{N} \sum_i \frac{(\sum_j \chilj \delta_j)_i}{(\chic + \sum_j
\chilj)_i} \\
f&=&\fth + \fnth = \frac{1}{N} \sum_i \frac{(\sum_j \chilj)_i}{(\chic + \sum_j
\chilj)_i} \quad < 1! \\
\fnth&=& f - \fth,
\eeqa
and after some simple algebraic manipulations, the following relation is
obtained:
\beq
\chilm = \chim\,(\fth + \fnth).
\eeq
With this equation it is easy to show that the mean source-function 
(\ref{meansource}) allows for a correct thermalization, if
$\etact \rarr \chict B_\nu$ and $\<J_\nu\> \rarr B_\nu$. In this
case, the mean source-function becomes $\<S_\nu\> = B_\nu$, q.e.d.

\begin{figure*}
\resizebox{\hsize}{!}
   {\includegraphics{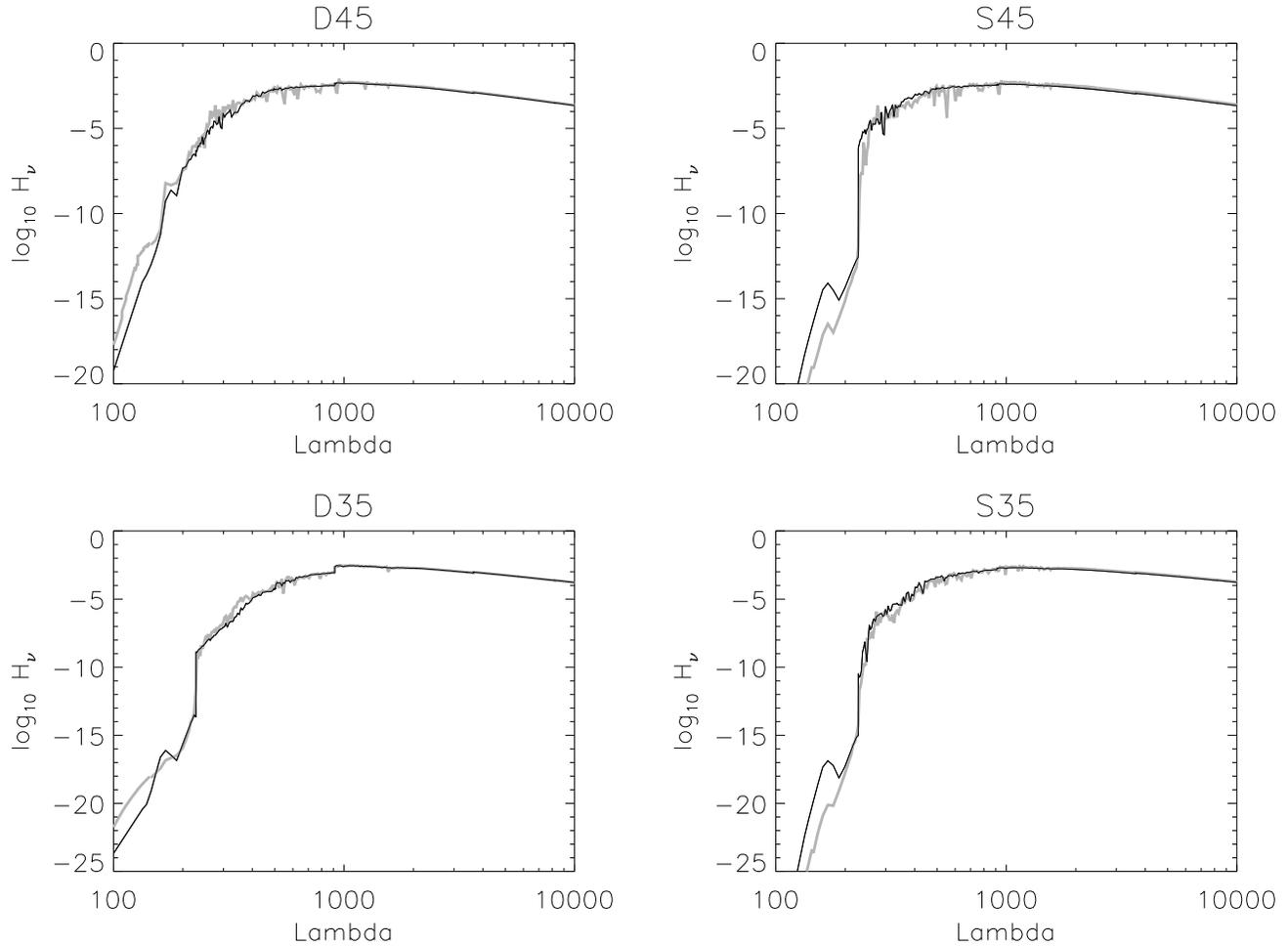}}
\caption{{\sc fastwind} vs. {\sc wm}-Basic (grey): 
comparison of emergent fluxes for
two dwarf and two supergiant models at 35 and 45~kK (for parameters, cf.
\citealt{Paul01}). In order to allow for a meaningful comparison, the high resolution
frequency grid provided by {\sc wm}-Basic has been re-mapped 
while keeping the corresponding flux-integrals conserved.} 
\label{comp_adiflux}
\end{figure*}

\smallskip
\noindent
We now need to incorporate the effects of the velocity
field into our approach. Due to the method to average the
opacity, we cannot simply shift the lines with respect to the
stellar frame. Consider, e.g., one strong line to be present without
any other interfering lines. In ``reality'' and in the observer's frame, the
absorption part of this line becomes broader as a function of velocity, i.e.,
the larger the velocity the more flux is blocked (of course, a significant
part is reemitted due to scattering). If we simply shift our line(s)
as a function of velocity, almost nothing would happen,
since, as shown above, the  mean opacity/radiation field remains almost 
unaffected by one strong line, due to the possible escape via the (N-1)
unblocked sub-channels. Thus, in order to simulate the physical process, we
proceed in a different way. When the velocity shift becomes
larger than twice the average ``thermal'' width (including
micro-turbulence), we combine (in proportion to the local velocity) more and
more subchannels to increase the relative weight of the line in the mean
opacity. In particular, the line width (more precisely, the width of the
sub-channels) is set to the value 
\beq 
\Delta \nu = {\rm max}(2 \Delta
\nu_{\rm D}, \frac{{\rm v}(r)}{\lambda_0}). 
\eeq 
Although this procedure is highly approximative, it allows to deal with
the effects of ``line-shadowing'' and prevents any premature escape of
photons when the lines begin to overlap.

\begin{figure*}
\begin{minipage}{8.8cm}
\resizebox{\hsize}{!}
   {\includegraphics{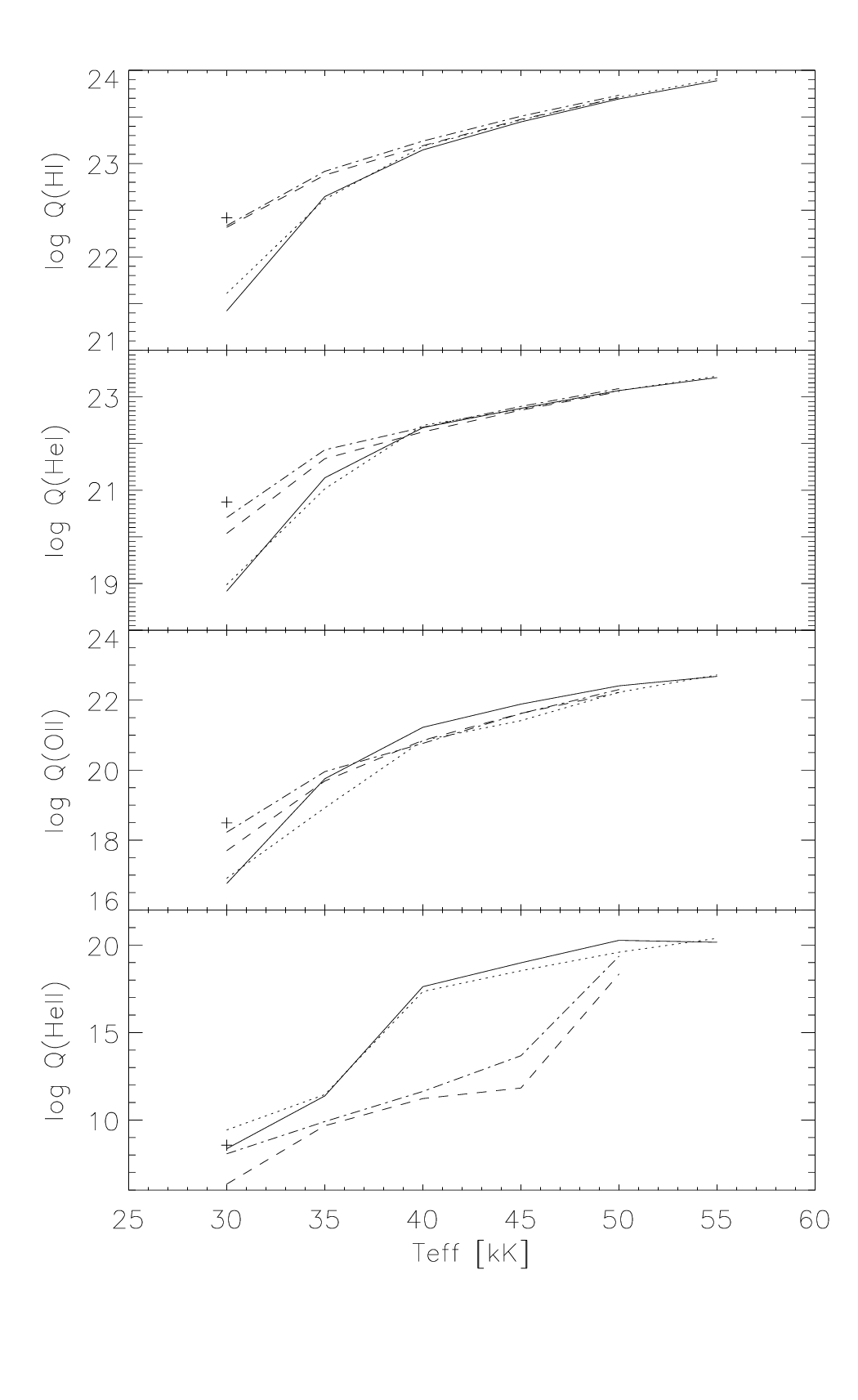}}
\end{minipage}
\hfill
\begin{minipage}{8.8cm}
   \resizebox{\hsize}{!}
      {\includegraphics{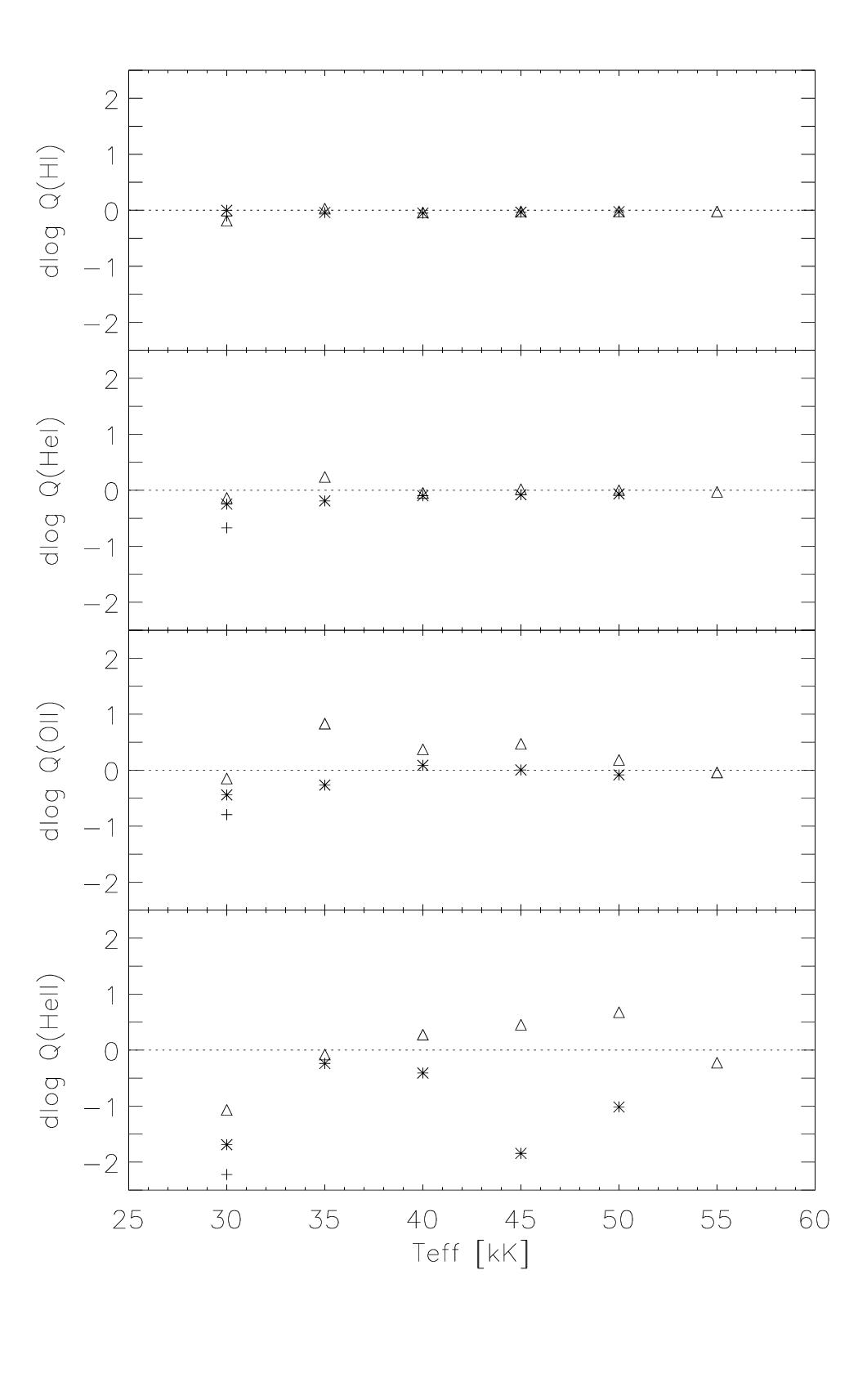}}
\end{minipage}
\caption{Comparison of ionizing photon number for the model grid provided by
\citet{Paul01}. Left panel: logarithm of Zanstra-integrals, $\log Q_x$,
(Eq.~\ref{zanstraint}), for \Hi, \Hei, \Oii and \Heii. Bold/dashed:
dwarfs/supergiants as calculated by {\sc wm}-Basic; dotted/dashed-dotted: 
results from {\sc fastwind}. Right panel: ratio of corresponding Zanstra
integrals, $\Delta \log Q_x = \log Q_x^{\rm WMB} - \log Q_x^{\rm FW}$ (WMB:
{\sc wm}-Basic, FW: {\sc fastwind}), for dwarfs (triangles) and
supergiants. For the supergiant model at 30,000~K (``S30''), we have used
the {\sc fastwind} model {\it without} photospheric line-pressure in both
figures. The corresponding results for the ``correct'' model, i.e., 
including photospheric line-pressure, are indicated by the ``+'' sign (see
text).} 
\label{qzcomp}
\end{figure*}

\subsection{Tests/Comparison with {\sc wm}-Basic}
\label{compwmbasic}

Before we test our approximate approach by comparing with alternative
calculations, let us mention two important consistency checks we have
performed.
\begin{enumerate}
\item[a)] The calculated models (and spectral energy distributions/line profiles) are,
almost, independent of the actual value of coarse grid cells, $N$, at least
if varied within a reasonable range. (We checked for values between $0.5 N$
to $2 N$, for $N=60$.) 
\item[b)] As long as the IR/radio-range is not considered, our simpler 
models with a temperature-structure calculated from Hopf-parameters 
and {\it all} background elements in approximate NLTE 
agree very well with complex models including a consistent T-structure. This check verifies
analyses performed with previous versions of {\sc fastwind}, e.g., 
\citet{h02} and \citet{Repo04}.
\end{enumerate}
In the following, we will compare the fluxes from our
models with those calculated by {\sc wm}-Basic by means of the O-star grid
presented by \citet{Paul01}\footnote{available via {\tt
http://www.usm.uni-muenchen.de/people/adi/\\Models/Model.html}}. These tests
should give reasonable agreement, since both codes use the same atomic data
base for the background-elements. A comparisons with results from 
{\sc cmfgen} will be discussed later on.

The parameters of the corresponding models (calculated without X-rays) can be
found in \citet[ Table~5]{Paul01}. Our models have been constructed as
closely as possible to the approach inherent to {\sc wm}-Basic, i.e., 
including a consistent temperature stratification (which will be
described in Sect.~\ref{tempstrat}) and Sobolev line-transfer. For the
velocity-field, we have used $\beta = 0.9$, which results in a stratifiction
very close to the one predicted by {\sc wm}-Basic (see below). The computation
time on a 2 GHz processor machine is of the order of 15 to 20 minutes per
model (typically 40 to 50 iteration cycles for a final convergence below
0.003 in {\it all} quantities, if the temperature is updated each 2nd cycle).

The grid comprises 6 ``dwarfs'' and 5 ``supergiants'' in the range between
30{\ldots} 55~kK (``D30''{\ldots} ``D55'' and ``S30''{\ldots} ``S50'',
respectively), and we have concentrated on the grid with solar
abundances, in order to deal with more prominent effects related to
line-blocking/blanketing. Fig~\ref{comp_adiflux} compares the emergent fluxes 
for some typical cases, two dwarf and two supergiant models 
at 35 and 45~kK. In order to allow for a meaningful comparison, we have
re-mapped the high resolution frequency grid provided by {\sc wm}-Basic 
while keeping the corresponding flux-integrals conserved.

\begin{figure*}
\begin{minipage}{8.8cm}
\resizebox{\hsize}{!}
   {\includegraphics{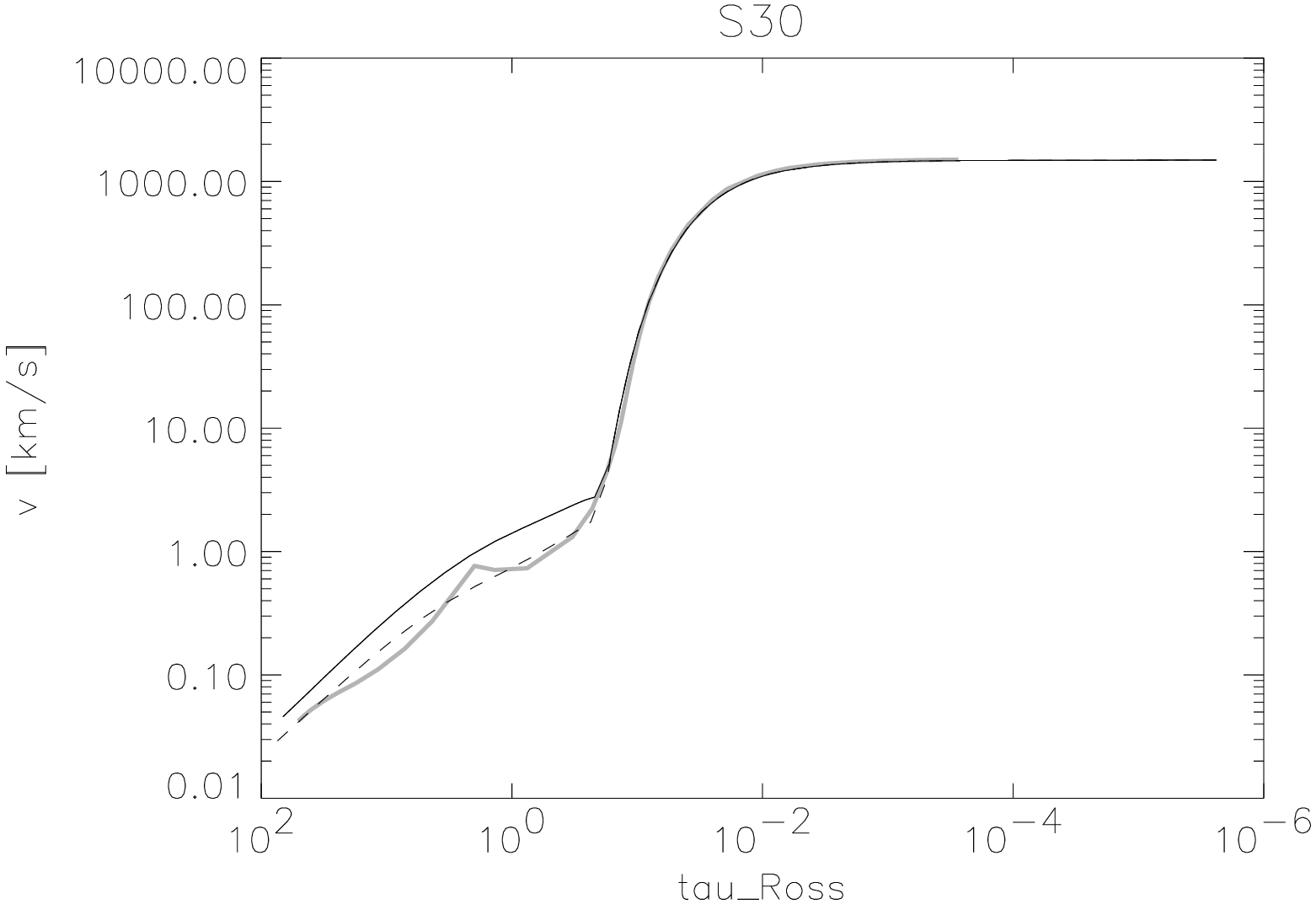}}
\end{minipage}
\hfill
\begin{minipage}{8.8cm}
   \resizebox{\hsize}{!}
      {\includegraphics{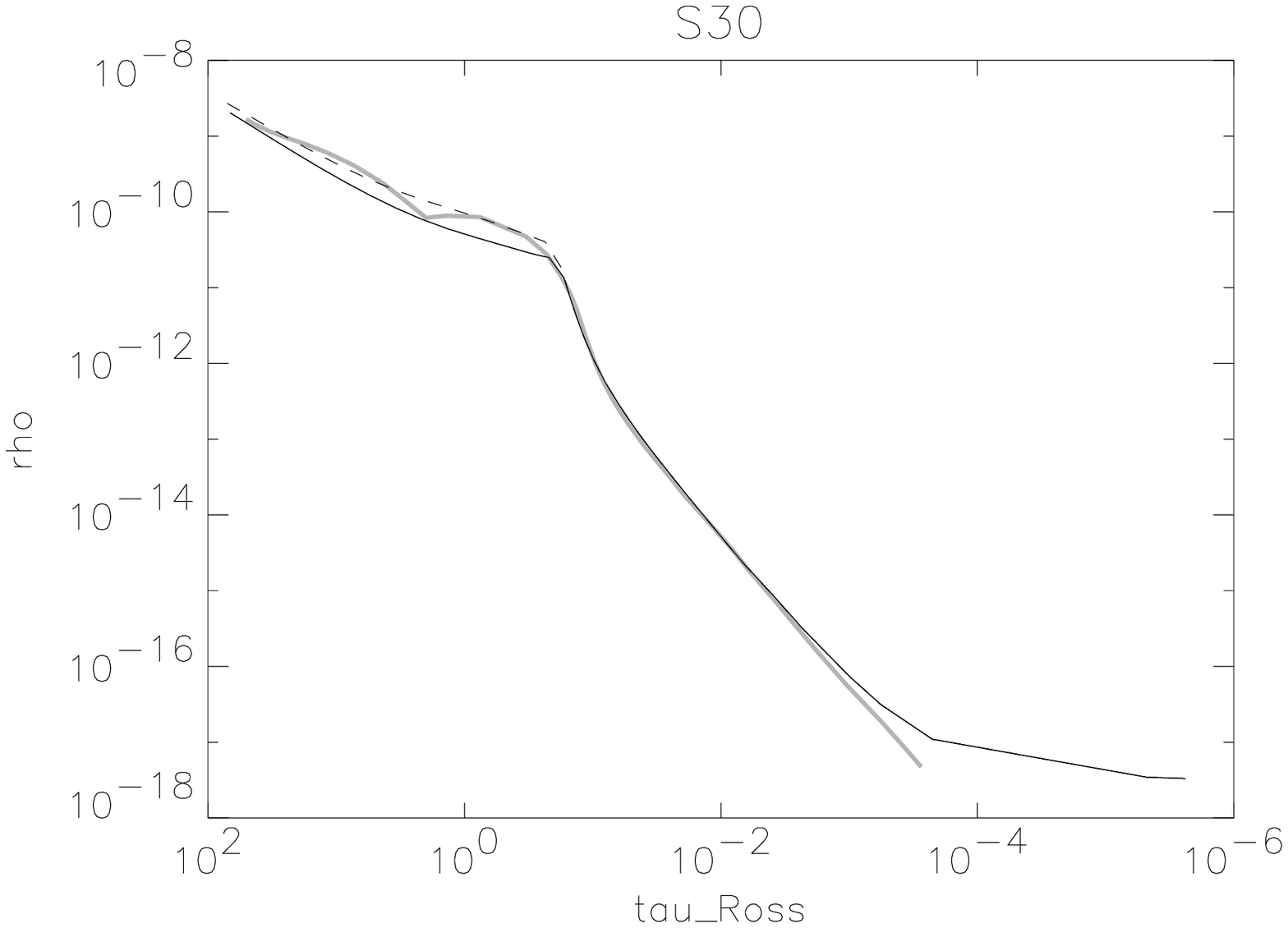}}
\end{minipage}
\caption{Comparison of velocity/density structure for model S30. Grey: {\sc
wm}-Basic; bold/dashed: {\sc fastwind} with/without photospheric 
line-pressure, respectively. } 
\label{s30}
\end{figure*}

Overall, the agreement is rather good; in particular, the range above
400~\AA\, is reproduced very well, except for some strong
absorption/re-emission features which are missing in our mean-opacity
approach. We have convinced ourselves that in all cases also the IR-fluxes
(not displayed here) agree perfectly, i.e., the IR flux-excess induced by
the wind is reproduced equally well in both codes. Major differences are
``only'' present in two regions: most models differ in fluxes below
200~\AA, although the strength of \Heii Lyman-jump itself is very similar. 
Mostly, this problem is related to the enormous bound-free
opacity provided by \Oiv (and \Fev or \Civ for the hotter or cooler objects,
respectively) leading to an optically thick wind from the outermost radius
point on (in our case, $\Rmaxe = 120 \Rstare$), so that the flux is rather
badly defined in this frequential region. As we will see from a comparison
with models calculated with {\sc cmfgen} (Figs.~\ref{comp_cmfgenflux}),
these models predict a third alternative for $\lambda < 200$~\AA, and even
the Lyman-jump is different. As a result, we consider the ionizing fluxes in
this wavelength range as not particularly reliable. Moreover, the influence
of X-rays becomes decisive, implying that any tool for nebula
diagnostics should use these numbers only with care. 

The second inconsistency is found in the region between 300 to 400~\AA.
Although this range poses no problem for supergiants, the flux-blocking
predicted by {\sc fastwind} {\it for dwarfs between 35 to 45~kK} is larger
than calculated by {\sc wm}-Basic, with a maximum discrepancy around 35~kK.
The reader might note that {\sc cmfgen} again produces somewhat different
results in this range: agreement with {\sc wm}-Basic is found for dwarfs,
whereas the fluxes emitted from supergiants are larger compared to both {\sc
fastwind} and {\sc wm}-Basic.

This dilemma becomes particularly obvious if we consider the corresponding 
Zanstra-integrals,
\beq
\label{zanstraint}
Q_x = \int_{\nu_x}^\infty \frac{H_\nu}{h\nu} \dd \nu,
\eeq
which are proportional to the emitted number of ionizing photons. In the
left panel of Fig.~\ref{qzcomp}, we compare the logarithm of $Q_x$,
evaluated for \Hi, \Hei, \Oii and \Heii, whereas the corresponding ratios,
$\Delta \log Q_x = \log Q_x^{\rm WMB} - \log Q_x^{\rm FW}$ (WMB: {\sc wm}-Basic, 
FW: {\sc fastwind}), are displayed in the right panel. Obviously
both codes predict the same numbers in the hydrogen Lyman and in
the \Hei\, continuum. As already discussed, the situation is much less
satisfactory for the \Heii\, continuum, where the differences are particularly
significant for supergiants.  Note, however, that the principal dependence
of $Q_{\rm HeII}$ on spectral type and luminosity class, which shows the
largest variation throughout the spectrum (lower left panel), is much more
consistent than one might expect on basis of the right panel alone. In
the \Oii continuum ($\lambda < $ 352~\AA), finally, the differences for the
dwarfs at intermediate spectral type are evident. 

Note that in this wavelength range the line-density is very large, and
differences in the treatment of the weakest background opacities might
explain the established disagreement. 
An argument in support of this hypothesis is given by the
fact that {\sc fastwind} recovers the results by {\sc wm}-Basic perfectly if
a line-list is used which has significantly less (overlapping) weaker lines 
in the considered interval. For a final statement, however, more tests are
certainly required. Note that a comparison with {\sc cmfgen} addressing this
point will not solve the problem, since the number of lines included in 
this code is mostly lower than described here, because {\sc cmfgen} uses
only those lines where the occupation numbers of {\it both} levels are
known, in contrast to our approach which uses also lines where the upper
level is lying too high to be included into the rate equations.

One last point we would like to mention concerns model S30. In a first
comparison, we immediately encountered the problem that particularly this
model provided fluxes which showed significantly less agreement at all
frequencies than the other models (indicated by the plus-signs in
Fig.~\ref{qzcomp}). Comparing the models themselves, it turned out that
temperature, density and velocity structure showed a severe mismatch in
photospheric regions (cf. Fig.~\ref{s30}, grey vs. black curves). After some
tests, we found that both models agree well if the photospheric line
pressure is neglected in {\sc fastwind} (grey vs. dashed curves in
Fig.~\ref{s30}). Most likely, this problem is related to the treatment of
the line pressure in {\sc wm}-Basic. Whereas the continuum forces are
calculated from correctly evaluated opacities, the line pressure,
independent of location, is calculated in terms of the force-multiplier
concept, utilizing the Sobolev approximation. Particularly, $g_{\rm
rad}^{\rm line} \propto t^{-\alpha}$, with ``depth parameter'' $t\propto
\rho/(\dd v/\dd r)$. Thus, $g_{\rm rad}$ decreases rapidly in photospheric
regions when the density is large and the velocity gradient small.

In those cases where the (static) line pressure is non-negligible in
photospheric regions, the chances are high that the above approximation 
leads to a too large effective gravity, i.e., too high densities. Actually,
this problem is already known for a long time and has been discussed in fair
detail in \citet[ particularly Fig.~6c]{PPK}. The reason that this problem
occurs only in S30 results from fact that the Eddington factor is
considerably higher than for almost all other models ($\Gamma = 0.52$).
Insofar, the photospheric line pressure has much more impact than for models
with either high gravity or low $\Gamma$. Moreover, at an effective
temperature of 30~kK, \Feiv with its enormous number of lines spread
throughout the spectrum is the dominant (or almost dominant) ionization
stage in the ``middle'' photosphere, thus, contributing a much larger amount
of static line pressure than for hotter temperatures, where \Fev or even
\Fevi are contributing.  

\begin{figure}
\resizebox{\hsize}{!} {\includegraphics{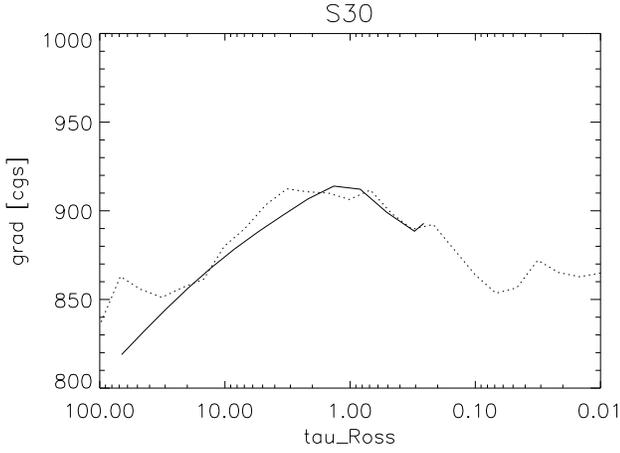}}
\caption{Comparison of {\it total} photospheric radiative acceleration for
model S30 (bold) vs. results from an analogous hydrostatic Kurucz-model
(dotted). Note that the gravitational acceleration for this model is 
1000~cms$^{-2}$, i.e., the radiative acceleration is very close to this
value and, thus, of extreme importance (cf. Fig.~\ref{s30}). The deviations
at largest depths are due to the fact that this model becomes (spherically) 
extended  in the lowermost photosphere, an effect which cannot be treated in
a plane-parallel approach (cf. Paper~I).}
\label{s30grad}
\end{figure}

Note that we have also compared our (cooler) models (from our grids as
described in Sect.~\ref{modelgrids} and from additional A-star models)
with corresponding
Kurucz models, where in most cases a very good agreement regarding the
photospheric radiative acceleration has been found, e.g., Fig.~\ref{s30grad}.
Only for models cooler than 9,000~K a mismatch becomes obvious, where
``our'' radiation pressure is too low, due to a number of missing \Feii
lines in the optical (improvements under way!).

In order to allow for a meaningful comparison concerning our approximate
line-blocking, in Fig.~\ref{qzcomp} we have used the results from our S30
model {\it without} photospheric line-force, whereas the results from the
``actual'' model (including $g_{\rm rad}^{\rm line}$) are indicated by ``+''.
Independently, however, Fig.~\ref{s30} (left panel) also shows
the validity of our treatment of the transition zone from photosphere to
wind (cf. Paper~I), since in this region both velocity fields agree
perfectly. (Remember that {\sc wm}-Basic solves the hydrodynamical
equations in a consistent way).

\section{Treatment of inverted levels}
\label{treatinv}

One of the more complex problems when solving the coupled equations of
statistical equilibrium and radiative transfer is the presence of population
inversions, which often occur in the outermost layers of hot expanding
stellar atmospheres. The amount of the overpopulation (i.e., $n_u/g_u >
n_l/g_l$) is usually small, but even in this case it invokes a number of
problems concerning the solution of the radiative transfer equation.
Particularly with respect to the usual concept of using source functions, a
problem occurs in the transition zone between ``normal'' population and
overpopulation, where the source function formally diverges. In addition,
factors like $\exp(-\tau)$ may produce numerical problems for $\tau<0$. In a
number of codes, this problem is ``solved'' by setting the upper level into
LTE with respect to the lower one or by other approximations. Since level
inversions are particularly present between levels responsible for IR-lines
and since {\sc fastwind} aims at a reliable solution also in these cases, we
cannot afford such approximations and have to solve the ``exact'' case 
which in turn has an influence on the degree of overpopulation itself. In
this section, we briefly describe how we have solved the problem in {\sc
fastwind} both with respect to the Sobolev approach and within the
CMF-transport.

\subsection{Treatment of inversions in the Sobolev approximation}

Since the Sobolev approach uses only {\it
local} quantities, a divergence of the source function is not possible,
except for the extremely unlikely case that upper and lower populations,
normalized to the appropriate statistical weight, are numerically
identical. Thus, we can retain the standard concept (optical depth and
source function) and follow the approach described in \citet{Tareschetal97}: 
in case of a level inversion, the interaction function 
$U(\tau_S,\beta_{\rm P})$\footnote{which describes the interaction between
line and continuum processes, where $\tau_S$ is the Sobolev optical depth
and $\beta_{\rm P}$ the ratio of continuum to line opacity in a frequency
interval corresponding to the thermal Doppler width, cf. \citet{HumRyb85,
PulsHum88}.} is split into two parts in order to avoid numerical problems,
\beq
\bar{U}=U_{1}+U_{2}, \qquad {\rm with} \qquad U_{1}=1-\beta.
\eeq
$\beta$ is the usual escape probability in Sobolev approximation
(Eq.~\ref{betasob}), which for the case of inversion is given by
\beq
\beta=\frac{\exp{|\tau_S|}-1}{|\tau_S|}
\eeq
and $U_{2}$ has been described in \citet[ Eq. A13]{Tareschetal97}.

For $|\beta_{\rm P}| \rightarrow \infty$, i.e., dominating continuum,
$U_{2}$ approaches zero. In the case of dominant line processes, on the other
hand, and $\tau_S < 0$, $U_{2}$ approaches ($\beta-1)$ and $\bar{U}$ goes to
zero. Thus, we recover the ``classical'' result by Sobolev, where the
influence of continua has been neglected.

In our approach, we have significantly extended the grid used by
\citet{Tareschetal97} from which $U_2$ is calculated by means of
interpolation. Due to the different behaviour of this function in different
regions of the $(\tau_S, \beta_{\rm P})$ plane, (four) different tables with
different degrees of resolution have been calculated. The boundaries of the
complete grid comprise the area between $-6 \le \log|\tau_S| < 2.8$ and $-6
\le \log|\beta_{\rm P}| < 6$. Beyond the boundaries, 
$U_2$ is calculated analytically (by either 
considering the appropriate limits or using a first order expansion). In
particular, it turns out that 
\beqa
U_2 &=& 0.5 |\tau_S|, \qquad \log |\tau_S| < -6 \quad \mbox
{independent of } \beta_{\rm P} \\
U_2 &=& \beta-1, \qquad \log |\beta_{\rm P}| < -6 \quad \mbox
{independent of } \tau_S \\
U_2 &=& \frac{1}{\sqrt{2\pi|\beta_{\rm P}|}}, \quad \log |\beta_{\rm P}| > 6, 
\quad \log |\tau_S| > -6
\eeqa
and the limits for large $\log |\tau_S| >2.8$ and $-6 \le \log|\beta_{\rm P}|
\le 6$ result from appropriate extrapolations from the pre-calculated tables.

\subsection{Treatment of inversions in the comoving frame}

In the CMF solution, the problem of source-function divergence is inevitable
when a population inversion occurs and the standard formalism is used. Even
if the local quantities are not diverging, there will be an implicit
divergence just between the two depth-points before and at the beginning of
overpopulation, which, due to the applied discretization, will not be handled
consistently. To avoid this problem, it is more suitable to work directly
with emissivities and opacities rather than with optical depths and source
functions. Thus, in the case of inversion, we solve the 
two coupled equations of radiative transfer in the comoving frame according to
\beqa
\frac{\partial u}{\partial z}-\frac{\partial v}{\partial x} & = &-\chi v \\
\frac{\partial v}{\partial z}-\frac{\partial u}{\partial x} & = & \eta-\chi u,
\eeqa
where $u$ and $v$ are the usual Feautrier variables, $x$ is the frequency
measured from the center of the line in Doppler width units, and $z$ is the 
depth variable along the impact parameter.  The opacity is $\chi = \chi_{c}(z) +
\chi_{L}(z,x)$ and the emissivity is $\eta=\eta(z,x)$.

In order to discretezise the equations with respect to $z$ and $x$, a
fully implicit scheme is used. As was shown by \citet[ Appendix
B]{Mihalasetal75} this method is {\it unconditionally} stable. 

\subsection{Tests}

A number of tests have been performed concerning both the Sobolev and the
CMF implementations. Most importantly, we have also tested models where the
above discretization of the CMF equations with respect to $z$ has been used
for {\it all} transitions, not only for the ``inverted'' ones, and found
satisfactory agreement with our standard implementation using a
discretization with respect to $\tau$.

After convincing ourselves that the algorithms are working in principal, we
have tested our improved methods by comparing them with older results (where in
case of inversion the upper level and the line source function where set to
zero). This comparison has been performed for the O-star grid described in
the previous section. The results were very satisfying, and a
number of convergence problems originating from the older treatment of
inverted populations are no longer present. 

The differences in the resulting H/He line profiles (both in the optical and
in the IR) turned out to be rather small, since for our grid parameters 
these lines are formed below those regions where the inversion sets in.
However, we like to point out that a consistent treatment might be important
for winds with more extreme mass-loss rates and for a number of metallic IR 
transitions with an inversion already occurring in photospheric regions.\footnote{a 
typical example is the Si{\sc iv} IR transition 4d $^2{\rm D}^e$ - 
4f $^2{\rm F}^o$}  

\section{Temperature stratification}
\label{tempstrat}

As has been previously mentioned, the present version of {\sc
fastwind} allows for the calculation of a consistent temperature
stratification, utilizing a flux-correction method in the lower wind and the
thermal balance of electrons (cf. \citealt{kubat99}) in the outer
part.\footnote{Note that adiabatic cooling resulting from wind expansion
is presently neglected in our models (work under way).} The region where
both methods are connected is somewhat dependent on mass-loss, but typically
lies at $\taur = 0.5$. Although the implementation of this method is
straightforward, and the contribution of individual processes have been
discussed in fair detail by \citet{drew85, drew89}, three points are worth
mentioning.

In order to calculate the appropriate heating/cooling rates resulting from
collisional bound-bound transitions, the population of excited levels is as
important as the population of ground and meta-stable ones. This can 
readily be seen from the fact that the {\it net} heating rate from a
collisional transition between lower level $l$ and upper level $u$ can be
expressed as 
\beq Q_{ul}-Q_{lu} = \left(n_u C_{ul} - n_l C_{lu}\right) h\nu_{lu} 
= n_l C_{lu} h \nu_{lu} \left(\frac{b_u}{b_l} -1\right), 
\label{lineheating}
\eeq 
with collisional rates $C_{ul}$ and NLTE departure coefficients $b_l, b_u$.
Thus, the {\it ratio} of departure coefficients controls whether a certain
transition heats or cools the plasma and its deviation from unity controls
the degree of energy transfer. Heating results from transitions with an
upper level being overpopulated with respect to the lower one, and cooling
vice versa. Thus, the occupation numbers of {\it all} ionic levels have to
be known with some precision, and we have to modify our approach when the
electron thermal balance is used to calculate the temperature profile. The
approximate NLTE solution as described in Sect.~\ref{nlteapprox} simply does
{\it not} yield the required occupation numbers of excited levels (except
those which are directly connected to the ground or meta-stable level), and any
brute force approximation would give incorrect heating/cooling rates.

To overcome this dilemma we incorporated a detailed solution of the
statistical equilibrium at least for those elements with large contributions
to the net heating rates (positive or negative). After some experiments it
turned out that the inclusion of the most abundant background elements 
C, N, O, Ne, Mg, Si, S, Ar, Fe, Ni (plus the explicit elements, of course) 
is sufficient to stabilize the results. For these elements
then, the complete rate-equations are solved with line transitions
treated in Sobolev approximation, whereas for the remaining ones the
approximate NLTE solution is employed.

The second point to be mentioned regards the flux-conservation of the final
models. The conventional approach to calculate the energy balance,
formulated in terms of radiative equilibrium, satisfies this constraint by
construction, at least in principle. (Most numerical codes, including
{\sc cmfgen} and {\sc fastwind}, calculate mean intensity and flux 
on different grids, which somewhat destroys the coupling between
radiative equilibrium and flux conservation). On the other hand, our
formulation in terms of the electron thermal balance is decoupled from the
latter requirement, at least regarding any {\it explicit} dependence. Note,
however, that there is an implicit coupling via the rate equations, assuring
that the constraints of electron thermal balance and radiative equilibrium
are physically equivalent (cf. \citealt{hil98} and \citealt{hil03}, where
further discussion concerning both methods and their correspondance is
given). Insofar, we can use the achieved flux-conservation as an almost
independent tool to check whether our models have been constructed
in a consistent way. In most of the cases considered so far we have found a
perfect conservation, but in the worst cases (below 5\% of all models) a
violation up to 1.5\% is possible. 

The third point to be discussed is mainly relevant for our specific approach
of modeling stellar atmospheres. Presently, and in accordance with the
majority of similar codes, we do {\it not} update the photospheric density
stratification once it has been calculated. Since the photospheric structure
equations are solved for the gas-pressure $P$ and the density is calculated
from the ratio $P/T$, the density is only as good as the initial ``guess''
for the temperature stratification. Moreover, an implicit dependence 
of the final temperature distribution on this initial guess is created.

Thus, it is still important to obtain a fair approximation for the latter 
quantity, which in our models is accomplished via the corresponding NLTE
Hopf-parameters (see Paper~I) which have to account for line-blanketing
effects. Meanwhile, we have accumulated a large set of these parameters from
our model-grid calculations (and, for cooler temperatures, from
corresponding Kurucz-models). If, on the other hand, the initial
(photospheric) temperature stratification were not appropriate, both 
occupation numbers and line profiles would be affected from the erroneous
density (although the flux would be conserved, see above). 

\begin{figure}
\resizebox{\hsize}{!}
   {\includegraphics{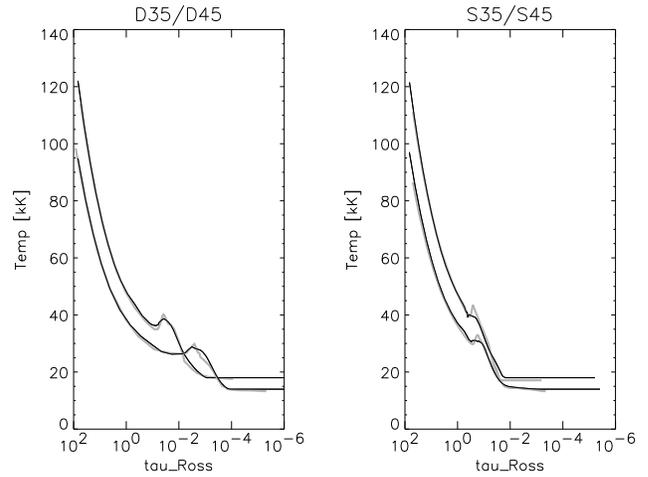}}
\caption{{\sc fastwind} (bold) vs. {\sc wm}-Basic (grey): 
comparison of temperature stratification for some of the models described in
Sect.~\ref{compwmbasic}.}
\label{comp_adit}
\end{figure}

In Fig.~\ref{comp_adit} we show some of our results in comparison with
results calculated by means of {\sc wm}-Basic, a code which also uses the
electron thermal balance. Obviously, the differences are tiny and visible
only for the temperature bumps of supergiants, which are predicted
to be more prominent by {\sc wm}-Basic. Note, however, that our solution
is more consistent with the results from {\sc cmfgen} (see
Fig.~\ref{comp_cmfgent}), which will be presented in the next section.

Comparing the computation time of models with and without consistent
temperature structure, we find a typical difference of a factor of two.
Interestingly, the number of iterations becomes only moderately larger 
(because of the fast convergence of the temperature when using the
electron thermal balance, see \citealt{kubat99}), and most of the additional
time is spent for solving the NLTE equations for the important back-ground
elements.

\begin{figure}
\resizebox{\hsize}{!}
   {\includegraphics{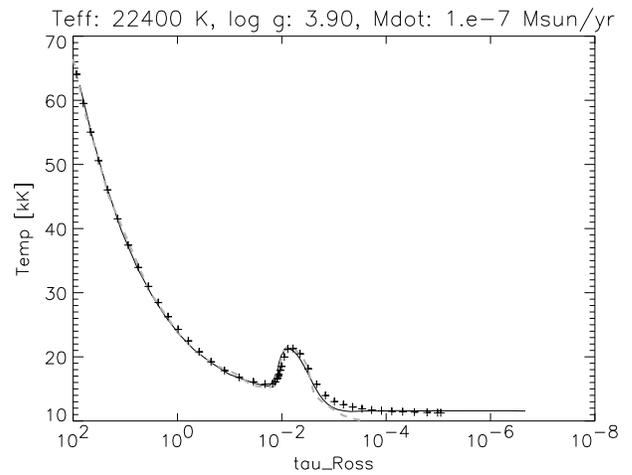}}
\caption{Extreme temperature-bump around 22,000~K: 
{\sc fastwind} (bold)  vs. {\sc wm}-Basic (grey, dashed) and {\sc
cmfgen} (crosses). See text.}
\label{compt22}
\end{figure}

We finish this section with an interesting finding and warning. After having
calculated a large number of models with our code, in certain domains of
\Teff we have found temperature bumps of {\it extreme} extent. In contrast
to ``normal'' bumps (arising from line-heating in the outer photosphere) 
which are of the order of 2,000~K or less for O-stars
(Fig.~\ref{comp_adit}), corresponding values at lower effective temperatures
might reach 5,000~K, as shown for an exemplary dwarf-model at \Teff =
22,400~K in Fig.~\ref{compt22}.

We like to stress the fact that this behaviour has been confirmed by
calculations performed by means of {\sc wm}-Basic and {\sc cmfgen}, kindly
provided by T.~Hoffmann and F.~Najarro on our request. This finding 
(identical results for an unexpected and somewhat strange effect) allows for
two conclusions. First, the effect is ``real'', at least in terms of the
applied physics (see below), and second, the results from different codes
using different techniques are strongly converging, which is very
promising and allows for an increasing trustworthiness of the results
themselves.

After some investigations, it turned out that the feature under discussion
originates from bound-bound heating by \Ciii\footnote{at this specific
temperature, bumps at other temperatures originate from different ions,
e.g., helium.}(which is a major ion at these temperatures), contributed by
few transitions connected to the ground-state (singlet), to the meta-stable
level (lowermost triplet state) and the transition between ground and
meta-stable level at roughly 1909 \AA. Note that the latter transition has
been identified to be of significant importance for the energy-balance in
the wind of P~Cyg, in that case as a cooling agent (cf. \citealt{drew85},
Fig.~3). In our case, however, the \Ciii ground-state is strongly
underpopulated in the transonic region (because of the same effect
under-populating the \Heii ground-state in hot stars, cf.
\citealt{Gabler89}), so that the bracket in Eq.~\ref{lineheating} becomes
very large and the heating-rate enormous, also because of the large
collisional strengths of these transitions. If, on the other hand, the
contributions by \Ciii are neglected at all, a temperature bump of only
moderate size is created.

The lesson we learn from this exercise is two-fold. First, only a couple of
lines (from one ion) can lead to a considerable heating in stellar
atmospheres, at least theoretically. Since this heating takes place in the
outer photosphere it will have a significant effect on the spectra, and we
can check this prediction observationally. However, we have also to consider
that the degree of heating (i.e., the extent of the temperature bump) depends
strongly on the corresponding collision strengths of the responsible
transitions (as a function of temperature), and before relying on our
results we have to carefully check for possible 
uncertainties.\footnote{Note that even some of the hydrogen collision
strengths have been revised recently, cf. \citet{Przybilla04}.}

\section{Comparison with CMFGEN}
\label{cmfgencomp}

\begin{figure}
\resizebox{\hsize}{!}
   {\includegraphics{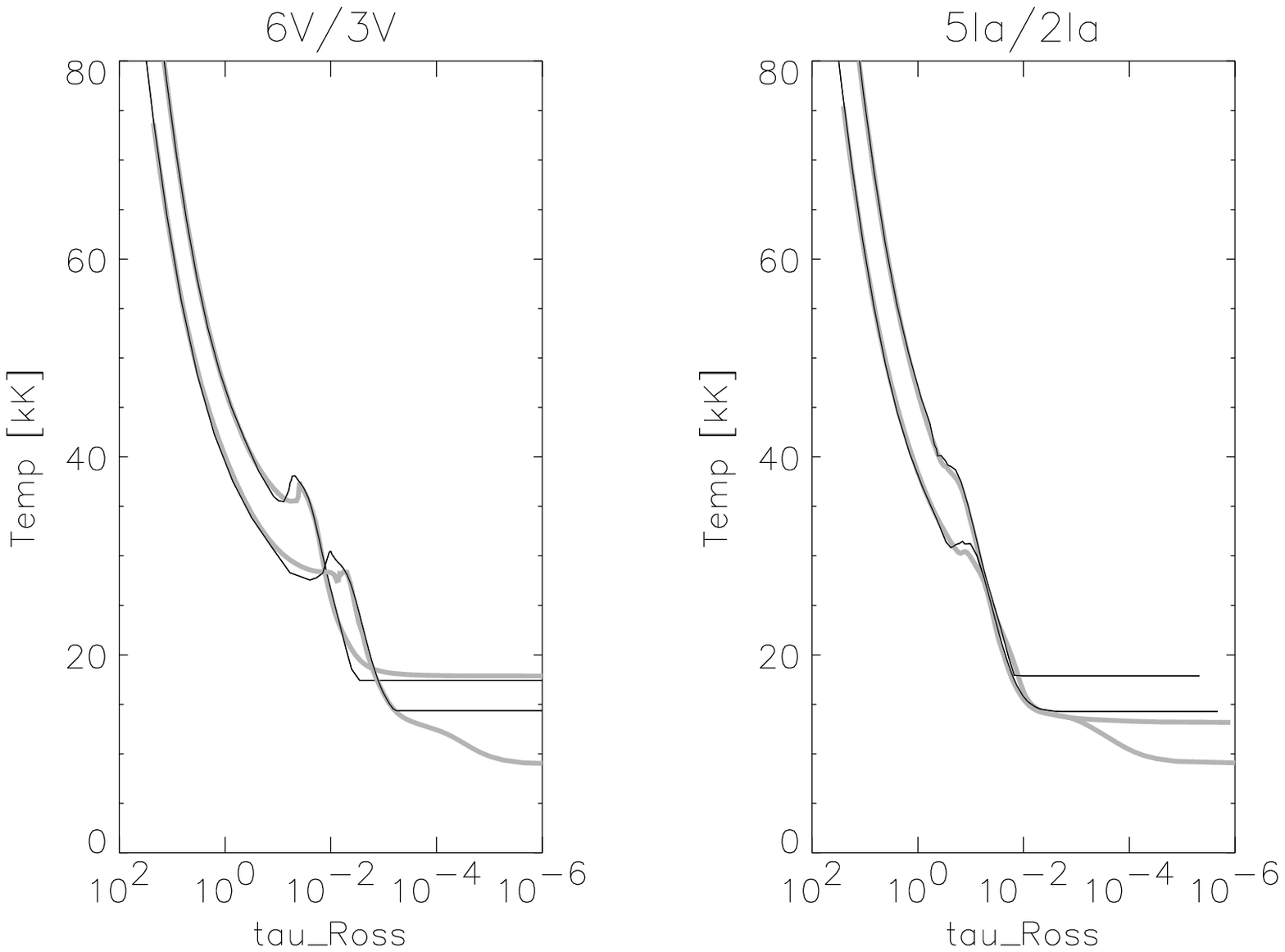}}
\caption{As Fig.~\ref{comp_adit}, but for {\sc fastwind} (bold) vs. {\sc
cmfgen} (grey, dashed). The stellar parameters are similar to the models
displayed in Fig.~\ref{comp_adit}, with \Teff(6V) = 35861~K, \Teff(3V) =
43511~K, \Teff(5Ia) = 35673~K and \Teff(2Ia) = 44642~K. {\sc cmfgen} results
from the model grid as calculated by \citet{len04}.}
\label{comp_cmfgent}
\end{figure}

\begin{figure}
\resizebox{\hsize}{!}
   {\includegraphics{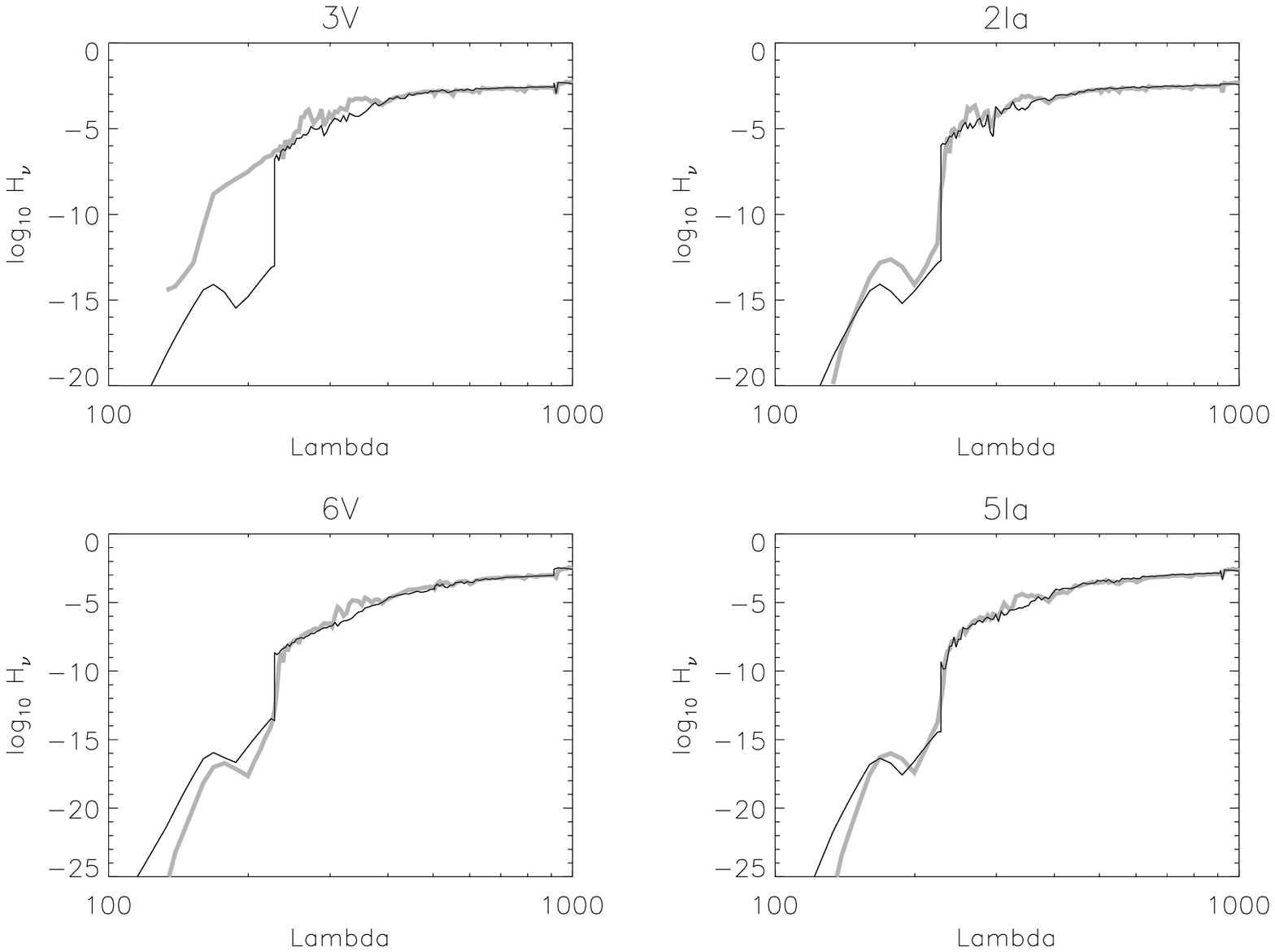}}
\caption{As Fig.~\ref{comp_adiflux}, but for
{\sc fastwind} vs. {\sc cmfgen} (grey). Effective temperatures as in 
Fig.~\ref{comp_cmfgent}. Only the EUV part is plotted, at larger wavelength
the results are extremely similar.}
\label{comp_cmfgenflux}
\end{figure}

In this section, we will compare the results from our models with
corresponding results from {\sc cmfgen}, with particular emphasis on the
optical H/He profiles which cannot be compared to results from {\sc
wm}-Basic, due to lack of comoving frame transport and adequate
line-broadening. For this purpose, we have used the {\sc
cmfgen-}simulations by \citet{len04}, who have provided a grid of dwarf,
giant and supergiant models (no clumping) in the O-/early B-star range. The
corresponding {\sc fastwind} models have been calculated with identical
parameters, and the explicit elements (H/He) have been treated with comoving
frame transport. Thus, the only ``physical'' difference in both calculations
concerns the photospheric density stratification, which is approximated by a
constant scale-height in {\sc cmfgen}, but described consistently
by {\sc fastwind} (cf. Sect.~\ref{lineblock}). 

\begin{figure}
\resizebox{\hsize}{!} 
   {\includegraphics{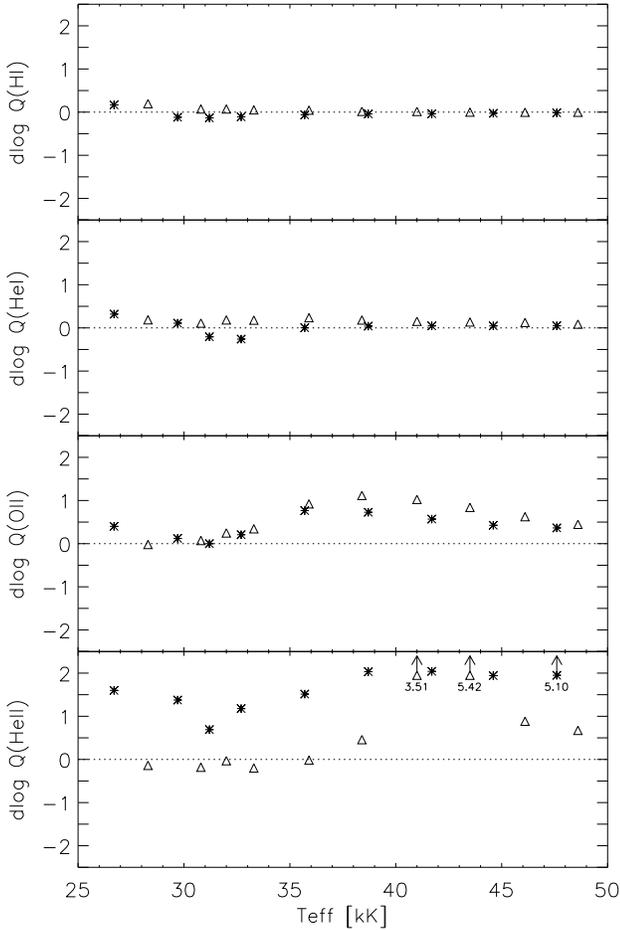}}
\caption{As Fig.~\ref{qzcomp}, right panel, but for {\sc cmfgen} vs.
{\sc fastwind} (positive values result from Zanstra-integrals being larger
in {\sc cmfgen}). Triangles: dwarfs; asterisks: supergiants. Note that the
x-axis extends only until \Teff=50,000~K. The three objects denoted by
arrows in the lowermost panel (\Heii) correspond to the dwarf models `3V'
and `4V' and the supergiant model `1Ia', respectively. The resulting
differences for $\Delta \log Q_{\rm HeII}$ are given beneath the arrows.} 
\label{qzcomp1_cmfgen}
\end{figure}

The corresponding temperature profiles are displayed in
Fig.~\ref{comp_cmfgent}, for two dwarf and two supergiant models with
parameters similar to our comparison with {\sc wm}-Basic. Remember that the
temperature structure is derived from radiative equilibrium in {\sc cmfgen},
whereas {\sc fastwind} uses the thermal balance of electrons in the outer
atmosphere. Overall, the differences are small, and the extent of the 
temperature bumps are comparable. The only disagreement is found in the
outer wind, where {\sc fastwind} uses an artificial cut-off ($T_{\rm
min}=$0.4 \Teff) in order to prevent numerical problems at lower effective
temperatures. We have convinced ourselves that this cut-off has no further
consequences for the models as described here, which neglect adiabatic cooling
in the outer wind anyway.

Fig.~\ref{comp_cmfgenflux} compares the corresponding EUV-fluxes, in analogy
to Fig.~\ref{comp_adiflux}. As already discussed in Sect.~\ref{compwmbasic},
the largest differences occur in the \Heii-continua. This effect can be
seen even clearer in Fig.~\ref{qzcomp1_cmfgen}, lowermost panel. Regarding 
the supergiants, the deviation is contrary to our comparison with {\sc
wm}-Basic. The {\sc wm}-Basic \Heii-fluxes were mostly lower than
those from {\sc fastwind}, whereas the {\sc cmfgen}-fluxes are larger, particularly
at the edges, so that the corresponding Zanstra integrals become larger as
well. Thus, the {\sc fastwind} results for $Q_{\rm HeII}$ lie roughly in the
middle of the results from {\sc cmfgen} and {\sc wm}-Basic, at
least for the supergiants. Again, we like to point out the extreme
sensitivity of the model predictions in this frequential range and warn the
reader about any uncritical use of corresponding results, e.g., with
respect to nebula modeling.

Regarding the dwarf models, both codes give more or less identical results
for the \Heii-continua for \Teff $<$ 36,000~K, whereas at hotter
temperatures extreme differences are found for the two models at \Teff =
41,000~K and 43,500~K, respectively. In contrast to both our predictions and
those from {\sc wm}-Basic the {\sc cmfgen}-models do not show any
\Heii-edge at all, cf.~Fig.~\ref{comp_cmfgenflux}, model ``3V''.

\begin{figure*}
\begin{minipage}{18cm}
\resizebox{\hsize}{!} 
  {\includegraphics{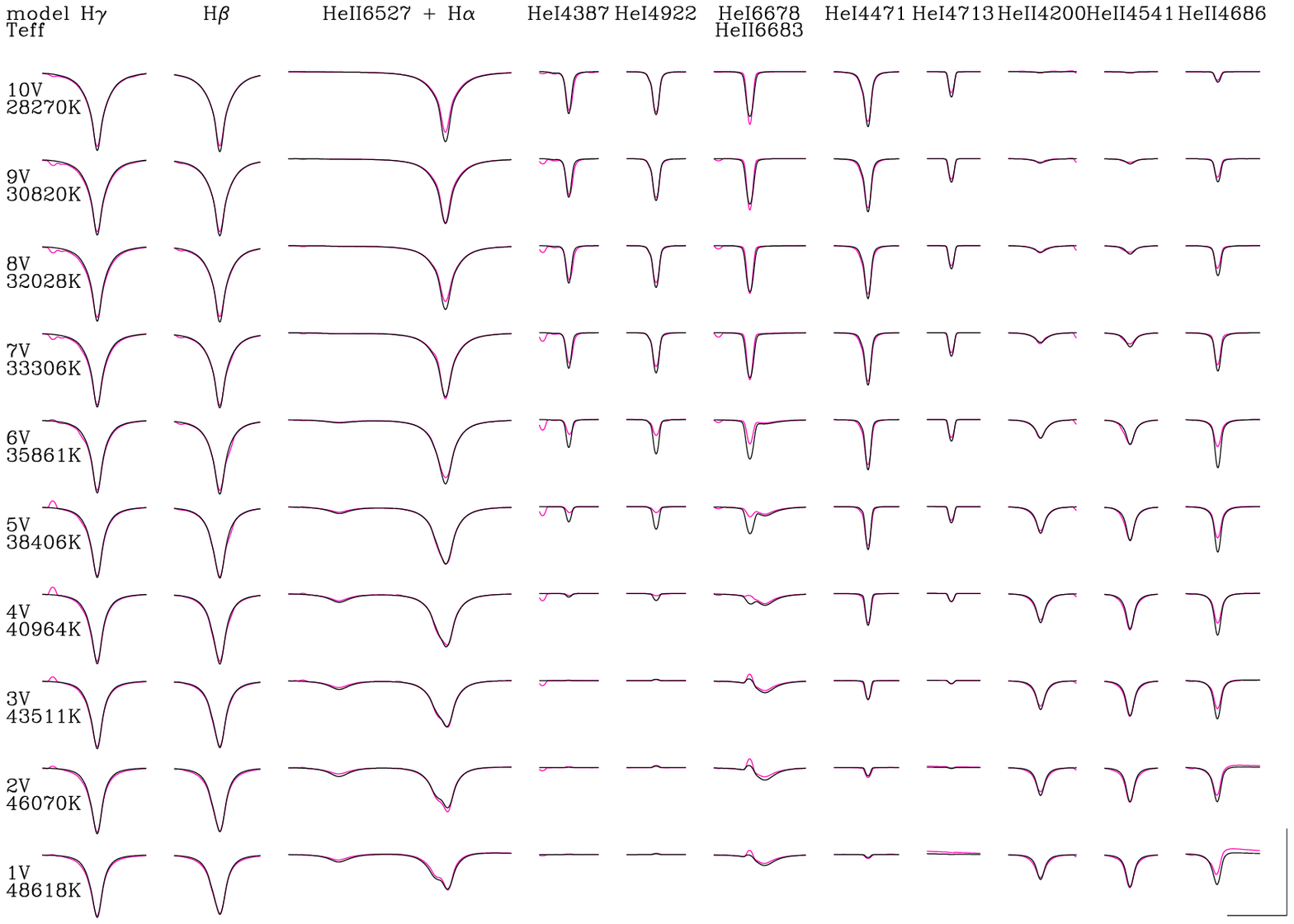}} 
\caption{{\sc fastwind} (black) vs. {\sc cmfgen} (magenta): comparison of strategic H/He
lines in the optical for the dwarf-models from the grid by \citet{len04}.
For both models, the lines have been degraded to a resolution of 10,000 and
rotationally broadened with $v\sin i$=80~\kms. \Hei
$\lambda \lambda$~4387, 4922 and 6678 are singlet lines, and \Hei $\lambda
\lambda$ 4471 and 4713 are triplets. The horizontal and vertical lines
in the bottom right corner indicate the scale used and correspond to 20~\AA\
in wavelength and 0.5 in units of the continuum, respectively (extending
from 0.65 to 1.15).} 
\label{optcompdw}
\end{minipage}
\hfill
\begin{minipage}{18cm}
\resizebox{\hsize}{!} 
  {\includegraphics{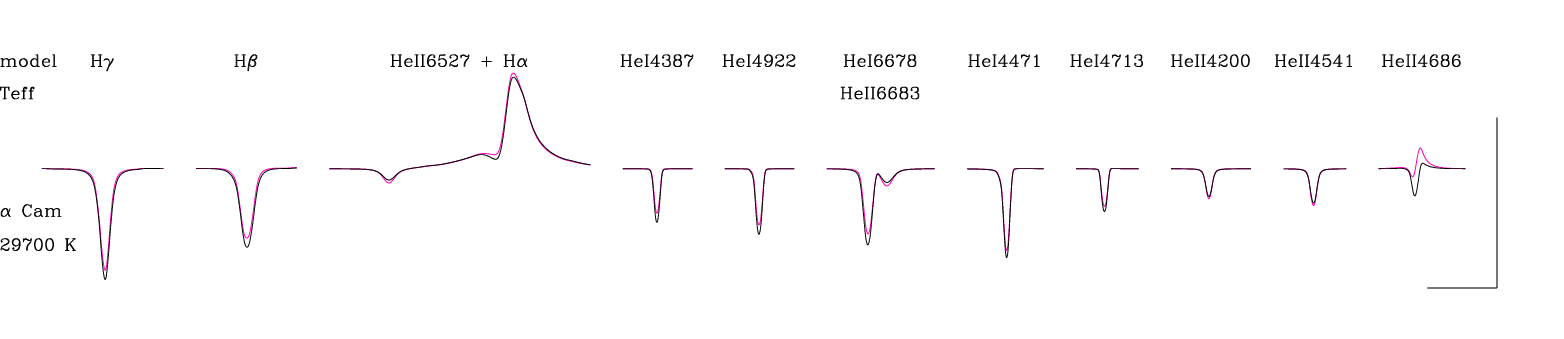}} 
\caption{Wind-strength parameter $Q$ as an optical depth invariant: H/He
profiles for the model of $\alpha$ Cam as determined by \citet{Repo04}, with
\Mdot=6.04\Mdu\, and \Rstar=32.5\Msun.  Overplotted in magenta are the
corresponding profiles for a model with identical $Q$-parameter
(Eq.~\ref{defQ}) but different mass-loss rate and radius
(\Mdot=3.3\Mdu\, and \Rstar=21.7\Msun).}
\label{compacam}
\end{minipage}
\end{figure*}

Concerning the \Oii-continua (actually, for the complete range within 
300~\AA\, $< \lambda <$ 400~\AA), the hotter models (\Teff $>$ 35,000~K)
show a higher flux-level in {\sc cmfgen}, for both the supergiants and the
dwarfs. We have already commented on this problem in
Sect.~\ref{compwmbasic} and speculated that this behaviour is related to
missing line-opacity. (\Oii itself plays no role at these temperatures). 
Of course, we cannot exclude a problem in our
approximate treatment of line-blocking. Finally, and in accordance with the
comparison with {\sc wm}-Basic, the agreement of the \Hi- and \Hei-continua
is almost perfect.

\begin{figure*}
\begin{center}
\begin{minipage}{16cm}
\resizebox{\hsize}{!}
   {\includegraphics{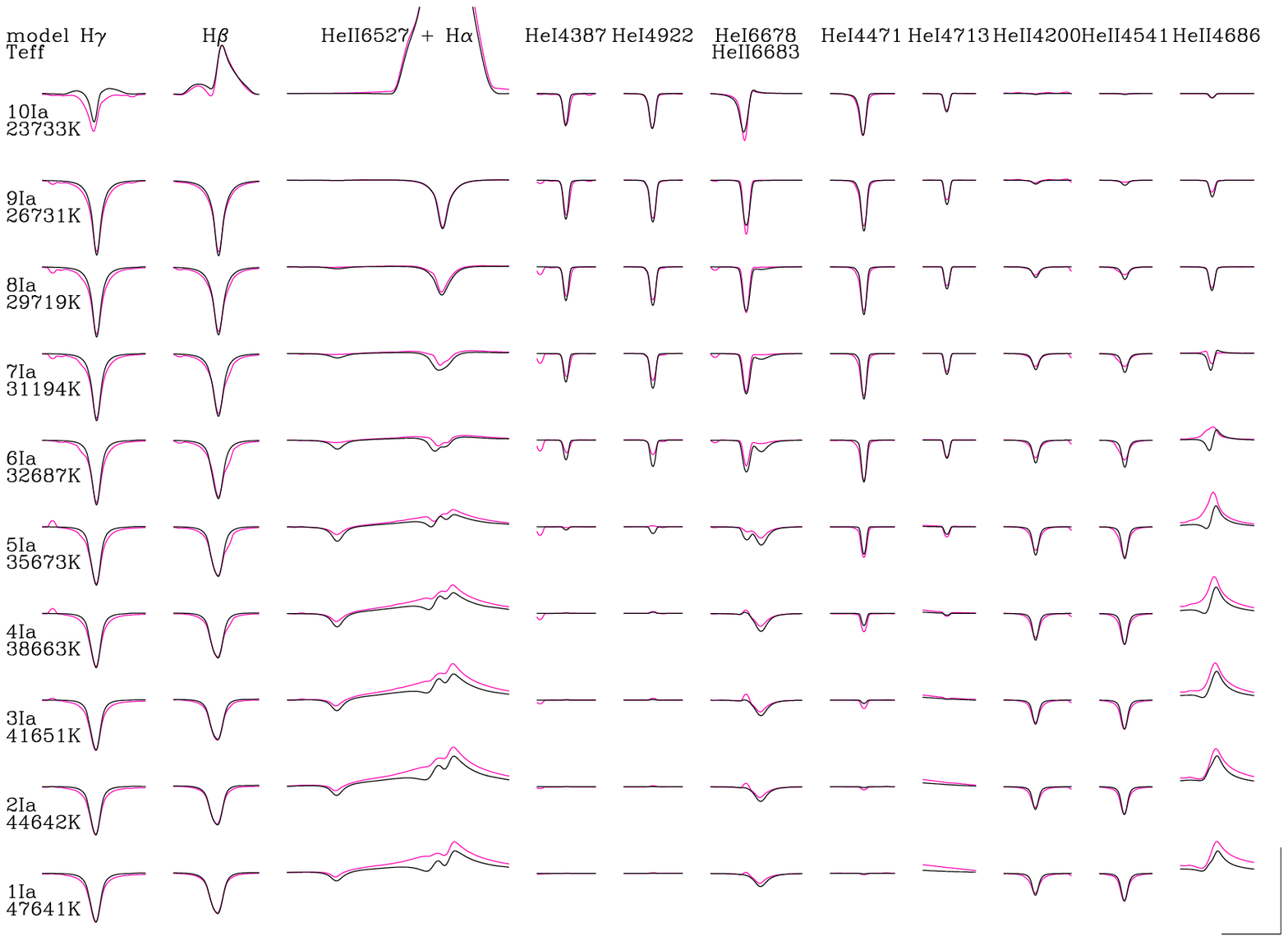}}
\end{minipage}
\hfill
\begin{minipage}{16cm}
   \resizebox{\hsize}{!}
      {\includegraphics{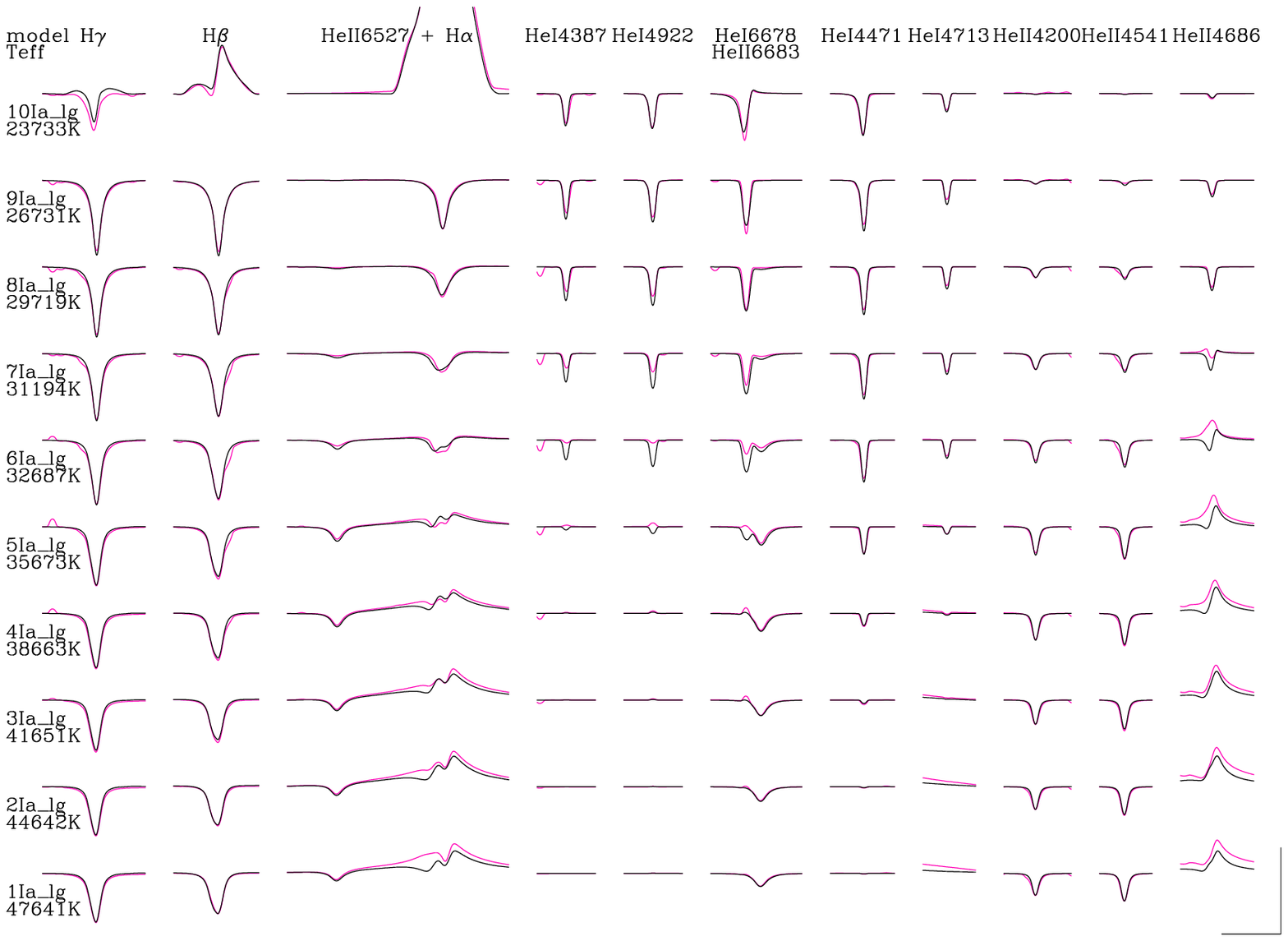}}
\end{minipage}
\end{center}
\caption{As Fig.~\ref{optcompdw}, however for the supergiants from the grid
by \citet{len04}. Upper panel: {\sc cmfgen} models with ``standard''
gravities.  Note that the differences in the wings of the
Balmer lines and in the \Ha-emission almost vanish if our results
are compared to the ``low gravity'' {\sc cmfgen} models in the lower panel
(see text).} \label{optcompsg}
\end{figure*}

Figs.~\ref{optcompdw} and \ref{optcompsg} are now the most interesting plots
in this section, displaying the strategic H/He lines in the optical ({\sc
cmfgen}-profiles in magenta). Regarding the dwarfs, the agreement of almost all
lines is excellent. The only differences are found for the line cores of
\Heii 4686, which are shallower in {\sc cmfgen} at almost all temperatures,
and for the \Hei\, singlets for models `4V' to`6V' with \Teff lying in the
range between 41,000~K and 36,000~K, respectively. (Note that for model `4V'
\Hei 4387 agrees well whereas \Hei 4922 and \Hei 6678 differ). Most prominent
are the differences for models `5V' and `6V' (the same is true for the giant
models not displayed here), where all singlet lines predicted by {\sc
cmfgen} are almost a factor of two smaller in equivalent width than those 
predicted by {\sc fastwind}. Most interestingly, however, the triplet
lines agree perfectly throughout the grid.

So far, the origin of this discrepancy could not be identified;
particularly, the atomic data used (incl. broadening functions) are very
similar, and also the ionizing continua (important for the
singlet-formation) agree very well, as shown above. One might speculate that
there is a connection to the flux differences around the \Heii\, resonance
line at 304~\AA\, or to possible discrepancies at the \Hei\, resonance
line(s), but this has to be checked carefully (investigations under way).
Further comments on this discrepancy will be given after we have discussed
the results for the supergiants.

The corresponding profiles are displayed in Fig.~\ref{optcompsg}, upper
panel. There, the situation is somewhat different from the dwarf case. At
first, we note that the deviations of the \Hei singlets are not as extreme
as before. Significant disagreement is found only for \Hei 4922 and 6678 (no
problem for \Hei 4387) in model ``5Ia'' (36,000~K), where these singlets are
weak anyway. For model ``6Ia'' the differences are moderate, much less than
the factor of two in equivalent width encountered above. Noticeable
differences are found for other lines though. At first, the hydrogen Balmer
line wings predicted by {\sc cmfgen} are much stronger, which would lead to
lower gravities if an analysis of observed spectra were performed. Second,
both \Ha and \Heii 4686 show stronger wind emission which would lead to
lower mass-loss rates compared to {\sc fastwind}. Note however that the 
wind emission in both lines is a strongly increasing function of mass-loss
(e.g., \citealt{puls96}), and an analysis of observed spectra 
would result in \Mdot-differences not exceeding the 20 to 30\% level.

The difference in the Balmer line wings points to a problem mentioned 
above, namely the assumption of a constant photospheric scale height
in {\sc cmfgen}. In order to obtain an impression in how far this
approximation (as well as the somewhat artificial transition from
photosphere to wind) has an influence on the resulting models and profiles,
\citet{len04} have calculated an additional set of ``low-gravity''
supergiants, where the gravity has been lowered by typically 0.1 to 0.2 dex
(model series ``\_lg'') with respect to their ``standard'' grid of
supergiants. Due to this manipulation, at least part of the effect of
photospheric radiation pressure $g_{\rm rad}$ is accounted for (although
this quantity is {\it not} constant throughout the photosphere), since the
profiles provide a measure of the {\it effective} gravity (i.e., $g_{\rm
grav}-g_{\rm rad}$) alone. 

In Fig.~\ref{optcompsg}, lower panel, we compare the {\sc fastwind} profiles
(identical to those from the upper panel, since our ``high gravity'' models
{\it do} include the photospheric $g_{\rm rad}$) with these low-gravity
models calculated by {\sc cmfgen}. Consequently, the photospheric densities
should be much more similar than in the previous case, at least in those
regions where the Balmer line wings are formed. Indeed, the differences in
H$_\gamma$ and H$_\beta$ have now vanished, and also the \Ha emission is
very similar, except for the hottest models on the blue side of the profile.
In some cases, the discrepancy for \Heii 4686 has become weaker as well. The
\Hei\,triplets have not changed (they seem to be almost independent on the
photospheric density in {\sc cmfgen}), whereas a strong influence on the
\Hei singlets is found. In the ``critical'' temperature region, they have
become significantly weaker, and a strong discrepancy also for the
low-gravity model `6Ia\_lg' is present again, by the same degree as we have
found for the dwarfs. 

In summary, we find a very good agreement with the optical spectra from
{\sc cmfgen} if the problem of different density stratifications is
accounted for. The only disturbing fact is the strong difference in the
\Hei singlets for dwarfs between 36,000 to 41,000~K and for supergiants
between 31000 to 35,000~K.

Although it is presently not clear which profiles are ``correct'' or whether 
the truth lies in the ``middle'', we like
to point out the following. In our analyses of Galactic O-stars
(\citealt{Repo04}), no problems were found to match both the observed
singlet and triplet lines in dwarfs. Concerning the supergiants, we actually
met a problem for almost {\it all} stars cooler than O6, namely the
well-known ``generalized dilution effect'' (see the discussion and
references in \citealt{Repo04}). Briefly, we could fit all \Hei\,
lines (singlets and triplets) in parallel with the \Heii\, lines, except for
\Hei 4471 (triplet) which was predicted to be too weak. One might argue that this
is a symptom of generally incorrect \Hei\, lines, and speculate that this
problem is related to the inconsistency seen here. Assuming that the
\Hei-singlets produced by {\sc fastwind} are erroneous it might then be
possible to fit all \Hei\, singlets and the $\lambda$ 4471 triplet at cooler
temperatures. In this case, however, we (and {\sc cmfgen}!) would meet the 
problem that the other triplet lines would be too strong and the \Heii lines
too weak! 

Presently, there is no other way out of this dilemma than to perform a
number of detailed comparisons, with respect to both the models and the
observations. Since the actual problem concerns the {\it ratio} of triplet
to singlet lines and the problem is most pronounced for dwarfs, it should
be possible to find a solution by comparing the theoretical predictions for
this ratio (in terms of equivalent widths) as a function of \Teff vs. the
observed ratio as a function of spectral type for a large sample of stars.
Such work is in progress now.

\begin{figure*}
\begin{minipage}{8.8cm}
\resizebox{\hsize}{!}
   {\includegraphics{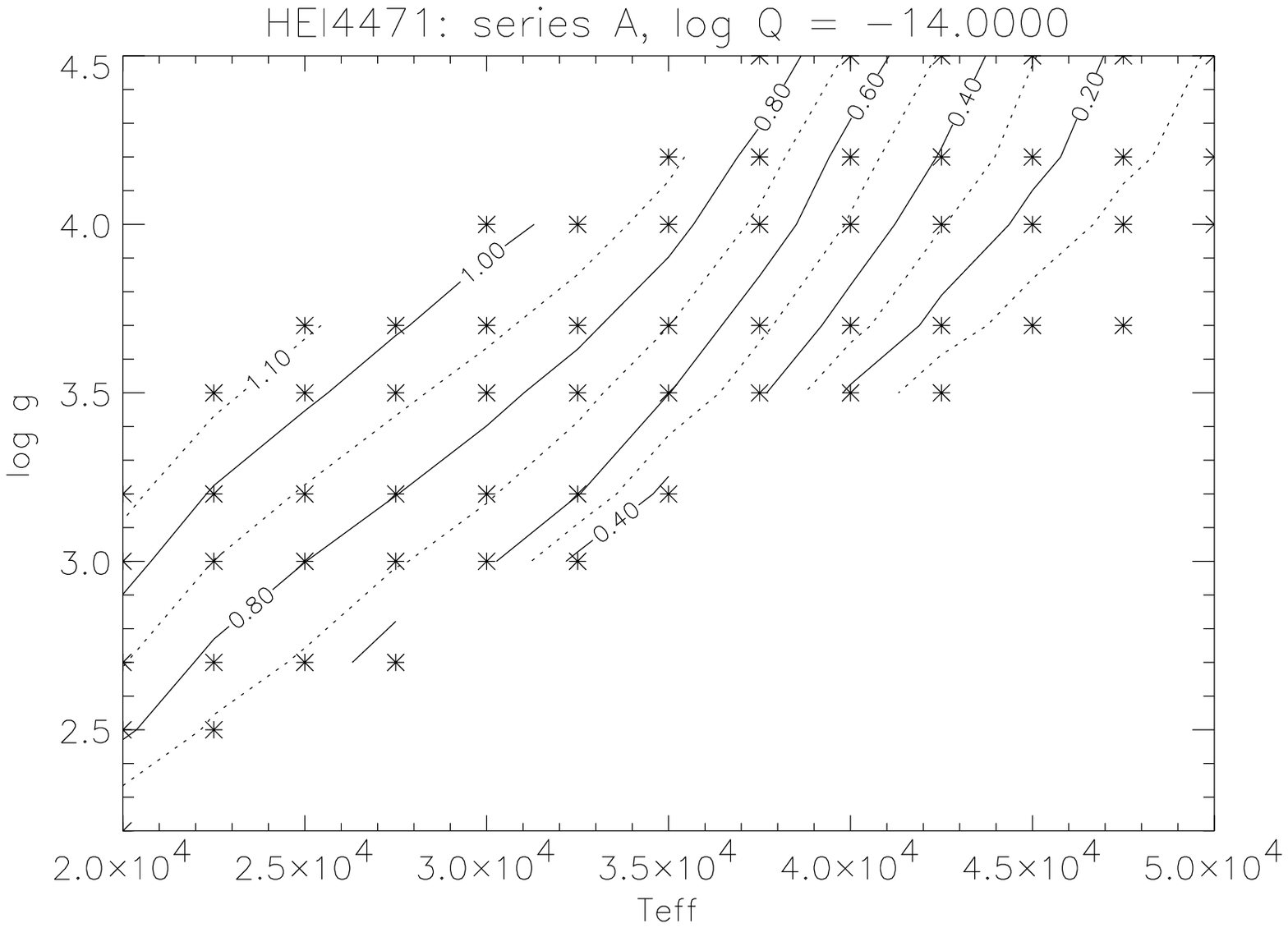}}
\end{minipage}
\hfill
\begin{minipage}{8.8cm}
   \resizebox{\hsize}{!}
      {\includegraphics{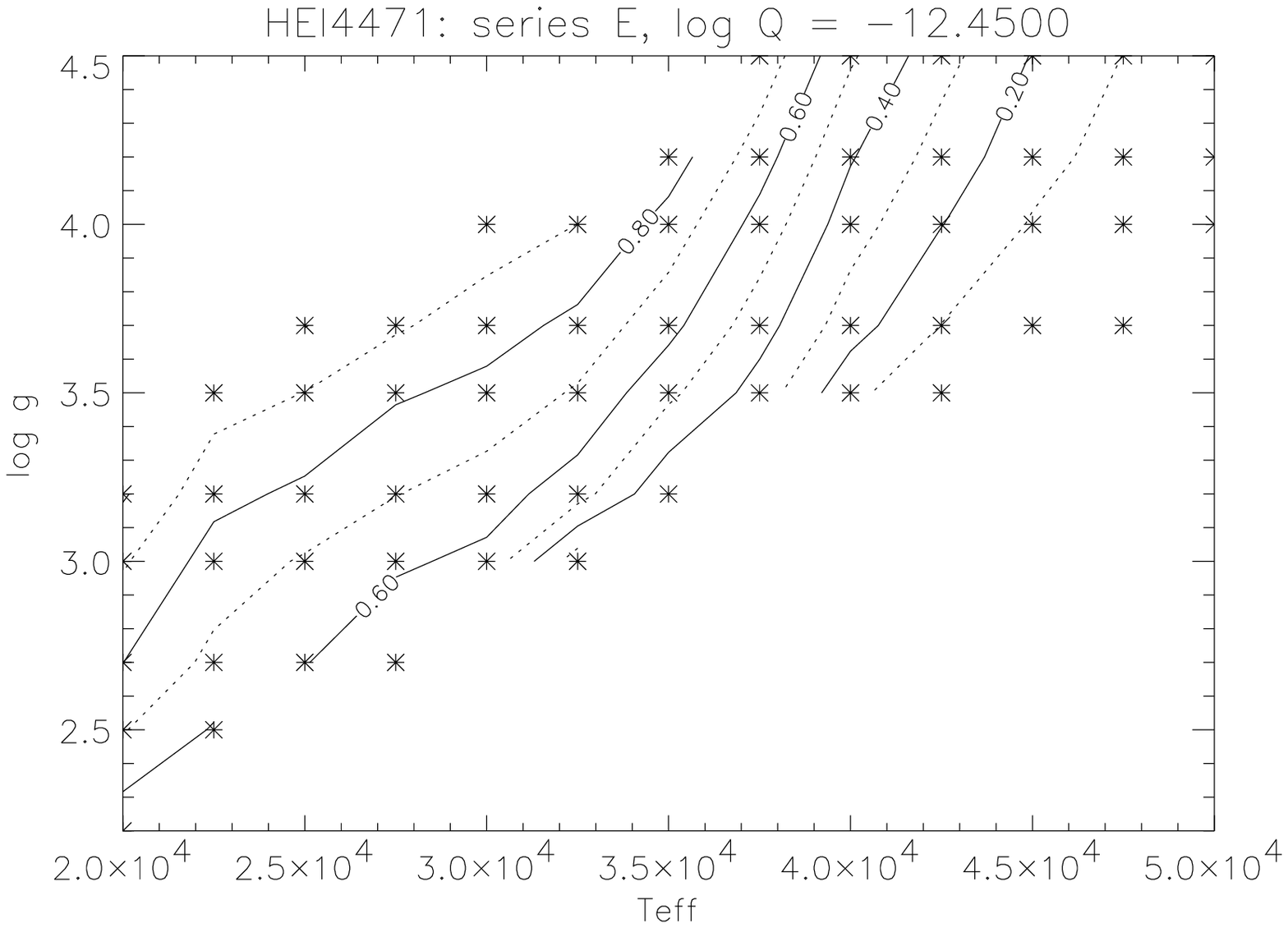}}
\end{minipage}
\caption{Iso-contours of equivalent widths for \Hei4471, as predicted by
{\sc fastwind}, using results from our model-grid with Helium abundance \Yhe
= 0.1 and solar metallicity for background elements. Left: negligible wind.
Right: typical O-supergiant wind. Note the effect of wind-emission, shifting
the iso-contours to the left. The locations of the corresponding models are 
indicated by asterisks.} 
\label{isoew}
\end{figure*}

\section{Model grids}
\label{modelgrids}

As already outlined in Sect.~\ref{intro}, the parameter space to be
investigated for the analysis of one object alone is large and almost
prohibitive for the {\it detailed} analysis of very large samples of stars
which have recently been collected (e.g., by means of the multi-object
spectrograph {\sc flames}). Alternatively, a somewhat coarser analysis by
means of the ``traditional'' model-grid method is still applicable if an
appropriate grid can be constructed. In this section, we will give some
suggestions for this objective and report on first progress.

Although the presence of a wind introduces a large number of additional
parameters to be considered in a fine fit (\Mdot, \vinf, $\beta$ and
\Rstar), there is a fortunate circumstance which allows for the construction
of such model-grids with only {\it one} more parameter compared to grids
from hydrostatic, plane-parallel models, at least if we do not aim at the
analysis of specific (UV) resonance lines. 

As has been shown by, e.g., Puls et al. (1996, see also 
\citealt{Schmutzetal89, deKoteretal98} for diversifications), the
wind-emission from recombination dominated transitions (so-called
$\rho^2$-lines) remains rather unaffected from the specific choice of the
individual values of \Mdot, \vinf and \Rstar as long as the wind-strength
parameter $Q$ (also denoted as the ``optical depth invariant''),
\beq
Q=\frac{\Mdote}{(\vinfe \Rstare)^{\frac{3}{2}}},
\label{defQ}
\eeq
does not vary. In this case then, most of the other lines also preserve
their shape. An example is given in Fig.~\ref{compacam}, where we have
varied the mass-loss rate of a model of $\alpha$ Cam (cf. \citealt{Repo04})
by a factor of two (and accordingly the radius by $2^{1.5}$) without almost
any effect on the resulting H/He spectrum.

This behaviour (i.e., spectrum (and emergent fluxes!)  
depend almost exclusively on $Q$ and not on its individual 
constituents) follows from the fact that
\begin{itemize}
\item $\rho^2$-dependent line processes (e.g., recombination
lines and resonance lines from ions one stage below the major
one\footnote{e.g., \Siiv in most hot stars}) scale with $Q$ in the wind regime,
\item the wind density scales with $\Mdote/(\vinfe \Rstare^2)$ (continuity
equation), and
\item (resonance) lines from major ions scale with $\Mdote/(\vinfe^2
\Rstare)$, e.g. \citet{hamann81}.
\end{itemize}
Thus, the common power of ``1.5'' with respect to \vinf and \Rstar used in
$Q$, which refers to the scaling of $\rho^2$ lines, is also the best
compromise to deal with the other physical parameters affecting a stellar
model (most importantly, the line-blocking which depends both on
density and line opacity).

Exploiting this knowledge, we have constructed a set of nine model-grids
for the analysis of H/He profiles with three different helium abundances, 
\Yhe = 0.1, 0.15 and 0.2, and three different background metallicities, z =
1.0,0.5 and 0.2 (cf. Sect.~\ref{atomdat}), respectively. Each grid with
given helium abundance and metallicity is three-dimensional with respect to
the parameters \Teff, \logg\, and $\log Q$, and the grid-spacing is roughly
equidistant. The individual values for parameters incorporated into $\log Q$
(which are actually needed to calculate a specific model) and additional
ones have been assumed according to present knowledge:
\begin{itemize}
\item \Rstar from ``empirical'' values, as a function of spectral type (\Teff) 
and luminosity class (\logg). 
\item \vinf as a function of photospheric escape velocity $v_{\rm esc}$,
\beq
\vinfe = C(\Teffe) \cdot v_{\rm esc},
\eeq 
in accordance with the results collected by \citet{kp00}.
\item Velocity exponent $\beta$ from empirical values (see also \citealt{kp00}
and references therein), with $\beta = 0.9$ (as a compromise) for O-stars
and increasing values towards later types.
\item \Mdot from $\log Q$, \Rstar and \vinf as specified above.
\item micro-turbulence $v_{\rm turb}$ = 15 \kms throughout the grid as a
compromise between O and B stars.
\end{itemize}
Our present grids comprise the range 20,000~K $\le$ \Teff $\le$ 50,000~K with
$\Delta$ \Teff = 2,500~K, \logg between 2.2 $\le$ \logg $\le$ 3.2 at \Teff =
20,000~K and 4.0 $\le$ \logg $\le 4.5$ at \Teff = 50,000~K. 

The position of all models can be inferred from Fig.~\ref{isoew}. With
respect to $\log Q$ we have used values with -14.0 $\le$ $\log Q$ $\le$
-11.4 ($\Delta \log Q$ = 0.35 in most cases), where the lowest value
corresponds to an almost negligible wind and the highest one to almost 
Wolf-Rayet conditions. 

The denotation is such that we specify a letter for
the wind density (``A'' to ``H'', with densities $\log Q$ = -14.0,
-13.5,-13.15,-12.8,-12.45,-12.1,-11.75,-11.4, respectively, if \Mdot is 
calculated in \Msun/yr, \vinf in \kms and \Rstar in \Rsun).
Effective temperature and gravity are denoted by two ciphers each. Thus,
model ``E2730'' refers to  $\log Q$ = -12.45, \Teff =
27,500~K and \logg = 3.0. Typical O-type supergiants correspond to series
``E'', and typical B-type supergiants to series ``D''.

For all these models we have calculated  H/He profiles and equivalent widths
in the optical and the IR. Thus, by simply over-plotting observed vs.
simulated spectra one finds an immediate guess for the parameters
\Teff, \logg, \Yhe and wind-strength if the background metallicity is
specified and the theoretical profiles have been convolved accounting for
rotational broadening and resolution. In this way, the coarse analysis of one
star is possible within a couple of minutes and might be fine-tuned by
calculating specific models (particularly with respect to $\beta$ if
inferable from the emission line shapes).

In addition, a plot of various iso-contours of calculated equivalent widths
gives deeper insight into certain dependencies. As an example,
Fig.~\ref{isoew} shows the effect of wind emission on \Hei 4471. Further
examples, particularly with respect to the spectral type classification
criterium of O-star, $\log W' = \log ({\rm ew} 4471) - \log ({\rm ew} 4541)$,
are given in \citet{massey05}. 

We intend to make these grids publicly available in the near future when 
the problem regarding the \Hei singlets has been solved.

\section{Summary and outlook}
\label{summary}

In this paper we have described all updates applied to our previous version 
of {\sc fastwind} (Paper~I), regarding the approximative treatment of metal
line-blocking/blanketing and the calculation of a consistent temperature
structure. 

The problem of line-blocking has been tackled in two steps.  First, the
occupation numbers of back-ground elements are calculated by an
approximative solution of the corresponding equations of statistical
equilibrium with the option that the most abundant elements are treated 
almost ``exactly'', i.e., by means of the Sobolev transport for line
processes. Compared to alternative approaches (cf. Sect.~\ref{nlteapprox})
our method allows for the treatment of different spin systems, radially and 
frequency dependent radiation temperatures and a consistent ALI-iteration
scheme. We have tested our solutions by comparing the approximative results
with results from exact solutions and have not found any major
discrepancies.

The occupation numbers derived in this way are subsequently used to
calculate the line-blocked radiation field, again, in an approximative way.
To this end, we have formulated suitable means for the opacities (in analogy
to Rosseland means but for frequency intervals not larger than 1,000{\ldots}
1,500 \kms) and emissivities (two-level-atom approach), and the resulting
pseudo-continuum of overlapping lines is treated by means of a conventional
continuum radiative transfer. Specific problems inherent
in our approach (regarding a rigorous statistical description) have been
pointed out and might lead to inaccurate solutions in a few cases.
Investigations to improve our approach are presently under way in our group,
as discussed in Sect.~\ref{lineblock}.

Our new version of {\sc fastwind} allows for the calculation of a consistent
temperature structure by applying a flux-correction method in the lower
atmosphere and the electron thermal balance in the outer one. Regarding
optical H/He lines, no major differences have been found compared to our
previous NLTE Hopf-function method (cf. Paper~I and \citealt{Repo04}).

Due to the approximations applied and as intended, the performance of our
code is very fast. The total computational time (starting all models from
scratch) is of the order of 30 minutes on a PC with a 2 GHz processor if
only H and He lines are considered as explicit ions, whereas the inclusion
of other elements (e.g., \citealt{urban04}) into the ``explicit'' treatment
requires additional 5 to 10 minutes each.

The new methods have been extensively tested by comparing with results from
{\sc wm}-Basic and {\sc cmfgen}, concerning temperature stratification,
fluxes, number of ionizing photons and  
optical\footnote{IR-lines will be presented in a forthcoming paper
(Repolust et al., in prep. for A\&A), with a similar agreement between {\sc
fastwind} and {\sc cmfgen} as for the
optical ones.} H/He profiles (comparison with {\sc cmfgen} only).

We have highlighted the importance of photospheric line-pressure, which is
incorporated into the {\sc fastwind} models and neglected in the standard
version of {\sc cmfgen}, if not coupled to the plane-parallel code {\sc
tlusty} (see Sect.~\ref{intro}). Particularly, we have found indications
that the use of the Sobolev approximation (within the force-multiplier
concept) in {\sc wm}-Basic can lead to an underestimate of this quantity, as
already predicted by \citet{PPK}. On the other hand, the density/velocity
stratification resulting from our approach (smoothly connecting the
quasi-static photosphere and a $\beta$-law wind) agrees surprisingly well
with the hydrodynamic structure as calculated from a consistent solution if 
$\beta$ is not too different from the ``canonical'' value of 0.8{\ldots} 1.0.

All three codes predict almost identical temperature structures and
fluxes for $\lambda >$ 400~\AA, whereas at lower wavelengths certain discrepancies
are found. Compared to {\sc wm}-Basic (using an identical line list for the
background elements), our {\it supergiant} models differ only in the \Heii\,
continua, where the {\sc fastwind}-fluxes are somewhat larger, but still
lower than the corresponding fluxes from {\sc cmfgen}. Since fluxes and
corresponding numbers of ionizing photons react extremely sensitive to subtle
model differences in this wavelength regime, we consider any uncritical use of
these quantities as being dangerous. 

Major discrepancies are also found in the range 300~\AA\, $<$ $\lambda$ $<$ 
400~\AA, i.e., in the \Oii\, continuum and at the \Heii 304 resonance line.
Compared to both {\sc wm}-Basic and {\sc cmfgen}, our {\it dwarf} models
produce less flux in this region (more blocking or less re-emission), whereas
the {\it supergiant} models of {\sc fastwind} and {\sc wm}-Basic agree very
well. The supergiant models of {\sc cmfgen}, on the other hand, show much
less blocking which might point to some missing opacity. Again, we like to
stress that the \Hi\, and \Hei\, continua agree very well in all three codes.

For the optical H/He lines, the coincidence between {\sc fastwind}
and {\sc cmfgen} is remarkable, except for the \Hei\, singlets in the
temperature range between 36,000 to 41,000~K for dwarfs and between 31,000
to 35,000~K for supergiants, where {\sc cmfgen} predicts much weaker singlets.
Up to now, the origin of this discrepancy could not be identified, but work 
is under way to solve this problem.

Although it is reassuring that the different codes agree well with respect
to most of their predictions, this is only part of the story. One
particularly disturbing fact concerns the present mismatch between the
parameters obtained from an analysis in the optical and the UV,
respectively. In the majority of cases, the UV gives lower effective
temperatures, i.e., of the order of 2,000 to 4,000~K, if one compares the analyses
of Galactic stars performed by \citet{bg02} and \citet{gb04} with results
from \citet{Repo04} ({\sc wm}-Basic vs. {\sc fastwind}) and the corresponding
work for Magellanic Cloud stars by \citet{Hillier03} and
\citet{bouret03} (partly including also the optical range) with the results
from \citet{massey04, massey05} ({\sc cmfgen} vs. {\sc fastwind}).
(Interestingly, the work by \citealt{crow02} ({\sc cmfgen}) indicates 
higher temperatures for MC supergiants than derived by \citealt{massey05}.)

Part of this discrepancy (if combined UV/optical analyses are compared) 
might be related to the \Hei\, singlet vs. triplet problem as discussed
above. Note, however, that this would account only for discrepancies in
certain domains of the \Teff space and would typically result in maximum 
differences of the order of 2,000~K, as has been found from a number of test
calculations performed by one of us (J.P) and F.~Najarro (using {\sc
cmfgen}), which will be reported on in a forthcoming publication.
Moreover, the temperature scale for O-type dwarfs as derived by
\citet{martins02} using {\sc cmfgen} and concentrating on the classification
criterium \Hei4471 (triplet) vs. \Heii4541 is actually 1,000 to 2,000~K
{\it hotter} than the calibration by \citet{Repo04}.

In a recent paper, \citet{martins04} have discussed the uncertainties in
\Teff which is obtained by relying on different diagnostic tools in the UV,
analyzing four SMC-N81 dwarfs of spectral types O6.5 to O8.5. From
the specific values derived from the UV-color index, the
ionization balance of O{\sc iv}/{\sc v} and Fe{\sc iv}/{\sc v} and the 
N{\sc v}1238/1242 and C{\sc iii}1426/1428 doublets, respectively, 
they quote a typical uncertainty of $\pm$ 3,000~K in \Teff, 
which might easily account for part of the discrepancies with the optical.

Unfortunately, it is rather difficult to compare the differences obtained so
far in a strict one-to-one case, simply because the corresponding samples
hardly overlap. In particular, a large fraction of the objects which have
been analyzed by means of {\sc cmfgen} are somewhat extreme, comprising
either supergiants with (very) dense winds (\citealt{crow02}) or dwarfs with
very thin winds (\citealt{martins04}). The analysis of SMC stars by
\citet{bouret03}, on the other hand, covers only a sample of 6 dwarfs, in
contrast to the larger sample by \citet{massey04, massey05}, and, therefore,
it is not clear in how far selection effects do play a role. Finally, it is
interesting to note that at least for one object in common, the O4I(f) star
$\zeta$ Pup (HD\,66811), the different analyses give almost identical
results (\citealt{crow02}, \citealt{Repo04} and Pauldrach et al., in prep.
for A\&A, analyzing the UV by means of {\sc wm}-Basic).

Thus, we conclude that the present status of hot star parameters is not as
clear as we would like it to be. Actually, we need to understand a number of
additional physical processes and their influence on the derived parameters.
Most important are the direct and indirect effects of the line-driven wind
instability, i.e., the formation and interaction of clumps and shocks
leading to X-ray emission and enhanced EUV-flux in the wind (e.g.,
\citealt{Feld97, Paul01}). Although incorporated to some extent into present
codes, there are simply too many questions to be answered before we can
consider these problems as solved. To give only two examples: We do not know
the spatial distribution of the ``clumping factor'', and also the X-ray
emission is only on the verge of being understood (e.g.,
\citealt{krameretal03, oskinova04}). 

Before these effects can be treated in a realistic way, we suggest to
primarily rely on diagnostic tools which are least ``contaminated'', i.e.,
to concentrate on weak lines formed in the stellar photospheres (except, of
course, the mass-loss indicators which will always be affected by
clumping). Future investigations of O-type stars performed by
{\sc fastwind} will have to utilize not only H and He but also metal lines,
as already incorporated into the analysis of B-stars (cf. Sect~\ref{intro}).
Particularly, one of the most important tools will be nitrogen with its
strong sensitivity even at higher temperatures where \Hei\, begins to fail.
Work in this direction is under way.

\begin{acknowledgements}
We like to thank a number of colleagues for their enduring willingness to
discuss problems and provide assistance. The most important colleague in
this respect was and is Dr. Keith Butler, the living compendium in atomic
physics, since without his support it would have been extremely difficult to
finish this project. Particular thanks to Dr. Adi Pauldrach for providing 
his atomic data base (tailored for {\sc wm}-Basic). Many thanks also to Drs.
Paco Najarro and Tadziu Hoffmann for performing a number of test
calculations with {\sc cmfgen} and {\sc wm}-Basic, respectively, and their
stimulating discussions concerning NLTE-effects. Finally, we would like to
thank both our anonymous referee and Drs. Phil Massey, Alex de Koter and
John Hillier for valuable comments on the manuscript.

J.P. appreciates support by NATO Collaborative Linkage Grant No. PST/CLG
980007. M.A.U. acknowledges financial support for this work by the Spanish
MCyT under project PNAYA2001-0436, R.V. acknowledges support from the
University of La Plata by a FOMEC grant (Pr. 724/98), and T.R. gratefully
acknowledges financial support in form of a grant by the International
Max-Planck Research School on Astrophysics (IMPRS), Garching.

\end{acknowledgements}

\end{document}